\documentclass[lettersize,journal]{IEEEtran}
\usepackage{amsmath,amsfonts}
\usepackage{algorithmic}
\usepackage{array}
\usepackage[caption=false,font=normalsize,labelfont=sf,textfont=sf]{subfig}
\usepackage{textcomp}
\usepackage{adjustbox}
\usepackage{stfloats}
\usepackage{graphicx}
\usepackage{transparent}
\usepackage{float}
\usepackage{multirow}
\usepackage{tabularx}
\usepackage{longtable}
\usepackage[hidelinks]{hyperref}
\usepackage{booktabs}
\usepackage{array}
\usepackage{makecell}
\usepackage[most]{tcolorbox}
\usepackage{pifont}
\usepackage{url}
\usepackage{verbatim}
\usepackage{graphicx}
\usepackage{xltabular}
\usepackage{arydshln}
\usepackage{multirow}
\usepackage{booktabs}
\usepackage{stfloats}
\tcbuselibrary{breakable}
\usepackage[table]{xcolor} % 用于背景色

\definecolor{groupgray}{RGB}{242,242,242}
\begin{document}
\title{Software Development Life Cycle Perspective: A Survey of Benchmarks for Code Large Language Models and Agents}

\author{Kaixin~Wang,
        Tianlin~Li,
        Xiaoyu~Zhang,
        Chong~Wang,
        Weisong~Sun,
        Yang~Liu,~\IEEEmembership{Member,~IEEE},
        Aishan Liu,
        Xianglong Liu,~\IEEEmembership{Member,~IEEE},
        Chao Shen,~\IEEEmembership{Fellow,~IEEE,}
        and~Bin~Shi% <- 注意这里不要加逗号，直接接脚注
\IEEEcompsocitemizethanks{\protect
% 第一部分：Xi'an Jiaotong University 的作者
\IEEEcompsocthanksitem Kaixin Wang, Chao Shen and Bin Shi are with Xi'an Jiaotong University, Xi'an, China.
(E-mail: kxwang@stu.xjtu.edu.cn; chaoshen@mail.xjtu.edu.cn; shibin@xjtu.edu.cn.) \protect
% 北航
\IEEEcompsocthanksitem Tianlin Li, Aishan Liu and Xianglong Liu are with Beihang University, Beijing, China.
(E-mail: tianlin001@buaa.edu.cn; liuaishan@buaa.edu.cn; xlliu@buaa.edu.cn.)
\protect
% 第二部分：Nanyang Technological University 的作者
\IEEEcompsocthanksitem Xiaoyu Zhang, Chong Wang, Weisong Sun, and Yang Liu are with Nanyang Technological University, Singapore.
(E-mail: xiaoyu.zhang@ntu.edu.sg; chong.wang@ntu.edu.sg; weisong.sun@ntu.edu.sg; yangliu@ntu.edu.sg.)
}%
\protect
\thanks{Corresponding authors: Xiaoyu Zhang and Bin Shi.}}

% \author{IEEE Publication Technology Department
% \thanks{Manuscript created October, 2020; This work was developed by the IEEE Publication Technology Department. This work is distributed under the \LaTeX \ Project Public License (LPPL) ( http://www.latex-project.org/ ) version 1.3. A copy of the LPPL, version 1.3, is included in the base \LaTeX \ documentation of all distributions of \LaTeX \ released 2003/12/01 or later. The opinions expressed here are entirely that of the author. No warranty is expressed or implied. User assumes all risk.}}

\markboth{Journal of \LaTeX\ Class Files,~Vol.~18, No.~9, S}%
{How to Use the IEEEtran \LaTeX \ Templates}

% \markboth
% {A Survey of Benchmarks for CodeLLMs and
% Agents across the Software Development Life Cycle}
% {Wang \MakeLowercase{\textit{et al.}}: A Survey of Benchmarks for CodeLLMs and Agents}

\maketitle

\begin{abstract}

Code large language models (CodeLLMs) and agents are increasingly being integrated into complex software engineering tasks spanning the entire Software Development Life Cycle (SDLC). Benchmarking is critical for rigorously evaluating these capabilities. 
However, despite their growing significance, there remains a lack of comprehensive reviews that examine these benchmarks from an SDLC perspective.
To bridge this gap, we propose a tiered analysis framework to systematically review 178 benchmarks from 461 papers, comprehensively characterizing them from the perspective of the SDLC.
Our findings reveal a notable imbalance in the coverage of current benchmarks, with approximately 61\% focused on the software implementation phase in SDLC, while requirements engineering and software design phases receive minimal attention at only 5\% and 3\%, respectively. 
% Additionally, anti-contamination strategies are largely absent from current benchmarks, leading to an increased risk of data leakage.
Furthermore, current benchmarks lack effective anti-contamination strategies, posing significant risks of data leakage and potentially inflated performance assessments.
Finally, we identify key open challenges in current research and outline future directions to narrow the gap between the theoretical capabilities of CodeLLMs and agents and their practical effectiveness in real-world scenarios.
All survey materials and data are published in our repository (\url{https://github.com/whappily/sdlc-benchmark-survey}).
\end{abstract}

% Additionally, Python emerges as the dominant programming language across the reviewed benchmarks.

% \ccsdesc[500]{Do Not Use This Code~Generate the Correct Terms for Your Paper}
% \ccsdesc[300]{Do Not Use This Code~Generate the Correct Terms for Your Paper}
% \ccsdesc{Do Not Use This Code~Generate the Correct Terms for Your Paper}
% \ccsdesc[100]{Do Not Use This Code~Generate the Correct Terms for Your Paper}

%%
%% Keywords. The author(s) should pick words that accurately describe
%% the work being presented. Separate the keywords with commas. 
% \keywords{Do, Not, Us, This, Code, Put, the, Correct, Terms, for, 
%   Your, Paper}
% \keywords{Code Large Language Models, Software Development Life Cycle, Benchmarks, Software Engineering}

% \received{20 February 2007}
% \received[revised]{12 March 2009}
% \received[accepted]{5 June 2009}


\maketitle

\section{Introduction}

Code Large Language Models (CodeLLMs), particularly agents, are reshaping software engineering. Compared with traditional approaches, they offer two distinctive capabilities. First, they exhibit strong code understanding and reasoning, enabling tasks ranging from function completion to full repository generation~\cite{luowizardcoder,qian2024chatdev}. Second, they tightly integrate natural language and programming languages, supporting instruction-driven debugging and optimization~\cite{chen2023teaching}. Together, these strengths have accelerated adoption in both academia and industry~\cite{hou2024large,hui2024qwen2,hossain2024togll}. 
For example, GitHub Copilot~\cite{github2023}, powered by GPT-4, has been reported to improve developer productivity, enabling users to code 55\% faster~\cite{github2023report}.
Similarly, AI-powered editors such as Cursor~\cite{cursor2025} leverage contextual understanding to support code generation and debugging.
These advancements have moved well beyond code generation and are now rapidly transforming every stage of the Software Development Life Cycle (SDLC) \cite{dong2025survey}.

Through decades of software engineering practice, the SDLC has gradually emerged as a core paradigm \cite{royce1987managing}. The SDLC rigorously delineates activities spanning all phases from requirements engineering to software maintenance, substantially promoting the standardization of software engineering. With the rapid advancement of codellm, particularly the emergence of agentic workflows, their capabilities have progressively extended from isolated code generation to supporting activities across the entire SDLC, giving rise to numerous benchmarks \cite{chen2021evaluating, jimenez2024swebench, li2024evocodebench}. 
However, it remains unclear \textbf{whether existing benchmarks comprehensively evaluate real-world software engineering capabilities across the SDLC.}

This motivates us to examine two key dimensions.
First, coverage scope: software engineering spans the entire SDLC, yet it remains unclear whether existing benchmarks adequately cover all phases.
Second, assessment depth: it remains uncertain whether current benchmarks can faithfully measure the capabilities required in realistic settings across the SDLC.

Existing surveys remain insufficient to address these challenges.
As summarized in~\autoref{tab:t1}, prior studies either focus on traditional machine learning models~\cite{zhang2024unifying,gonzalez2024towards} or do not provide a comprehensive, benchmark-centric analysis~\cite{hou2024large, yang2024ecosystem, zheng2025towards, yang2025code}.
Even recent surveys on CodeLLMs and agents~\cite{zheng2024survey,dong2025survey} largely restrict their scope to implementation and testing, overlooking other SDLC phases.
Concurrent studies~\cite{cao2025how, hu2025assessing} primarily adopt task clustering and emphasize benchmark meta-properties (e.g., data quality), but do not systematically analyze benchmarks through the lenses of coverage scope and assessment depth.

\begin{figure*}[t]
  \centering
  \includegraphics[width=0.9\textwidth]{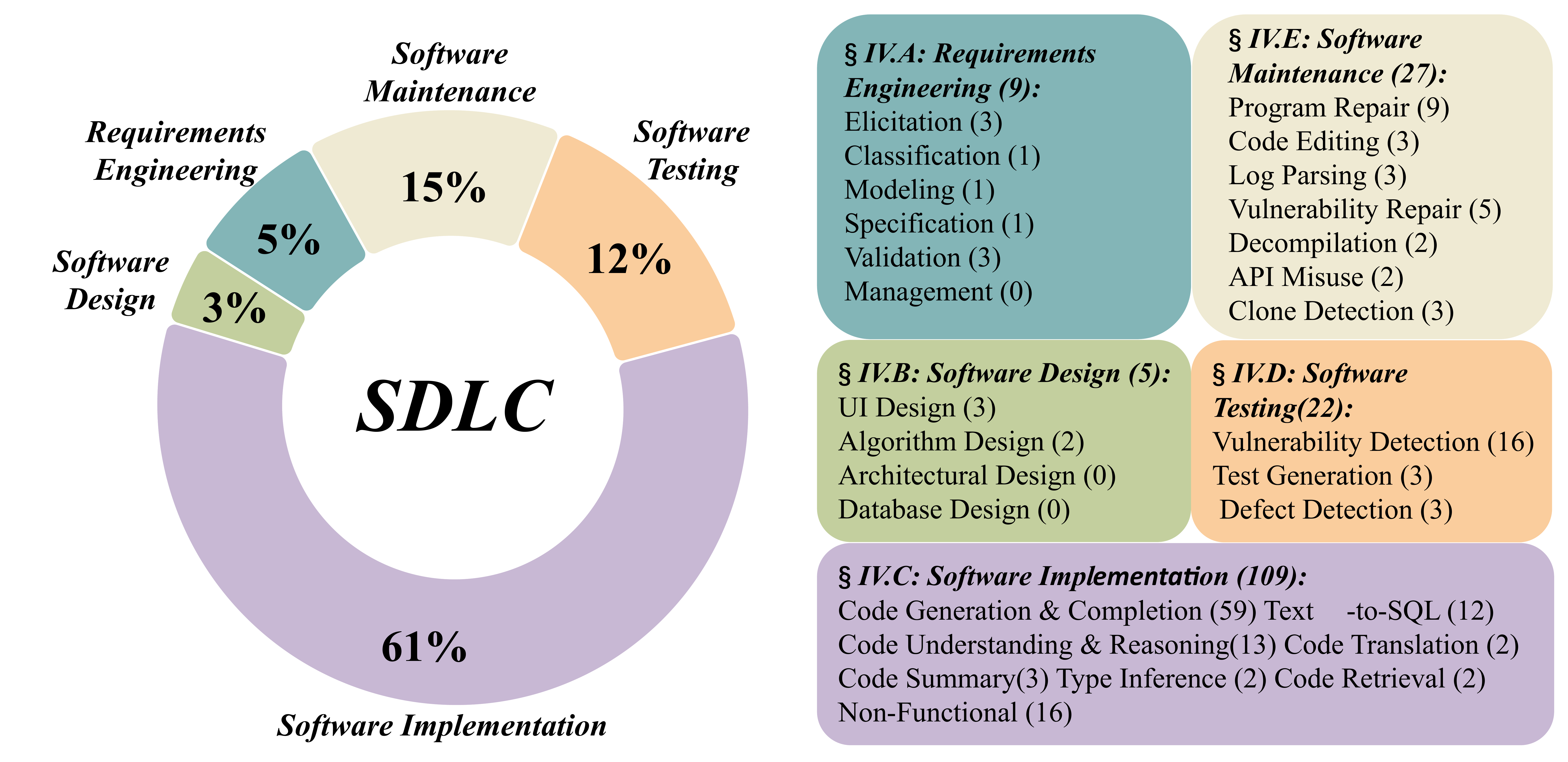}

 % \caption{Distribution of LLM tasks across SDLC stages. Percentages are calculated based on the total of 178 benchmarks. Note that 7 cross-stage benchmarks (approx. 4\%) are not shown in this specific stage breakdown. Taxonomy of CodeLLM benchmarks across the SDLC. This figure categorizes existing CodeLLM benchmarks according to different phases of SDLC. The numbers indicate how many benchmarks correspond to each phase or task.}

\caption{Taxonomy of CodeLLM benchmarks mapped to the SDLC. This Categorization covers 178 benchmarks, where numbers represent the quantity per phase or task. Note: 6 cross-stage benchmarks (approx. 4\%) are omitted from this visualization.}

  \label{fig:figure1}
\end{figure*}

To bridge this gap, this paper focuses on evaluating real-world software engineering capabilities from an SDLC perspective.
As show in \autoref{fig:figure1}, 
we systematically categorize existing benchmarks by SDLC phase and propose a tiered framework for fine-grained analysis.
This SDLC-oriented perspective reveals cross-phase structural imbalances and highlights mismatches between existing benchmarks and practical engineering needs. The main contributions of this paper are as follows:

\begin{itemize}
\item We systematically review 178 benchmarks and map them to five SDLC phases, revealing substantial structural imbalances.
\item We propose a tiered analysis framework with both general and phase-specific dimensions, and use it to assess how well existing benchmarks align with real-world software engineering needs.
\item We identify key deficiencies in existing benchmarks and, based on these findings, outline targeted future directions to align benchmarks with realistic engineering scenarios.
\item We curate and open-source a standardized knowledge base with detailed metadata for all 178 benchmarks, providing a foundational resource for future research on CodeLLMs and agents.
\end{itemize}

The remainder of this paper is organized as follows.
\autoref{Background} introduces the background of Large Language Models for Software Engineering (LLM4SE) and the SDLC.
\autoref{STUDY_DESIGN} describes the research questions, literature collection protocol, and the proposed tiered analysis framework.
\autoref{Functional Benchmarks} reviews benchmarks for each SDLC phase, and \autoref{Statistics} provides a statistical analysis of the collected benchmarks.
\autoref{Future Directions} summarizes the main findings and discusses future research directions.
\autoref{Threats} discusses threats to validity (construct, internal, and external), and \autoref{Conclusion} concludes the paper.

% 同期综述工作如 Cao et al.~\cite{cao2025how} 和 Hu et al.~\cite{hu2025assessing} 主要侧重于基准测试的元属性（如数据质量、构建规范），且采用基于 SE 任务聚类的分类方式，尽管涉及部分智能体内容，但完全缺乏独立的 Agent 分析维度。相比之下，本文超越了元属性的范畴，聚焦于软件工程的的实际能力维度。我们依据 SDLC 对基准进行严格分类，并构建了分层分析框架进行深度评估。 这一视角精准揭示了被任务导向视角所掩盖的 SDLC 阶段结构性失衡，以及基准与真实工程需求之间的匹配差距。此外，我们在分析框架中确立了‘智能体交互 (Agentic Interaction)’的独立分析维度，从而深刻揭示了当前基准在智能体时代的适用性与局限性。

\section{Background}

\begin{table*}[t]
\caption{Comparison Between Existing Surveys and Our Work.}
\centering
\tabcolsep=3pt
\begin{tabular}{ccccccccc}
\toprule
\multirow{3}{*}{Survey} & \multicolumn{2}{c}{Scope} & \multirow{3}{*}{\begin{tabular}[c]{@{}c@{}}Focal on\\ Benchmark\end{tabular}} & \multicolumn{5}{c}{SDLC Phases} \\ \cmidrule{2-3} \cmidrule{5-9}
 & CodeLLM & Agent &  & \begin{tabular}[c]{@{}c@{}}Requirements\\ Engineering\end{tabular} & Design & \begin{tabular}[c]{@{}c@{}}Software\\ Implementation\end{tabular} & Testing & \begin{tabular}[c]{@{}c@{}}Software\\ Maintenance\end{tabular} \\
 \midrule
Zhang et al.~\cite{zhang2024unifying} & $\times$ & $\times$ & \checkmark & \checkmark & $\times$ & \checkmark & \checkmark & \checkmark \\
Gonzalez et al.~\cite{gonzalez2024towards} & $\times$ & $\times$ & $\times$ & \checkmark & \checkmark & \checkmark & \checkmark & \checkmark \\
Hou et al.~\cite{hou2024large} & \checkmark & $\times$ & $\times$ & \checkmark & \checkmark & \checkmark & \checkmark & \checkmark \\
Yang et al.~\cite{yang2024ecosystem} & \checkmark & $\times$ & $\times$ & $\times$ & $\times$ & \checkmark & $\times$ & $\times$ \\
Zheng et al.~\cite{zheng2025towards} & \checkmark & $\times$ & $\times$ & \checkmark & \checkmark & \checkmark & \checkmark & \checkmark \\
Zheng et al.~\cite{zheng2024survey} & \checkmark & $\times$ & \checkmark & $\times$ & $\times$ & \checkmark & \checkmark & $\times$ \\

Dong et al.~\cite{dong2025survey} & $\times$ & \checkmark & $\times$  & \checkmark & $\times$ & \checkmark & \checkmark & \checkmark \\

Yang et al.~\cite{yang2025code} & \checkmark & \checkmark & $\times$  & $\times$ & $\times$ & \checkmark & \checkmark & \checkmark \\

{\bf \textit{Ours}} & \checkmark & \checkmark & \checkmark & \checkmark & \checkmark & \checkmark & \checkmark & \checkmark \\
\bottomrule
\end{tabular}
\label{tab:t1}
\end{table*}

% {dong2025survey}

\label{Background}
This section presents the background on the SDLC and LLM4SE, laying the groundwork for the rest of the paper.
\subsection{LLMs for Software Engineering}

LLMs for Software Engineering Large Language Models (LLM4SE) are fundamentally reshaping the software engineering~\cite{lo2023Trustworthy}. 
Beyond code generation~\cite{xu2022systematic}, these models demonstrate remarkable proficiency in a broader range of SDLC tasks, such as automated program repair~\cite{xia2023automated}, vulnerability detection~\cite{zhou2024large}, and test oracle generation~\cite{hossain2024togll}.
Crucially, the emergence of agentic frameworks has further extended these boundaries, enabling models to interact with external environments and manage multi-step engineering processes autonomously~\cite{jin2024llms}. This evolution positions LLM4SE not merely as a productivity tool, but as a transformative force across the industry~\cite{haque2024llms}. However, it remains unclear whether existing benchmarks adequately align with these evolving capabilities to support a holistic evaluation across the SDLC.

% Despite these advancements, it remains unclear whether existing benchmarks provide sufficient coverage scope and assessment depth to comprehensively evaluate these capabilities across the entire SDLC.

\subsection{Software Development Life Cycle}

% \begin{figure*}[t]
%   % \Description{The phases and their core tasks within the SDLC.}
%   \includegraphics[width=0.9\textwidth]{pictures/sdlc.pdf}
%    \caption{The phases and their core tasks within the SDLC.}
%   \label{fig:sdlc}
%   \vspace{-6pt}
% \end{figure*}

In the 1960s, software development often lacked formal engineering discipline, which led to unmanageable complexity and ambiguous requirements.
To address these challenges, the SDLC was proposed as a systematic framework.
By defining distinct phases, it helps teams allocate resources more effectively and maintain consistent software quality.

Although development paradigms vary (e.g., from Waterfall to Agile), they generally encompass the fundamental phases of the SDLC \cite{royce1987managing, hou2024large}.
The SDLC typically starts with requirements engineering to elicit and document specifications, followed by design to translate these specifications into architectural blueprints.
Implementation constructs the system, testing validates conformance to requirements, and maintenance ensures long-term stability through monitoring and updates.
Despite the importance of applying LLMs throughout this cycle, the literature still lacks a comprehensive benchmark survey spanning the full SDLC, which limits holistic assessments of LLM capabilities in software engineering.

\section{Study Design}
\label{STUDY_DESIGN}

\subsection{Review  Questions}

This survey comprehensively analyzes benchmarks for CodeLLMs across the SDLC, identifies key limitations in the current landscape, and outlines directions for future research. 
Specifically, this survey addresses the following research questions:
\begin{itemize}

\item RQ1: Do existing benchmarks comprehensively evaluate real-world software engineering capabilities across the entire SDLC? (answered in~\autoref{Functional Benchmarks})

\item RQ2: What are the distributional characteristics of current benchmarks, and what structural imbalances exist? (answered in~\autoref{Statistics})

\item RQ3: What are the potential future directions for benchmarking CodeLLMs and agents? (answered in~\autoref{Future Directions})
% \xy{Do not use `key direction'. This is just our forecasting, you should use `potential future direction'*}
% \ltl{future direction is necessary}
\end{itemize}
\subsection{Collection Strategy}

\begin{figure*}[t]
  % \Description{Research process for collecting and analyzing benchmarks. This process includes initial search, literature screening, keyword expansion, and snowball sampling.}
  \centering
  \includegraphics[width=0.9\textwidth]{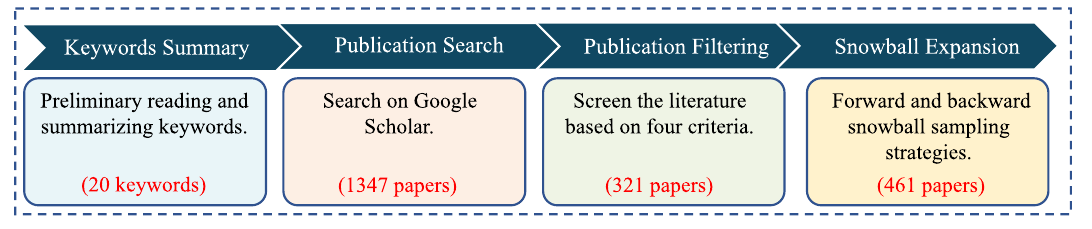}
  \vspace{-3mm}
  \caption{Research process for literature collection. This process includes keywords summary, publication search, publication filtering, and snowball expansion.}
  \label{fig:collection}
\end{figure*}

To ensure a comprehensive and systematic coverage of research at the intersection of large language models (LLMs) and software engineering, we adopt a rigorous literature collection protocol.
As illustrated in~\autoref{fig:collection}, our process consists of four primary stages.
The details are as follows:
% \xy{check the tense in this paper, use present tense*}
% \xy{Plz add keywords about agent and modify numbers in this section}

\begin{enumerate}
\item {\it Keywords Summary:} 
We conducted an initial scan of the literature and derived 20 keyword combinations, including: 
``Code Large Language Model", 
``Code LLMs",
``AI4SE", 
``LLM4SE",
``AI for SE", 
``Code Agents Benchmarks",
``Programming Agents Benchmarks",
``LLM Agents for Software Engineering",
``AI for Requirements Engineering", 
``AI for Software Design", 
``AI for Software Implementation", 
``AI for Software Testing", 
``AI for Software Maintenance", 
``Programming Benchmarks", 
``Code Benchmarks", 
``Code Generation", 
``Code Understanding", 
``Code Completion", 
``Automated Programming", 
``Automated Testing".

\item {\it Publication Search:} We conducted automated literature searches using predefined keywords across four major academic databases: IEEE Xplore, ACM Digital Library, Elsevier ScienceDirect, and Springer. This set of sources provides broad coverage of core research on LLMs and software engineering. In total, we retrieved 1,347 relevant papers.

\item {\it Publication Filtering:}  We performed a manual selection of the retrieved articles based on the following criteria, resulting in 321 articles.
\begin{itemize}
    \item Research on the application of LLMs in software engineering. 
    % (parameters greater than 1B) 
    \item Priority is given to papers accepted by influential journals (e.g., TOSEM, TSE, IJCN, JMLR, TDSC, TOCHI) and conferences (e.g., ICSE, ASE, FSE, ACL, ICML, IJCAI, ISSTA, NeurIPS, PLDI, SIGKDD, S\&P, CHI).
    \item For arXiv papers, selection is prioritized based on community impact (e.g., citation counts and GitHub stars) and paper quality. 
    \item Papers published from 2022 onward. Following a commonly accepted definition of LLMs~\cite{zhao2023survey}, we include literature published since 2022.
    \item Papers proposing code-related benchmarks are directly included in the scope of this survey.
% 如果论文提出了code相关benchmark，则直接纳入所调研的文献
    
% \ltl{give a reason.}
% 该人工过滤流程由两名具备软件工程背景的研究者独立完成，并在存在分歧时通过讨论协商达成一致，以保证筛选结果的客观性与一致性
    
\end{itemize}

\item {\it Snowball Expansion:} Because keyword-based searches may overlook relevant studies, we conducted both forward and backward snowball sampling starting from the initial set.
This process added 140 additional papers, improving coverage and completeness.
Ultimately, we collected 461 papers that met the inclusion criteria.

\end{enumerate}

% \xy{Does your filtering require human invention? If so, please provide a brief introduction of manual instructions.}
We performed a strict quality assessment of the included papers.
Each paper was independently reviewed by two co-authors with expertise in both software engineering and LLMs.
Disagreements were resolved through discussion with a third co-author.
This procedure helps ensure that the included studies are of high quality and relevance.

% 评估用于降低数据泄露风险的机制。鉴于大语言模型预训练语料的庞大规模，基准测试必须采用动态更新、时间截断或强效改写等策略，以确保模型表现源于真正的推理能力而非机械记忆
% 如图~\ref{fig:taxonomy_pyramid}所示

\subsection{Tiered Analysis Framework for CodeLLM Benchmarks}

We propose a tiered analysis framework that integrates general dimensions for full-lifecycle assessment with stage-specific dimensions tailored to the unique demands of each SDLC phase.

\subsubsection{General Dimensions}
General dimensions provide shared evaluation criteria applicable across all SDLC stages, reflecting core capabilities required by modern CodeLLMs and agents.

\textbf{Anti-Contamination Strategy:} 
This dimension captures the mechanisms that benchmarks adopt to prevent test data from being included in LLM training corpora.
As pretraining datasets rapidly expand, mitigating data leakage has become a critical challenge.
Prior work highlights the risks of static benchmarks and shows that public datasets are often ingested during LLM pretraining, which can compromise evaluation validity~\cite{zhou2023don,zhou2025LessLeakBench}.

% \xy{you should use other better literature to support this dimension. `existing work does something' is not a good support. You shold say : `existing work has highted the importance of anti-contamination strategy...'*}

% \textbf{Anti-Contamination Strategy:} 
% This dimension focuses on the strategy implemented by benchmarks to prevent test data from being included in LLM training corpora. 
% As pretraining datasets undergo rapid expansion, preventing data leakage has become a critical challenge. Many existing benchmarks rely on static, public datasets that are inadvertently ingested during LLM pretraining~\cite{naman2024livecodebench,wang-etal-2023-recode}.
% Consequently, high performance on these benchmarks may stem from rote memorization rather than genuine reasoning capabilities.

% 针对软件开发生命周期中各阶段的差异，我们提出一个结构化评估框架，该框架由覆盖全周期的通用维度与适配各阶段的特定维度共同构建。

% \subsubsection{通用维度}
% 通用维度定义了适用于软件开发生命周期所有阶段的共性评估标准，反映了现代代码大语言模型与智能体所需的核心能力。

% 防数据泄露措施： 随着大语言模型预训练规模不断扩大，防止数据泄露已成为一项重要挑战。当前面向代码的大语言模型评测基准多源于公开数据，极易被纳入大型语言模型的预训练语料中，从而导致观测到的性能提升可能来自模型对已有内容的记忆，而非真正的逻辑推理能力\cite{naman2024livecodebench,wang-etal-2023-recode}。 

\textbf{Authenticity:} 
This dimension distinguishes between synthetic and real-world data sources.
Because synthetic benchmarks can oversimplify practical settings and inflate performance, real-world data is indispensable for evaluating LLMs in complex production environments~\cite{lai2023ds,yu2024codereval}.

\textbf{Interaction Mode:}
This dimension categorizes interaction paradigms into three levels:
\textbf{Single-turn}, where the model produces an output from a static prompt;
\textbf{Iterative}, where multi-turn dialogue or feedback enables stepwise refinement; and
\textbf{Agentic}, where the system performs autonomous planning, tool use, and environment exploration~\cite{liu2024agentbench}.

% \textbf{Evaluation Paradigm:}
% This dimension specifies the evaluation methods used at different phases of a benchmark. In non-coding phases such as requirements engineering and software design, it distinguishes automated/objective metrics from human judgment. While the latter can assess complex qualities, it is highly subjective and difficult to reproduce ~\cite{li2024Devbench}. For code-related phases, this dimension distinguishes text matching from dynamic execution. Because code can take many surface forms while preserving the same semantics, relying solely on text matching cannot guarantee functional correctness. Therefore, the validation paradigm should be built around dynamic execution~\cite{chen2021evaluating}.

\textbf{Evaluation Paradigm:}
This dimension characterizes the validation methods used across SDLC phases.
For non-coding stages, it distinguishes between automated metrics and human judgment; the latter often suffers from subjectivity and limited reproducibility~\cite{li2024Devbench}.
For code-related stages, it distinguishes dynamic execution from text matching, as static comparisons cannot reliably verify functional correctness~\cite{chen2021evaluating}.

\subsubsection{Stage-Specific Dimensions}

% To better capture the unique requirements across different phases of the SDLC, we introduce stage-specific dimensions that complement the general metrics:
% \xy{If there is no comparison, do not use `better'. Make your paper presentation formal.*}

% \textbf{Non-code stages (Requirements Engineering and Software Design)}: In requirements engineering, documentation often spans specialized fields such as law and medicine, where technical terminology creates barriers and ambiguity. Consequently, benchmarks must incorporate datasets from diverse domains to evaluate a LLMs' domain-specific comprehension~\cite{bruel2019formality,nahri2025extracting}  (\textbf{\textit{\#Application Domain}}). Second, to address diverse requirement formats (e.g., text, diagrams, and use cases), benchmarks should include multimodal inputs (\textbf{\textit{\#Input Format}}). Additionally, to mirror the temporal evolution of requirements, benchmarks need to evaluate the LLMs' capacity to manage changes and maintain consistency~\cite{wu2024versicode} (\textbf{\textit{\#Requirement Evolution}}). For software design, we focus on the "task type" dimension (\textbf{\textit{\#Task Paradigm}}) to distinguish between retrieval-based and generative tasks. Existing software design benchmarks predominantly target well-defined tasks such as classification and retrieval. This limitation fails to capture the core nature of software design as a creative problem-solving process~\cite{zheng2024opencodeinterpreter,falessi2010applying}.
To address the unique requirements across different phases of the SDLC, we introduce stage-specific dimensions that complement the general metrics:

Non-code stages (Requirements Engineering and Software Design): 
Requirements engineering often involves artifacts from specialized domains (e.g., law and medicine), which can be highly technical and ambiguous.
Accordingly, benchmarks should span diverse application domains to assess LLMs' domain-specific understanding~\cite{bruel2019formality,nahri2025extracting} (\textbf{\textit{\#Application Domain}}).
To cover heterogeneous artifacts, benchmarks should also support multimodal inputs such as text and diagrams (\textbf{\textit{\#Input Format}}).
Moreover, to reflect the evolving nature of requirements, benchmarks should evaluate a model's ability to handle requirement changes and maintain consistency~\cite{wu2024versicode} (\textbf{\textit{\#Requirement Evolution}}).
For software design, we distinguish retrieval-based and generative paradigms (\textbf{\textit{\#Task Paradigm}}).
Existing benchmarks largely focus on well-defined tasks (e.g., classification), which fails to capture software design as a creative process~\cite{zheng2024opencodeinterpreter,falessi2010applying}.

For code-related stages (Implementation, Testing, and Maintenance), we emphasize the granularity of contextual information (\textbf{\textit{\#Context Scope}}).
Modern CodeLLMs should extend beyond function-level edits and demonstrate repository-level understanding~\cite{jimenez2024swebench}.
Moreover, for software maintenance, we highlight its evolutionary nature (\textbf{\textit{\#Evolution}}).
Unlike implementation, maintenance typically involves fixing, adapting, and evolving an existing system with complex code history and context, rather than building from scratch~\cite{thai2025swe}.

\section{Benchmark Taxonomy by the SDLC}
\label{Functional Benchmarks}

% To address RQ1, we conduct an in-depth analysis of benchmarks across the five phases of the software development life cycle (SDLC): requirements engineering, software design, software implementation, testing, and software maintenance. We categorize these benchmarks according to their primary evaluation objectives to systematically analyze their distinctive characteristics, strengths, and limitations. Acknowledging that phase boundaries can be fluid in modern DevOps and Agile practices, we classify benchmarks spanning overlapping activities based on their primary evaluative focus, while those inherently targeting multi-stage coverage are allocated to the "Cross-Phase" section. The subsequent analysis is structured around the specific tasks of each phase, providing a detailed examination of the corresponding benchmarks.

To answer RQ1, we categorize benchmarks into five SDLC phases according to their primary evaluation objectives. We assign benchmarks that span overlapping activities to the phase they primarily evaluate, and place inherently multi-stage benchmarks in the cross-phase section. Within each phase, we further subdivide the benchmarks by specific tasks

% \ltl{to address RQ1, we **}
% 随着大型语言模型快速发展，基准数据污染（BDC）成为了一个重要问题。BDC是指在大语言模型的训练过程中，模型意外地接触到了评估基准数据，导致在评估阶段表现过于夸大或不准确。
% 评估过程中应考虑时效性

% 
% 测试基准的唯一目的，就是评估模型发现未知缺陷的能力。缺陷揭示效度。

% % 回答这个基准能用吗？
% % 这个基准有多好

% 通用：合格性,可复现性，客观性  时效性
% 1.Requirement Engineering：多模态 需求与演化
% 3.software Development Benchmark： 真实性
% 4. Software Testing Benchmark         .缺陷导向性。基准应包含已知的真实缺陷，使测试的有效性取决于其能否触发这些缺陷，而非仅验证程序在正常情况下的行为
% 5. Software Maintenance 长上下文.

\subsection{Requirement Engineering}

\begin{table*}[htbp]
  \centering
  \scriptsize % 字体保持小号，否则全称会导致表格溢出
  \setlength{\tabcolsep}{3pt} % 紧凑列间距
  \caption{Analysis of Requirements Engineering Benchmarks. Note: We identified no public benchmarks for Requirements Management.}
  \label{tab:req_benchmarks_full}
  \begin{adjustbox}{width=\textwidth}
  
  % 定义11列对齐方式
  \begin{tabular}{cc cc cccc ccc}
  \toprule
  
  % --- 表头 ---
  \multirow{3}{*}{\textbf{Tasks}} & 
  \multirow{3}{*}{\textbf{Benchmarks}} & 
  \multirow{3}{*}{\textbf{Langage}} & 
  \multirow{3}{*}{\textbf{Year}} & 
  \multicolumn{4}{c}{\textbf{General Dimensions}} & 
  \multicolumn{3}{c}{\textbf{Stage-Specific Dimensions}} \\
  \cmidrule(lr){5-8} \cmidrule(lr){9-11}
  
   & 
   & 
   & 
   & 
  \textbf{\makecell{Anti-Contamination\\Strategy}} & 
  \textbf{Authenticity} & 
  \textbf{\makecell{Interaction\\Mode}} & 
  \textbf{\makecell{Evaluation\\Paradigm}} &
  \textbf{\makecell{Application\\Domain}} & 
  \textbf{\makecell{Input\\Format}} & 
  \textbf{\makecell{Requirements\\Evolution}} 
  \\

\midrule

% ---------------- Requirements Elicitation ----------------
\multirow{3}{*}{\textbf{\makecell[c]{Requirements\\Elicitation}}} 
& Pure~\cite{ferrari2017pure}       & NL & 2017 & None & Real-world & Single & Static & Multi & Text & \ding{55} \\
& Jain's~\cite{jain2023transformer} & NL & 2023 & Perturbation & Real-world & Single & Human+Static & Multi & Text & \ding{55} \\
& CPSBENCH~\cite{jin2024Evaluation} & NL & 2024 & None & Real-world & Single & Static & Cyber-Physical & Text & \ding{55} \\
\midrule

% ---------------- Requirements Classification ----------------
\textbf{\makecell[c]{Requirements\\Classification}} 
& PROMISE~\cite{Sayyad2005}         & NL & 2005 & None & Real-world & Single & Static & Multi & Text & \ding{55} \\
\midrule

% ---------------- Requirements Modeling ----------------
\textbf{\makecell[c]{Requirements\\Modeling}} 
& Ferrari's~\cite{ferrari2024Model} & NL & 2024 & None & Real-world & Single & Human & Multi & Text & \ding{51} \\
\midrule

% ---------------- Requirements Specification ----------------
\textbf{\makecell[c]{Requirements\\Specification}} 
& Krishna's~\cite{krishna2024Using} & NL & 2024 & None & Synthetic & Iterative & Human & Education & Text & \ding{55} \\
\midrule

% ---------------- Requirements Validation ----------------
\multirow{4}{*}{\textbf{\makecell[c]{Requirements\\Validation}}} 
& Reinpold's~\cite{reinpold2024exploring} & NL & 2024 & None & Synthetic & Single & Static & Smart Grid & Text & \ding{55} \\
& REQuestA~\cite{ezzini2023ai}      & NL & 2022 & None & Real-world & Single & Static & Multi & Text & \ding{55} \\
& rSDE-Bench~\cite{hu2024self}      & Python           & 2024 & None & Synthetic & Single & Exec & \makecell{Web/Game\\Development}& Text & \ding{55} \\

% \midrule
% \textbf{\makecell[c]{Requirements\\Management}} 
% & \textit{None Identified} & - & - & - & - & - & - & - & - & - \\

\bottomrule
  \end{tabular}
  \end{adjustbox}
\vspace{-4mm}
\end{table*}

% The essence of requirements engineering is to achieve a precise understanding and formal definition of the requirements established by the users or stakeholders. 
% Requirements engineering consists of several key tasks. First, in the requirements elicitation task, customer needs are systematically collected, organized, and clarified. Next, requirements classification divides the collected needs into functional and non-functional requirements. 
% Then, in the requirements modeling task, these needs are transformed into formal models using modeling languages like UML and SysML~\cite{yang2019integrating}. 
% Finally, in the requirements specification and validation task, the requirements are converted into formal specification documents, which detail the system's functions, performance, and interface requirements. Validation is then conducted to ensure that the requirements accurately reflect the stakeholders' expectations~\cite{tihanyi_new_2024}.
% As illustrated in Table \ref{tab:req_benchmarks}, to construct a scientifically rigorous evaluation framework, we first established three foundational dimensions: \textbf{Compliance}, \textbf{Reproducibility}, and \textbf{Objectivity}—defined as the elimination of human evaluation bias through the use of deterministic metrics. Building upon this foundation, and incorporating the specific characteristics of Requirements Engineering (RE), we introduced the following critical dimensions:

The essence of requirements engineering is to achieve a precise understanding and formal definition of the requirements established by the users or stakeholders.

\subsubsection{Requirements Elicitation} 

\textbf{PURE}~\cite{ferrari2017pure} is the most widely used benchmark for requirements acquisition. It spans multiple domains and contains 79 publicly available, real-world natural language requirement documents; however, it is limited to plain text.
Later benchmarks move to specialized domains, such as software engineering contracts (\textbf{Jain et al.}~\cite{jain2023transformer}) and cyber-physical systems (\textbf{CPSBENCH}~\cite{jin2024Evaluation}).
Ultimately, a common key limitation across all these benchmarks is their failure to support the evolution of requirements.

\subsubsection{Requirements classification}
\textbf{PROMISE}~\cite{Sayyad2005} is a widely used benchmark for requirements classification, built from 15 projects.
However, it is limited to plain text, does cannot evaluate requirements evolution, which constrains its suitability for modern LLM evaluation.

\subsubsection{Requirements Modeling} 
\textbf{Ferrari's}~\cite{ferrari2024Model} benchmark uses 28 industrial documents across multiple domains, supports structured artifacts, and partially considers requirements evolution.
However, its evaluation relies on subjective human scoring, limiting objectivity and reproducibility.

% \textbf{Ferrari's}~\cite{ferrari2024Model} 基准在内容上覆盖了多个关键维度：来自不同应用领域的28份工业文档提供了领域适应性，支持结构化数据，并罕见地尝试引入需求演化。然而，尽管内容框架颇具价值，该基准最严重的缺陷在于其评估依赖人为打分，缺乏客观性。 这一根本性问题使得其无法得到一个可靠的评估结果，从而极大地限制了其作为一个标准化基准的实用性。

% \textbf{Ferrari's}~\cite{ferrari2024Model} benchmark consists of 28 industrial requirement documents from various application domains, with requirement models manually constructed based on these documents.

\subsubsection{Requirements Specification}
\textbf{Krishna's}~\cite{krishna2024Using} benchmark evaluates requirements specification using manually written documents aligned with IEEE standards~\cite{IEEEStd}.
However, it is narrowly scoped to university club management systems and relies on subjective human scoring, limiting generalizability, objectivity, and reproducibility.

% \textbf{Krishna}~\cite{krishna2024Using} 基准用于评估需求规范的制定，其采用符合IEEE标准~\cite{IEEEStd}的手工编写文档，并支持结构化数据。然而，该基准存在两大关键缺陷：首先，其应用场景高度单一，仅限大学俱乐部管理系统，导致领域适应性极差；更根本的问题在于，其依赖主观的人工打分，使得评估缺乏客观性与可复现性。尽管其在数据格式上有所考量，但上述局限使其难以成为一个可靠且通用的评估基准。

% \textbf{Krishna's}~\cite{krishna2024Using} benchmark is used for requirements specification, adopting a manually crafted software requirements specification in accordance with IEEE standards~\cite{IEEEStd}.

\subsubsection{Requirements verification}
In requirements verification, existing benchmarks remain fragmented and lack a unified framework.
\textbf{rSDE-Bench}~\cite{hu2024self} adopts a code-centric paradigm by converting requirements into programming tasks and evaluating them via unit tests, mainly in web and gaming.
In contrast, \textbf{REQuestA}~\cite{ezzini2023ai} frames verification as knowledge QA across three domains, while \textbf{Reinpold}~\cite{reinpold2024exploring} targets smart grids and supports structured artifacts.
Despite these differences, none evaluate requirements evolution, limiting their applicability to iterative development.

\begin{tcolorbox}[colframe=gray!140, colback=gray!10, coltitle=white, title=Observation from Requirement Engineering, breakable]

\textbf{Observation 1 (referred to as Obs. 1):}  
Benchmark coverage for requirements engineering remains incomplete.
Public benchmarks targeting \textit{requirements management} are largely absent, largely because capturing dynamic requirement evolution and conflict resolution in a benchmark setting remains challenging.
Moreover, benchmarks for \textit{modeling and specification} often rely on small-scale case studies and lack objective, reproducible evaluation metrics, which hinders standardized automated assessment.

\textbf{Obs. 2:} Existing evaluation methods remain misaligned with realistic, iterative requirements workflows. Current benchmarks predominantly focus on textual natural language requirements, with extremely limited support for real-world artifacts such as tables, diagrams, and SysML models. Moreover, reliance on subjective human scoring in some benchmarks undermines reproducibility. Furthermore, interaction is mostly restricted to single-turn exchanges with almost no requirement evolution, and the lack of anti-contamination measures risks data leakage, undermining the validity of evaluation results.

\end{tcolorbox}

\subsection{Software Design}

Software design is a pivotal phase that converts stakeholder needs into a structured blueprint~\cite{kamal2020risk}. As shown in Table~\ref{tab:design_benchmarks}, it encompasses architecture design, user interface (UI) design, algorithm design, and database design.

\subsubsection{UI Design} 
% UI Design refers to the process of designing UI for software applications or websites. Current UI Design benchmarks predominantly focus on the mobile development domain. 

% 多是gui检索  而不是真正的设计生成  GUI检索是UI/UX设计的一个子任务或辅助工具，但远不能代表UI/UX设计的全过程和核心目标。
% 当前所谓的“UI设计基准”实际上主要评估的是 “GUI理解与检索” 能力，这只是UI/UX设计工作流中的一个辅助环节。真正的生成性、创造性的UI/UX设计基准仍然是一个有待开发的重要领域。
% 人机协作模式：如何设计有效的架构师-LLM协作工作流

UI design concerns creating user interfaces for software or websites; however, existing benchmarks largely evaluate auxiliary skills such as GUI understanding and retrieval rather than generative design.
For example, \textbf{Rico}~\cite{deka2017rico} supports GUI search, \textbf{SCapRepo/ScreenRepo}~\cite{weiGUingMobileGUI2024} emphasize interface description and retrieval, and \textbf{Screen2Words}~\cite{wang2021screen2words} focuses on mobile UI summarization.
These tasks are easier to score due to fixed references, but they underrepresent the creative aspects central to UI design.

\subsubsection{Algorithm Design} 

% 探索LLM与人类专家的协同设计模式

% Algorithm design is the process of solving a problem by creating a series of steps or rules. 

Algorithm design emphasizes creating novel and efficient solutions, not merely implementing known functions.
\textbf{CLRS-30}~\cite{velickovic2022clrs} mainly evaluates understanding and generalization over classical algorithms, whereas \textbf{BLADE}~\cite{van2025blade} explores models' ability to synthesize new algorithms.
However, the difference of evaluation metrics across these benchmarks precludes direct comparison, underscoring the need for a standardized evaluation framework.

% 算法设计的本质在于创造新颖高效的解决方案，而非仅仅实现预设功能。CLRS-30~\cite{velickovic2022clrs}主要测试经典算法的理解与泛化能力，但未能未能触及真正的算法创新。BLADE~\cite{van2025blade}则迈出了重要一步，探索了模型在合成新算法方面的潜力。然而，这两个基准采用独立的评估标准，揭示了该领域缺乏系统化、可比较的评估框架。

% 算法设计是通过创建一系列步骤或规则来解决问题的过程。和实现一个已知功能的代码生成不同，算法设计的目标是发明一个未知的、高效的解决方案。
% The \textbf{CLRS-30}~\cite{velickovic2022clrs}benchmark的目标是检测对已知经典算法的理解和泛化，\textbf{BLADE}~\cite{van2025blade}则检测llm发现与合成未知的新算法的能力。两个benchmark都是使用各自的评价指标，这显示了这个领域尚未建立系统化、标准化的评估体系。更为严重的是，现有基准普遍脱离复杂多变的工业实践。亟需建立统一、全面、实用的LLM4AD评估基准.

% The \textbf{CLRS-30}~\cite{velickovic2022clrs} benchmark is designed to assess the algorithm design abilities of LLMs. The benchmark includes 30 fundamental algorithms selected from a classic textbook, covering key topics such as sorting, searching, dynamic programming, and others. 
% In contrast, \textbf{BLADE}~\cite{van2025blade} is more challenging. It aims to assess the ability of LLM to automatically generate optimization algorithms and improve existing ones.

% 相比之下，BLADE更具挑战性，旨在评估大语言模型自动生成优化算法和在现有算法的基础上进行改进的能力。

% 代码生成的目标是实现一个已知的、明确的功能。比如，“用Python写一个函数，计算两个矩阵的乘积”。这里，“矩阵乘法”的规则是明确的，LLM的任务是准确地实现它。
% 自动算法设计的目标是发明一个未知的、高效的解决方案。比如，“设计一个算法来优化这个复杂的黑箱函数”。甚至没有人知道“最优”的算法应该长什么样，LLM的任务是去探索和创造。

\subsubsection{Architectural \& Database Design} 
Architectural design specifies the system’s high-level structure (components and their interactions), while database design defines how data are organized and accessed.
We found no publicly available benchmarks for either task.

% The survey does not identify relevant benchmarks.

% 目前的表格虽然列出了 Evolution，但略显单薄。如果你想让评估更上一层楼，可以在正文分析或表格备注中强调：
% 区分“真设计”与“辅助设计”：批评现有的 UI 基准大多是检索（Retrieval），呼吁更多生成式（Generative）基准。
% 强调“空白领域”：表格底部的 Architectural & Database Design 为空 ，这本身就是一个强有力的评价——说明目前的 Benchmark 生态严重偏科，这是未来最大的机会点（Direction 2）。
% 客观性悖论：在文中讨论 Objectivity 列（全是勾）背后的代价——为了追求客观性，现有的基准牺牲了设计的“创造性”维度，把任务简化成了检索或分类任务 。

% 软件设计基准测试存在显著的覆盖缺口，且缺乏成熟的评估范式。} 该领域的基准测试数量远少于软件工程的其他阶段。用户界面设计基准仅限于图形用户界面检索等辅助任务，算法设计仅处于初步探索阶段，而架构设计与数据库设计则完全空白。这种现状揭示了一个根本性的方法论困境：软件设计的创造性与主观性，与标准化评估所追求的客观性之间存在冲突。这一核心矛盾是制约软件设计基准测试发展的主要瓶颈。

% \ding{183} \textbf{设计流程的迭代演进性。} 软件设计通常需经过多轮持续优化才能达到理想效果 \cite{zheng2024opencodeinterpreter, falessi2010applying}。一个有效的基准测试必须支持多轮交互与逐步精化，其核心原因在于：软件设计本身就是一个通过反馈循环进行探索和决策的动态过程。评估一个静态的设计快照只能反映模型的初始构想能力，而无法衡量其根据新信息、变更的需求或发现的缺陷进行动态调整和优化的关键能力。因此，支持迭代的基准能够更真实地模拟设计演进的全过程，从而评估模型从初步构想到产出成熟方案的完整潜力。

% Observation #3：软件设计基准测试严重匮乏。 除 UI 与算法设计的初步尝试外，架构及数据库设计等核心领域仍属空白。现有 UI 测试多局限于静态检索或分类，难以评估 LLM 的生成式创造力，无法触及设计本质。

% Observation #4：缺乏多轮交互与迭代优化机制。 软件设计通常需经过多轮持续优化才能达到理想效果。现有基准仅关注静态“设计快照”，仅能捕捉初始构想，忽略了模型的迭代优化能力。

\begin{tcolorbox}[colframe=gray!140, colback=gray!10, coltitle=white, title=Observation from Software Design, breakable]
\textbf{Obs. 3:} Benchmarks for software design remain scarce.
Beyond early efforts in UI and algorithm design, core areas such as software architecture and database design remain largely unexplored, largely because it is difficult to translate open-ended, creative reasoning into standardized and objective evaluation tasks.
Moreover, existing UI design benchmarks predominantly emphasize retrieval or summarization and therefore struggle to assess the generative creativity of large language models.

\textbf{Obs. 4: }
Existing benchmarks largely lack support for iterative interaction modes.
Software design typically requires continuous, multi-round refinement to reach an ideal result; however, current benchmarks focus on static design snapshots.
As a result, they capture only initial design proposals while failing to evaluate a model’s ability to perform iterative optimization.

\end{tcolorbox}

% Observation #3. 软件设计基准面临严重的覆盖不足与评估范式缺失。 该领域的基准数量远少于其它软件工程阶段。UI设计基准仅局限在GUI检索等辅助任务，算法设计仅有初步探索，而架构与数据库设计则完全处于空白状态。这一现状揭示了深层方法论困境：软件设计的创造性和主观性，与标准化评估追求的客观性之间存在根本矛盾。这一核心矛盾是阻碍软件设计基准发展的主要瓶颈。

% \textbf{Obs. 3.}
% 软件设计领域的benchmark极为匮乏。算法设计benchmark数量稀少，架构设计，ui设计，数据库设计则完全缺乏相关基准。这可能是由于软件设计高度依赖创造性和主观决策，导致软件设计评价难以建立标准的评估体现（例如算法设计的两个基准），这可能是软件设计benchmark设计的根本瓶颈。

% 各设计任务的基准覆盖存在不均衡现象。现有基准过度集中于用户界面设计，而架构设计与数据库设计任务则缺乏专项基准。用户界面设计主要侧重检索与描述功能，专门针对直接界面设计的基准体系尚属空白。

% \textbf{Obs. 3.} Benchmark coverage is imbalanced across design tasks. Current benchmarks focus heavily on UI design, while architectural and database design tasks lack dedicated benchmarks. UI design primarily emphasizes retrieval and description, with a lack of benchmarks specifically focused on direct UI design.\\
% UI设计关注于检索和描述，直接进行ui设计的benchmark不足

 % Existing algorithm design benchmarks mainly focus on specific algorithm tasks and lack cross-domain applicability.

% 软件设计的基准测试缺乏及时的数据更新和防污染机制，评估结果可能受到数据泄露的影响。UI设计的基准测试无法反映设计过程的迭代演进，且仅局限于移动应用领域。现有的算法设计基准主要集中于特定的算法任务，缺乏跨领域的适用性。

% Software Design的benchmark缺乏即使数据更新，可能存在llm预训练预料，评估可能受到知识污染的影响。ui设计benchmark缺乏设计流程的迭代演进性，而且局限于移动应用领域。现有算法设计基准主要聚焦于特定算法任务，缺乏跨领域适用性。

\subsection{Software Implementation}

\begin{table*}[t]
  \centering
  \small
  \setlength{\tabcolsep}{4pt} % 调整列间距
  \caption{Analysis of Software Design Benchmarks. Note: We identified no public benchmarks for Architectural or Database Design.}
  \label{tab:design_benchmarks}
  \begin{adjustbox}{width=\textwidth}
  % 定义 9 列
  \begin{tabular}{c l c c c c c c c} 
  \toprule
  \multirow{3}{*}{\textbf{Tasks}} & 
  \multirow{3}{*}{\textbf{Benchmarks}} & 
  \multirow{3}{*}{\textbf{Language}} & 
  \multirow{3}{*}{\textbf{Year}} & 
  \multicolumn{4}{c}{\textbf{General Dimensions}} & 
  \multicolumn{1}{c}{\textbf{Stage-Specific}} \\ % 修改为跨1列
  \cmidrule(lr){5-8} \cmidrule(lr){9-9}
   & 
   & 
   & 
   & 
  \textbf{\makecell{Anti-Contamination\\Strategy}} & 
  \textbf{Authenticity} & 
  \textbf{\makecell{Interaction\\Mode}} & 
  \textbf{\makecell{Evaluation\\Paradigm}} &
  \textbf{\makecell{Task\\Paradigm}} % 移动到这里
  \\
  \midrule

  % ---------------- UI Design ----------------
  \multirow{4}{*}{\textbf{UI Design}} 
    & Rico~\cite{deka2017rico} & Image & 2017 
    & None & Real-world &Single & Static 
    & Retrieval \\ % Task Paradigm
    
    & Screen2Words~\cite{wang2021screen2words} & NL & 2021 
    & None & Real-world & Single & Static 
    & Summarization \\
    
    & \makecell[l]{SCapRepo \&\\ScreenRepo~\cite{weiGUingMobileGUI2024}} & Image & 2024 
    & None &Real-world & Single & Static 
    & Retrieval \\
  \midrule

  % ---------------- Algorithm Design ----------------
  \multirow{2}{*}{\textbf{\makecell{Algorithm\\Design}}} 
    & CLRS-30~\cite{velickovic2022clrs} & Python & 2022 
    & None & Synthetic & Iterative & Static 
    & Reasoning \\
    
    & BLADE~\cite{van2025blade} & Python & 2025 
    & None & Mixed & Iterative & Exec 
    & Generation \\
  % \midrule

  % % ---------------- Empty Categories ----------------
  % \textbf{\makecell{\makecell{Architectural \&\\ Database Design}}} 
  %   & \textit{None Identified} & - & - 
  %   & - & - & - & - 
  %   & - \\

  \bottomrule
  \end{tabular}
  \end{adjustbox}
  \vspace{-5mm}
\end{table*}

\begin{table*}[t]
  \centering
  
  \tiny % 使用最小字号
  \setlength{\tabcolsep}{5pt} % 紧凑列间距
  \caption{Analysis of Software Implementation Benchmarks (Code Generation \& Completion; * indicates benchmarks applicable to agents).}
  \label{tab:CodeGeneration}
  \begin{adjustbox}{width=0.97\textwidth}

  % 9 列定义
  \begin{tabular}{c c c c c c c c c} 
  \toprule
  \multirow{4}{*}{\textbf{Scenario}} &
  \multirow{4}{*}{\textbf{Benchmarks}} & 
  \multirow{4}{*}{\textbf{Language}} & 
  \multirow{4}{*}{\textbf{Year}} & 
  \multicolumn{4}{c}{\textbf{General Dimensions}} & % 现第5-8列
  \multicolumn{1}{c}{\textbf{Stage-Specific}} \\ % 现第9列
  \cmidrule(lr){5-8} \cmidrule(lr){9-9}
   & & & & 
  \textbf{\makecell{Anti-Contamination\\Strategy}} & 
  \textbf{Authenticity} & 
  \textbf{\makecell{Interaction\\Mode}}& 
  \textbf{\makecell{Evaluation\\Paradigm}} &
  \textbf{\makecell{Context\\Scope}} \\ % 移到了最后
  \midrule
    
  % ================= General SE (20 rows) =================
  \multirow{20}{*}{\textit{\textbf{General SE}}} 
    & HumanEval\cite{chen2021evaluating} & Python & 2021 & None & Hand-written & Single & Exec & Function \\
    & HumanEval+\cite{liu2024your} & Python & 2023 & None & Hand-written & Single & Exec & Function \\
    & MBPP\cite{austin2021program} & Python & 2021 & None & Hand-written & Single & Exec & Function \\
    & MBPP+\cite{liu2024your}& Python & 2023 & None & Hand-written & Single & Exec & Function \\
    & HumanEval-ET\cite{DongCodeScore2024}& Python & 2024 & None & Hand-written & Single & Exec & Function \\
    & MBPP-ET\cite{DongCodeScore2024}& Python & 2024 & None & Hand-written & Single & Exec & Function \\
    & BigCodeBench\cite{zhuo2024bigcodebench} & Python & 2024 & Perturbation & Synthetic & Single & Exec & Function \\
    & ClassEval\cite{du2023classeval} & Python & 2023 & None & Hand-written & Single & Exec & Class \\
    & ConCode\cite{iyer2018mapping} & Java & 2018 & None & Real-world & Single & Match & Class \\
    & CoderEval\cite{yu2024codereval} & Python/Java & 2024 & Rewrite & Real-world & Single & Exec & Repo \\
    & EvoCodeBench\cite{li2024evocodebench} & Python & 2024 & Updates & Real-world & Single & Exec & Repo \\
    & HumanEvo\cite{zheng2024towards} & Python/Java & 2024 & Update & Real-world & Single & Exec & Repo \\
    & CoNaLa\cite{yin2018learning} & Python/Java & 2018 & None & Real-world & Single & Match & Snippet \\
    & PragmaticCode\cite{agrawal2023guiding} & Java & 2023 & Time-cutoff & Real-world & Single & Exec+Match & Repo \\
    & FEA-Bench\cite{li2025fea} & Python & 2025 & None & Real-world & Single & Exec & Repo \\
    & DevEval\cite{li2024deveval} & Python & 2024 & None & Real-world & Single & Exec & Repo \\
    & CodeAgentBench*\cite{zhang2024codeagent} & Python & 2024 & None & Real-world & Agentic & Exec & Repo \\

  \midrule

  \multirow{3}{*}{\textit{\textbf{Code Completion}}} 
    & RepoBench\cite{liu2023repobench} & Python/Java & 2023 & Time-cutoff & Real-world & Single & Exec & Repo \\
    & RepoEval\cite{zhang2023repocoder} & Python & 2023 & Time-cutoff & Real-world & Iterative & Match & Repo \\
    & CrossCodeEval\cite{ding2024crosscodeeval} & Multi & 2023 & Time-cutoff & Real-world & Single & Match & Repo \\

  \midrule
  % ================= Competitions (5 rows) =================
  \multirow{5}{*}{\textit{\textbf{Competitions}}} 
    & APPS\cite{Hendrycks2021MeasuringCC} & Python & 2021 & None & Real-world & Single & Exec & Problem \\
    & CodeContests\cite{li2022competition} & Multi & 2022 & Time-cutoff & Real-world & Single & Exec & Problem \\
    & LiveCodeBench\cite{naman2024livecodebench} & Python & 2024 & Updates & Real-world & Single & Exec & Problem \\
    & LeetCode Contest\cite{guo2024deepseek} & Python & 2024 & Time-cutoff & Real-world & Single & Exec & Problem \\
    & CODEELO\cite{quan2025codeelo} & C++/Python & 2025 & Online & Real-world & Single & Exec & Problem \\

  \midrule
  % ================= Multimodal (8 rows) =================
  \multirow{8}{*}{\textit{\textbf{Multimodal}}} 
    & ReDraw\cite{moran2018machine} & Java & 2018 & None & Real-world & Single & Exec+Match & File \\
    & MMCode\cite{li2024MMCodea} & Python & 2024 & None & Real-world & Single & Exec & Problem \\
    & Plot2Code\cite{wu2024Plot2Code} & Python & 2024 & None & Real-world & Single & Exec+LLM-Judge & File \\
    & Spider2-v*\cite{cao2024Spider2v} & Multi & 2024 & None & Real-world & Agentic  & Exec & Repo \\
    & Web2Code\cite{NEURIPS2024_cb66be28} & HTML/CSS & 2024 & None & Mixed & Single & LLM-Judge & File \\
    & Design2Code\cite{si2025Design2Code} & HTML & 2025 & None & Real-world & Iterative & Exec+Human & File \\
    & MatPlotBench\cite{yang2024MatPlotAgent} & Python & 2024 & Perturbation & Real-world & Single & LLM-Judge & Program \\
    & BabelBench\cite{wang2024BabelBench} & Python & 2024 & None & Hand-written & Agentic & Exec+Match & Problem \\

  \midrule
  % ================= Multilingual (6 rows) =================
  \multirow{6}{*}{\textit{\textbf{Multilingual}}} 
    & HumanEval-X\cite{zheng2023codegeex} & Multi & 2023 & None & Hand Translated & Single & Exec & Function \\
    & MathQA-X\cite{Athiwaratkun2022MultilingualEO} & Multi & 2023 & None & Synthetic & Single & Exec & Function \\
    & MBXP\cite{Athiwaratkun2022MultilingualEO} & Multi & 2023 & None & Synthetic & Single & Exec & Function \\
    & \makecell{Multilingual\\HumanEval}\cite{Athiwaratkun2022MultilingualEO} & Multi & 2023 & None & Synthetic & Single & Exec & Function \\
    & MCEval\cite{chai2024mceval} & Multi & 2024 & None & Hand-written & Single & Exec & Problem \\
    & MultiPL-E\cite{cassano2022multipl} & Multi & 2022 & None & Synthetic & Single & Exec & Function \\

  \midrule
  % ================= Domain-specific (16 rows) =================
  \multirow{16}{*}{\textit{\textbf{Domain-specific}}} 
    & DS-1000\cite{lai2023ds} & Python & 2023 & Perturbation & Real-world & Single & Exec & Program \\
    & RoboScript\cite{chen2024roboscript} & Python & 2024 & None & Hand-written & Iterative & Exec & Task \\
    & PathBench\cite{hsueh2022systematic} & Python & 2022 & None & Mixed & Single & Exec & Task \\
    & CoBRA\cite{mayer2024cobra} & - & 2024 & None & Real-world & Single & Exec & Task \\
    & BioCoder\cite{tang2024biocoder} & Python/Java & 2024 & None & Real-world & Single & Exec & Function \\
    & MLAgentBench*\cite{huang2023BENCHMARKING} & Python & 2023 & Time-cutoff & Real-world & Agentic & Exec & Repo \\
    & ML-bench*\cite{tang2024MLbench} & Python Bash & 2024 & Rewrite & Real-world & Agentic & Exec & Repo \\
    & Robotouille\cite{gonzalez-pumariega2025robotouille} & - & 2025 & None & Synthetic & Agentic & Exec & Task \\
    & R-benchmark\cite{miah2024user} & R & 2024 & None & Real-world & Iterative & Human & Snippet \\
    & RTL-repo\cite{allam2024rtl} & Verilog & 2024 & Time-cutoff & Real-world & Single & Match & Repo \\
    & VerilogEval\cite{liu2023verilogeval} & Verilog & 2023 & None & Real-world & Single & Exec & Function \\
    & ChiBench\cite{sumitani2024chibench} & Verilog & 2024 & None & Real-world & Single & Match & File \\
    & FVEval\cite{kang2024fveval} & SV & 2024 & None & Mixed & Single & Formal Verif. & Task \\
    & Deep-Bench\cite{daghighfarsoodeh2025Deepbench} & Python & 2025 & None & Real-world & Single & Exec & Function \\
    & LessLeak-Bench\cite{zhou2025LessLeakBench} & Multi & 2024 & - & Real-world & Single & Mixed & - \\
    & StudentEval\cite{babe2023StudentEval} & Python & 2023 & None & Hand-written & Single & Exec & Function \\
    
   \midrule
   % ================= Others (4 rows) =================
   \multirow{4}{*}{\textit{\textbf{Others}}} 
    & JuICe\cite{agashe2019juice} & Python & 2019 & None & Real-world & Single & Match & Cell \\
    & Exec-CSN\cite{xie2024codebenchgen} & Python & 2024 & None & Real-world & Single & Exec & Function \\
    & DomainEval\cite{zhu2024domaineval} & Python & 2024 & None & Real-world & Single & Exec & Function \\
    & Race\cite{zheng2024Beyond} & Python & 2024 & None & Real-world & Single & Exec+Static & Function \\

  \bottomrule
  \end{tabular}
  \end{adjustbox}
\vspace{-5mm}
\end{table*}

Software implementation is the core phase of the SDLC, where software design is translated into executable code. Benchmarks for code generation tasks are presented in ~\autoref{tab:CodeGeneration}, while other implementation-related tasks are summarized in ~\autoref{tab:elseCodeBenchmarks}.

\subsubsection{Code Generation and Completion}

The objective of code generation is to automatically produce code from natural language descriptions.
As shown in~\autoref{tab:CodeGeneration}, we categorize existing benchmarks by application scenario into six classes: \textit{General SE}, \textit{Competitive Programming}, \textit{Multimodal}, \textit{Multilingual}, \textit{Domain-specific}, and \textit{Other}.

% 代码生成旨在将自然语言描述转化为可执行代码。该领域的基准测试通常依据任务粒度被划分为不同层级，涵盖函数级、类级和仓库级。如表xxx所示，我们根据应用场景将现有基准分为六大类：通用软件工程（General SE）、程序竞赛、多模态、多语言代码生成,领域特定和其他任务。

% 此外，还存在专注于特定领域的基准测试，包括竞赛编程、多语言代码生成、领域特定任务等。

% The objective of code generation is to automatically generate code based on natural language descriptions. In this field, benchmarks are categorized into various levels according to the granularity of the code generation tasks, such as function-level, class-level, and repository-level benchmarks. Furthermore, there are benchmarks focused on specific areas, including competitive programming, multilingual code generation, domain-specific tasks, and others.

% 早期基准测试如HumanEval、MBPP和APPS提供了带有测试用例的独立代码片段静态集合，虽对初期模型开发具有价值，但面临数据集污染以及缺乏真实场景上下文和依赖关系的局限性。EvalPlus等研究通过变异测试增强了静态基准的鲁棒性，但仍未能完全反映真实世界的代码复杂性，也无法实现代码仓库的自动构建。多语言基准测试如HumanEval-X、Aider多语言基准和AutoCodeBench虽然能评估跨语言能力，但存在数据泄露风险。聚焦仓库级上下文或依赖关系的基准测试包括RepoBench、CrossCodeEval、R2E、DevEval、BigCodeBench和CODEAGENT。WebBench引入了连续的真实世界网页开发任务，虽然捕捉了实际交互的某些方面，但往往缺乏充分测试

\textit{General SE.}
General software engineering (SE) benchmarks are among the most widely used evaluation standards and are intended to assess LLMs’ fundamental coding capabilities.
We group these benchmarks by generation granularity into three levels: function-level, class-level, and repository-level. 
Function-level benchmarks focus on generating or completing individual functions.
\textbf{HumanEval}~\cite{chen2021evaluating} and \textbf{MBPP}~\cite{austin2021program} are seminal benchmarks that established a strong foundation for function-level evaluation. HumanEval contains 164 human-written Python programming problems, each with an average of 9.6 test cases. MBPP targets basic programming tasks and includes approximately 974 lower-complexity problems. 
However, these benchmarks suffer from inadequate test coverage and a lack of realism. 
\textbf{HumanEval+}~\cite{liu2024your} and \textbf{MBPP+}~\cite{liu2024your} expand the test suites by roughly 80$\times$, substantially improving coverage of edge cases. Dong et al.~\cite{DongCodeScore2024} propose \textbf{HumanEval-ET} and \textbf{MBPP-ET}, which leverage LLMs to predict execution outcomes, enabling functional correctness evaluation even when test cases are unavailable. In addition, \textbf{BigCodeBench}~\cite{zhuo2024bigcodebench} broadens evaluation to settings that involve multi-library function calls, thereby moving code evaluation closer to practical use cases.

Class-level benchmarks focus on designing classes and implementing their attributes and methods for object-oriented programming. \textbf{ClassEval}~\cite{du2023classeval} provides 100 class-level code generation examples to assess LLM performance on class-level generation tasks.

To bridge the gap between traditional implementation benchmarks and real-world development scenarios, a number of repository-level benchmarks have emerged in recent years. These benchmarks have progressively expanded evaluation dimensions, improving the authenticity and reliability of assessments. Early repository-level benchmarks such as \textbf{CoNaLa}~\cite{yin2018learning} and \textbf{ConCode}~\cite{iyer2018mapping} were not designed for LLMs and lacked sufficiently comprehensive test suites, which limited their applicability.

Subsequent work has broadly progressed along two directions: (i) expanding the contextual information available to models, and (ii) simulating real-world software development processes.  
For context expansion, \textbf{CoderEval}~\cite{yu2024codereval} extends evaluation from independent functions to non-independent function generation, which requires understanding local, in-file context (e.g., class members and method calls). \textbf{PragmaticCode}~\cite{agrawal2023guiding} further expands context dependencies to a cross-file, global scope (e.g., API protocols). To enhance realism, \textbf{Devevall}~\cite{li2024deveval} aligns benchmarks with real-world repositories across multiple dimensions, such as code and dependency distributions. \textbf{CodeAgentBench}~\cite{zhang2024codeagent} evaluates full-repository settings by incorporating documentation, environments, and dependencies, aiming to assess AI agents’ end-to-end problem-solving ability under realistic development workflows.
In parallel, another line of research focuses on simulating real-world software development processes. \textbf{HumanEvo}~\cite{zheng2024towards} introduces an evolution-aware mechanism to evaluate adaptability to projects over time. \textbf{FEA-Bench}~\cite{li2025fea} targets incremental new-feature development scenarios. \textbf{EvoCodeBench}~\cite{li2024evocodebench} designs dynamic update mechanisms and domain-specific classification methods to mitigate data leakage and improve the reliability of domain-specific evaluations. 
% and \textbf{ML-bench}~\cite{tang2024MLbench}

\textit{Code Completion} aims to assist developers by predicting the next line of code or completing code snippets given the surrounding context.
\textbf{RepoBench}~\cite{liu2023repobench}, \textbf{RepoEval}~\cite{zhang2023repocoder}, and \textbf{CrossCodeEval}~\cite{ding2024crosscodeeval} all target repository-level code completion. RepoBench proposes a structured evaluation framework with three subtasks: RepoBench-R (cross-file retrieval), RepoBench-C (context-aware completion), and RepoBench-P (end-to-end workflow evaluation). RepoEval evaluates completion at three granularities—line level, API-call level, and function-body level—and verifies functional correctness by executing unit tests. In contrast, CrossCodeEval focuses on cross-file completion; it covers four mainstream languages and is designed to assess LLMs’ ability to model cross-file dependencies in multilingual settings.

\textit{Competitive Code Generation benchmarks.}
These benchmarks draw problems from competitive programming platforms to evaluate LLMs’ performance on contest-style code generation. An early benchmark, \textbf{APPS}~\cite{Hendrycks2021MeasuringCC}, introduced a large-scale dataset of roughly 10,000 problems sourced from platforms such as Codeforces and Kattis. It also provides more than 130,000 test cases and 230,000 reference solutions. However, as LLMs have rapidly advanced, such static datasets face substantial risks of data leakage. To mitigate this issue, later work proposes several improvements. \textbf{CodeContests}~\cite{li2022competition} and \textbf{LeetCode}~\cite{guo2024deepseek} adopt strict temporal splits to ensure that test data postdates training data, thereby reducing leakage. 
\textbf{LiveCodeBench}~\cite{naman2024livecodebench} implements a continuous evaluation framework that collects problems released after the model's training cutoff, thereby ensuring zero data leakage by design. 
Finally, \textbf{CODEELO}~\cite{quan2025codeelo} evaluates generated solutions by automatically submitting code to Codeforces, leveraging the platform’s execution environment and hidden test sets, thereby further preventing data contamination.

\textit{Multilingual benchmarks.}
To meet the need for evaluating code generation across multiple programming languages (PLs), benchmarks have evolved from initially supporting a single language (e.g., Python) to covering a broader range of PLs. \textbf{HumanEval-X}~\cite{zheng2023codegeex} manually translated 164 Python problems into 4 programming languages (C++, Java, JavaScript, Go), creating a multilingual code-generation benchmark. \textbf{MultiPL-E}~\cite{cassano2022multipl} leverages compilers to automatically translate HumanEval and MBPP into 18 PLs, enabling evaluation across a wide range of (including low-resource) languages. \textbf{MCEval}~\cite{chai2024mceval} further expands coverage to 40 PLs, substantially broadening the scope of multilingual evaluation.

\textit{Multimodal benchmarks.}
The rise of multimodal large language models (MLLMs) has spurred new programming paradigms and, consequently, a growing set of benchmarks to evaluate their capabilities.
In recent years, multiple benchmarks have been proposed to systematically assess code generation with multimodal inputs. \textbf{MMCode}~\cite{li2024MMCodea} introduced an early large-scale multimodal code benchmark that uses visual elements as cues for code generation, evaluating models’ understanding of mixed text--image inputs. Building on this, \textbf{BabelBench}~\cite{wang2024BabelBench} integrates images, text, and structured tables to assess comprehension of multimodal inputs. In specialized application domains, \textbf{Plot2Code}~\cite{wu2024Plot2Code} and \textbf{MatPlotBench}~\cite{yang2024MatPlotAgent} focus on data-visualization code generation, while \textbf{Web2Code}~\cite{NEURIPS2024_cb66be28} and \textbf{Design2Code}~\cite{si2025Design2Code} target front-end code generation from visual designs. For code agents, \textbf{Spider2-v}~\cite{cao2024Spider2v} further introduces multimodal interaction and real-world operations, serving as an early benchmark for multimodal agents that cover an end-to-end data engineering workflow (from storage to orchestration).

\textit{Domain-specific benchmarks.}
As code generation expands into specialized domains, benchmarks tailored to particular fields have emerged.
\textbf{DS1000}~\cite{lai2023ds} targets data science and systematically evaluates models’ code generation ability for data analysis tasks. \textbf{BioCoder}~\cite{tang2024biocoder} focuses on bioinformatics, while \textbf{MLAgentBench}~\cite{huang2023BENCHMARKING} assesses LLM-based agents’ ability to conduct end-to-end machine learning experiments. \textbf{DeepBench}~\cite{daghighfarsoodeh2025Deepbench} is designed to evaluate code generation in deep learning workflows.
\textbf{LessLeak-Bench}~\cite{zhou2025LessLeakBench} targets data leakage detection in CodeLLM benchmarks, and \textbf{StudentEval}~\cite{babe2023StudentEval} is intended for educational settings, providing 1,749 basic programming prompts.
In robotics, several benchmarks have been proposed, including \textbf{RoboScript}~\cite{chen2024roboscript}, \textbf{PathBench}~\cite{hsueh2022systematic}, \textbf{CoBRA}~\cite{mayer2024cobra}, and \textbf{Robotouille}~\cite{gonzalez-pumariega2025robotouille}, which support research on code generation for embodied AI.
Benchmarks for domain-specific languages have also emerged. For example, \textbf{R-benchmark}~\cite{miah2024user} is an early benchmark designed specifically for the R language. For hardware description languages, notable benchmarks include \textbf{RTL-Repo}~\cite{allam2024rtl}, \textbf{VerilogEval}~\cite{liu2023verilogeval}, and \textbf{ChiBench}~\cite{sumitani2024chibench} for Verilog, as well as \textbf{FVEval}~\cite{kang2024fveval} for SystemVerilog.

\textit{Other benchmarks.}
Some benchmarks target specific aspects of code generation.
\textbf{JuICe}~\cite{agashe2019juice} is an early benchmark that explores interactive code generation. To reduce the cost of manual benchmark construction, \textbf{Exec-CSN}~\cite{xie2024codebenchgen} leverages LLMs to automatically build high-quality and diverse code-generation benchmarks. \textbf{DomainEval}~\cite{zhu2024domaineval} focuses on cross-domain generalization, covering 2,454 topics across six major application areas to assess the breadth of LLM applicability. In addition, to go beyond traditional functional-correctness evaluation, \textbf{Race}~\cite{zheng2024Beyond} assesses code quality along four dimensions: readability, maintainability, correctness, and runtime efficiency.

% \textbf{JuICe}~\cite{agashe2019juice} is the first to explore interactive code generation.
% 针对人工构建评测基准过程繁琐的问题，
% \textbf{Exec-CSN}~\cite{xie2024codebenchgen} 利用
% 利用大语言模型来自动化地构建高质量、多样化的代码生成评测基准。
% % introduces a sandboxing process to ensure that generated code can run.
% 而 DomainEval~\cite{zhu2024domaineval} 则聚焦于跨领域泛化能力，覆盖六大应用方向下的 2,454 个主题，以检验模型的广泛适用性。
% 此外，为超越传统正确性评估的局限，Race~\cite{zheng2024Beyond} 进一步从可读性、可维护性、正确性及运行时效率四个维度综合评价代码质量。

% \textbf{Race}~\cite{zheng2024Beyond} assesses code quality from four key perspectives: readability, maintainability, correctness, and runtime efficiency.
% \textbf{DomainEval}~\cite{zhu2024domaineval} focuses on cross-domain generalization, covering 2,454 topics across six application areas.

\begin{table*}[t]
  \centering
  \vspace{-2mm}
  \tiny
  \setlength{\tabcolsep}{2pt}
  \caption{Analysis of Software Implementation Benchmarks (* indicates benchmarks applicable to agents).}
  \label{tab:elseCodeBenchmarks}
  \begin{adjustbox}{width=0.95\textwidth}

  \begin{tabular}{c c c c c c c c c}
  \toprule
  \multirow{3}{*}{\textbf{Task}} &
  \multirow{3}{*}{\textbf{Benchmarks}} &
  \multirow{3}{*}{\textbf{Language}} &
  \multirow{3}{*}{\textbf{Year}} &
  \multicolumn{4}{c}{\textbf{General Dimensions}} & % 现第5-8列
  \multicolumn{1}{c}{\textbf{Stage-Specific}} \\ % 现第9列
  \cmidrule(lr){5-8} \cmidrule(lr){9-9}
   & & & & 
  \textbf{\makecell{Anti-Contamination\\Strategy}} &
  \textbf{Authenticity} &
  \textbf{\makecell{Interaction\\Mode}} &
  \textbf{\makecell{Evaluation\\Paradigm}} &
  \textbf{\makecell{Context\\Scope}} \\ % 移至此处
  \midrule

  % ===== Text-to-SQL (12 rows) =====
  \multirow{12}{*}{\textbf{\textit{Text-to-SQL}}}
    & WikiSQL\cite{zhong2017seq2sql} & SQL & 2017 & None & Hand-written & Single & Exec & Statement \\
    & Spider\cite{yu2018spider} & SQL & 2018 & None & Hand-written & Single & Exec+Match & Database \\
    & Spider-Syn\cite{gan2021towards} & SQL & 2021 & None & Mixed & Single & Exec+Match & Database \\
    & Spider-DK\cite{gan2021exploring} & SQL & 2021 & None & Mixed & Single & Exec+Match & Database \\
    & Spider-Realistic\cite{deng2020structure} & SQL & 2021 & None & Mixed & Single & Exec+Match & Database \\
    & CSpider\cite{min-etal-2019-pilot} & SQL & 2019 & None & Translated & Single & Exec+Match & Database \\
    & BIRD\cite{li2024can} & SQL & 2023 & Time-cutoff & Real-world & Single & Exec & Database \\
    & KaggleDBQA\cite{lee2021kaggledbqa} & SQL & 2021 & None & Real-world & Single & Match & Database \\
    & SParC\cite{yu2019sparc} & SQL & 2019 & None & Mixed & Iterative & Match & Database \\
    & CoSQL\cite{yu2019cosql} & SQL & 2019 & None & Hand-written & Iterative & Match & Database \\
    & Ehrsql\cite{lee2022ehrsql} & SQL & 2022 & Shuffling & Hand-crafted & Single & Exec & Database \\
    & Termite\cite{ranaldi2024investigating} & SQL & 2024 & Encryption & Hand-crafted & Single & Exec & Database \\

  \midrule

  % ===== Code Understand & Reason (13 rows) =====
  \multirow{13}{*}{\textbf{\textit{\makecell{Code Understand \\\& Reason}}}}
    & Infibench\cite{li2025infibench} & Multi & 2024 & Paraphrase & Real-world & Single & Static & Program \\
    & CodeQA\cite{liu2021codeqa} & Python/Java & 2021 & None & Synthetic & Single & Match & Function \\
    & CS1QA\cite{lee2022cs1qa} & Python & 2022 & None & Real-world & Single & Match & Program \\
    & DQABench\cite{zheng2024Revolutionizing} & NL & 2024 & None & Mixed & Single & Match & Database \\
    & CRUXEval\cite{gu2024cruxeval} & Python & 2024 & None & Synthetic & Single & Exec & Function \\
    & CRUXEVAL-X\cite{xu2024cruxeval} & Multi & 2024 & None & Synthetic & Single & Exec & Function \\
    & CodeMMLU\cite{manh2024codemmlu} & Multi & 2024 & None & Synthetic & Single & Match & Snippet \\
    & REval\cite{chen2024evaluating} & Python & 2024 & None & Real-world & Single & Match & Function \\
    & CodeApex\cite{fu2023codeapex} & C++ & 2023 & Time-cutoff & Real-world & Single & Exec & Function \\
    & SpecEval\cite{ma2024speceval} & Java & 2024 & None & Hand-written & Single & Exec & Function \\
    & FAUN-Eval\cite{hu2024real} & Multi & 2024 & None & Real-world & Single & Static & File \\

  \midrule

  % ===== Code Translation (2 rows) =====
  \multirow{2}{*}{\textbf{\textit{Code Translation}}}
    & PolyHumanEval\cite{tao2024unraveling} & Multi & 2024 & None & Hand Translated & Single & Exec & Function \\
    & CodeTransOcean\cite{yan2023CodeTransOcean} & Multi & 2023 & None & Real-world & Single & Match+Exec & Program \\

  \midrule

  % ===== Code Summary (3 rows) =====
  \multirow{3}{*}{\textbf{\textit{Code Summary}}}
    & CODE-NN\cite{iyer2016summarizing} & C\#/SQL & 2016 & None & Real-world & Single & Match & Function \\
    & DeepCom\cite{hu2018deep} & Java & 2018 & None & Real-world & Single & Match & Function \\
    & TL-CodeSum\cite{hu2018summarizing} & Java & 2018 & None & Real-world & Single & Match & Function \\

  \midrule

  % ===== Type Inference (2 rows) =====
  \multirow{2}{*}{\textbf{\textit{Type Inference}}}
    & Lambdanet\cite{wei2020lambdanet} & TS & 2020 & None & Real-world & Single & Match & Project \\
    & IdBench\cite{wainakh2019Evaluating} & JS & 2019 & None & Real-world & Single & Match & Identifier \\

  \midrule

  % ===== Code Retrieval (2 rows) =====
  \multirow{4}{*}{\textbf{\textit{Code Retrieval}}}
    & DeepCS\cite{gu2018deep} & Java & 2018 & None & Real-world & Single & Human & Function \\
    & CodeSearchNet\cite{husain2019codesearchnet} & Multi & 2019 & None & Real-world & Single & Static & Function \\
    & CoSQA\cite{huang2021cosqa} & Python & 2021 & None & Real-world & Single & Match & Function \\
    & CoSQA+\cite{gong2024cosqa+} & Python & 2024 & None & Real-world & Single & Match & Function \\

  \bottomrule
  \end{tabular}
  \end{adjustbox}
\vspace{-5mm}
\end{table*}

\subsubsection{Text-to-SQL}

LLMs have become a key driver of progress in text-to-SQL research.
Early benchmarks such as \textbf{WikiSQL}~\cite{zhong2017seq2sql} primarily target single-table queries. Because these queries are relatively simple, WikiSQL is less representative of real-world needs in the era of LLMs.
\textbf{Spider}~\cite{yu2018spider} is the most widely used benchmark for text-to-SQL tasks. It contains 10,181 natural language questions and 5,693 complex SQL queries across 138 domains and 200 relational databases. Despite its scale, many queries rely on overly specific references to column names and values, which limits how well the benchmark reflects realistic user requests.

To increase realism and difficulty, several Spider variants have been proposed. \textbf{Spider-Syn}~\cite{gan2021towards} improves linguistic diversity via synonym replacement and structurally complex rewrites. \textbf{Spider-DK}~\cite{gan2021exploring} injects domain knowledge to better approximate natural user phrasing. \textbf{Spider-Realistic}~\cite{deng2020structure} removes explicit schema cues (e.g., column names) to better reflect real user queries. \textbf{CSpider}~\cite{min-etal-2019-pilot} is a Chinese translation of Spider that preserves the original schema structure and data distribution.

Beyond single-turn settings, \textbf{SParC}~\cite{yu2019sparc} and \textbf{CoSQL}~\cite{yu2019cosql} extend text-to-SQL evaluation to multi-turn dialogue. SParC emphasizes cross-domain generalization across 138 domains, whereas CoSQL focuses on simulating realistic database interactions through conversational workflows.
To further improve benchmark authenticity and practicality, \textbf{KaggleDBQA}~\cite{lee2021kaggledbqa} and \textbf{BIRD}~\cite{li2024can} adopt real-world, multi-table, cross-domain databases. KaggleDBQA introduces external documents as implicit knowledge sources and preserves domain-specific data types, formats, and natural language questions. BIRD scales up to 95 cross-domain databases (33.4 GB in total) and adds a metric for SQL execution efficiency.

Recent work has also explored domain adaptation and data timeliness. \textbf{EhrSQL}~\cite{lee2022ehrsql} is the first text-to-SQL benchmark tailored to the clinical domain. \textbf{Termite}~\cite{ranaldi2024investigating} targets data contamination by constructing an encrypted dataset.

\subsubsection{Code Understanding and Reasoning}
Code understanding and reasoning are core capabilities for CodeLLMs. Code understanding captures syntax and semantics, whereas reasoning supports logical inference and behavior prediction.

\textit{Question-Answering (QA).}
\textbf{Infibench}~\cite{li2025infibench} targets open-code question answering and is constructed from 234 real Stack Overflow questions. It covers 15 programming languages and five major topical areas.
Related benchmarks, including \textbf{CodeQA}~\cite{liu2021codeqa}, \textbf{CS1QA}~\cite{lee2022cs1qa}, and \textbf{DQABench}~\cite{zheng2024Revolutionizing}, study QA in different settings: CodeQA derives QA pairs from code comments to emphasize semantic understanding; CS1QA focuses on introductory programming education; and DQABench targets the database domain.

\textit{Code Reasoning.}
\textbf{CRUXEval}~\cite{gu2024cruxeval} evaluates reasoning via input/output prediction, comprising 800 Python functions with I/O pairs; \textbf{CRUXEval-X}~\cite{xu2024cruxeval} extends it to multiple languages.
\textbf{CodeMMLU}~\cite{manh2024codemmlu} adopts a multiple-choice format spanning over 50 domains and 10 programming languages.
In contrast, \textbf{REval}~\cite{chen2024evaluating} moves beyond I/O reasoning by emphasizing intermediate runtime behavior.

\textit{Code Understanding.}
\textbf{CodeApex}~\cite{fu2023codeapex} provides a bilingual evaluation suite covering multiple-choice questions, code generation, and code correction.
\textbf{SpecEval}~\cite{ma2024speceval} introduces formal program specifications (e.g., JML) and evaluates code understanding through four tasks: judgment, selection, filling, and generation.
\textbf{FAUN-Eval}~\cite{hu2024real} decomposes real-world code issues into three atomic tasks, namely code QA, fault localization, and code editing, to systematically assess fine-grained problem-solving ability.

\subsubsection{Code Translation}
Code translation focuses on converting code from one programming language to another while preserving functional equivalence.
\textbf{PolyHumanEval}~\cite{tao2024unraveling} extends HumanEval to 14 programming languages, providing manually curated equivalent solutions across languages and automatically generated test cases for evaluation.
To broaden language coverage further, \textbf{CodeTransOcean}~\cite{yan2023CodeTransOcean} introduces a large-scale dataset spanning over 40 programming languages and proposes an execution-based metric, DSR@K, to assess translation quality.

\subsubsection{Code Summarization}
Code summarization aims to generate natural-language descriptions for code snippets, functions, or complete programs.
Early work such as \textbf{CODE-NN}~\cite{iyer2016summarizing} targets C\# and SQL, pairing StackOverflow post titles with code snippets from accepted answers.
Later benchmarks, including \textbf{DeepCom}~\cite{hu2018deep} and \textbf{TL-CodeSum}~\cite{hu2018summarizing}, focus on Java and construct datasets from GitHub repositories. 
% 早期数据集CODE-NN~\cite{iyer2016summarizing}针对C#与SQL代码，其数据来自StackOverflow帖子标题与采纳答案中的代码片段配对。随后提出的DeepCom~\cite{hu2018deep}与TL-CodeSum~\cite{hu2018summarizing}则针对Java代码，数据源自GitHub代码库。

% \textbf{CODE-NN}~\cite{iyer2016summarizing} is an early dataset for C\# and SQL summarization, created by pairing StackOverflow post titles with code snippets from accepted answers. In contrast, \textbf{DeepCom}~\cite{hu2018deep} provides a larger scale dataset based on Java, sourced from GitHub repositories between 2015 and 2016. Building on this, \textbf{TL-CodeSum}~\cite{hu2018summarizing} introduces a more refined, method-level Java summarization benchmark, with data from GitHub repositories during the same time period.

\subsubsection{Type Inference}
Type inference aims to automatically determine the types of variables, functions, or expressions in source code.
\textbf{IdBench}~\cite{wainakh2019Evaluating} evaluates identifier embeddings via human-annotated semantic similarity scores.
\textbf{Lambdanet}~\cite{wei2020lambdanet} is built on 300 TypeScript codebases from GitHub, which contain abundant user-defined type annotations and variables.

\subsubsection{Code Retrieval}
Code retrieval focuses on identifying relevant code snippets or functions given a natural-language query.
\textbf{DeepCS}~\cite{gu2018deep}, an early effort in this area, constructed a Java code--query dataset but is limited to a single programming language.
Subsequently, \textbf{CodeSearchNet}~\cite{husain2019codesearchnet} became a widely used benchmark by covering six programming languages and providing approximately six million paired examples; however, its queries are still not fully authentic.
To improve query realism, \textbf{CoSQA}~\cite{huang2021cosqa} collects real user queries from Bing and annotates them with GitHub code snippets at scale.
Its enhanced version, \textbf{CoSQA+}~\cite{gong2024cosqa+}, targets the more challenging one-to-many query--code matching setting.

\begin{table*}[t]
  \centering
  \vspace{-2mm}
  % \tiny % 使用最小字号
  \setlength{\tabcolsep}{1.5pt} % 紧凑列间距
  \caption{Analysis of Software Implementation Benchmarks (Non-functional Benchmarks; * indicates benchmarks applicable to agents Note: We identified no publicly available benchmarks for Privacy or Human-AI Collaboration.).}
  % \xy{If you use tiny font in the table, make sure all this kind of tables use tiny. Make them the same setting. In addtion, privacy dimension can be removed*}}
  \label{tab:nofunction}
  \begin{adjustbox}{width=0.95\textwidth}

  % 9 列定义
  \begin{tabular}{c c c c c c c c c} 
  \toprule
  \multirow{3}{*}{\textbf{Property}} &
  \multirow{3}{*}{\textbf{Benchmarks}} & 
  \multirow{3}{*}{\textbf{Language}} & 
  \multirow{3}{*}{\textbf{Year}} & 
  \multicolumn{4}{c}{\textbf{General Dimensions}} & % 现第5-8列
  \multicolumn{1}{c}{\textbf{Stage-Specific}} \\ % 现第9列
  \cmidrule(lr){5-8} \cmidrule(lr){9-9}
   & & & & 
  \textbf{\makecell{Anti-Contamination\\Strategy}} & 
  \textbf{Authenticity} & 
  \textbf{\makecell{Interaction\\Mode}} & 
  \textbf{\makecell{Evaluation\\Paradigm}} &
  \textbf{\makecell{Context\\Scope}} \\ % 移至此处
  
  \midrule
  % ================= Robustness (1 row) =================
  \multirow{1}{*}{\textbf{\textit{Robustness}}}
    & ReCode\cite{wang-etal-2023-recode}       & Python      & 2023 & None         & Synthetic    & Single          & Exec & Function \\
    
  \midrule
  % ================= Security (3 rows) =================
  \multirow{3}{*}{\textbf{\textit{Security}}}
    & RedCode*\cite{guo2024redcode}              & Python/Bash & 2024 & None         & Synthetic    & Agentic     & Exec & File \\ 
    & RMCBench\cite{chen2024rmcbench}           & Multi   & 2024 & None         & Real-world   & Single          & LLM-based & Task \\ 
    & AdvBench\cite{zou2023universal}           & NL      & 2023 & None         & Synthetic    & Single          & Match & Prompt \\ 

  \midrule
  % ================= Copyright (2 rows) =================
  \multirow{2}{*}{\textbf{\textit{Copyright}}}
    & HMCorp\cite{xu2024distinguishing}         & Python/Java & 2024 & None         & Real-world   & Single          & Classification & Function \\ 
    & D$\alpha$-C8\cite{nguyen2024gptsniffer} & Java    & 2024 & None         & Mixed        & Single          & Classification & Snippet \\ 

  \midrule
  % ================= Efficiency (5 rows) =================
  \multirow{5}{*}{\textbf{\textit{Efficiency}}}
    & ECCO\cite{waghjale2024ecco}               & Python      & 2024 & Time-cutoff & Hand-written & Single          & Exec & Function \\ 
    & EffiBench\cite{huang2024Effibench}        & Python      & 2024 & None         & Hand-written & Single          & Exec & Function \\ 
    & Mercury\cite{du2024Mercury}               & Python      & 2024 & Updates      & Mixed        & Single          & Exec & Task \\ 
    & EvalPerf\cite{liu2024evaluating}          & Python      & 2024 & None         & Hand-written & Single          & Exec & Function \\ 
    & ENAMEL\cite{qiu2024efficient}             & Python      & 2025 & None         & Hand-written & Single          & Exec & Function \\ 

  \midrule
  % ================= Bias (3 rows) =================
  \multirow{3}{*}{\textbf{\textit{Bias}}}
    & FairCoder\cite{du2025FairCoder}           & Python/C    & 2025 & None         & Synthetic    & Single          & Match & Function \\ 
    & SocialBias-Bench\cite{ling2025Bias}       & Python      & 2025 & None         & Real-world   & Single          & Exec & Function \\ 
    & Zhang's\cite{zhang2025unveiling}          & Python      & 2025 & None         & Synthetic    & Single          & Match & Snippet \\ 

  \midrule
  % ================= Alignment (1 row) =================
  \multirow{1}{*}{\textbf{\textit{Alignment}}}
    & CodeArena\cite{yang2024evaluating}        & Multi   & 2024 & None         & Real-world   & Single          & LLM-Judge & Snippet \\ 

  \midrule
  % ================= Explainability (1 row) =================
  \multirow{1}{*}{\textit{\textbf{Explainability}}}
    & Galeras\cite{rodriguez2023benchmarking} & Python      & 2023 & Time-cutoff & Real-world   & Single          & Match & Function \\
    
  % \midrule
  % \textbf{\textit{\makecell{Privacy\&Human-AI\\ Collaboration}}}&\textit{None Identified}&-&-&-&-&-&-&-\\

  \bottomrule
  \end{tabular}
  \end{adjustbox}
\vspace{-5mm}
\end{table*}

\subsubsection{Non-functional Benchmarks}

\autoref{tab:nofunction} presents the non-functional benchmarks. 
Aligned with the software quality models in the ISO/IEC standard~\cite{iso25010} and recent multidimensional evaluations~\cite{zheng2024Beyond}, these benchmarks assess system-level performance, behavior, and constraints (e.g., efficiency and security), rather than focusing solely on functional correctness.

% \autoref{tab:nofunction} presents the non-functional benchmarks. These evaluations focus on system's performance, behavior, and constraints, rather than its specific functional correctness.
% \xy{What is the ssource of these non functional dimentsion? cite paper to support it.*}

\textit{Robustness.}
Robustness reflects a CodeLLM's ability to maintain stable performance under input perturbations and noise.
\textbf{ReCode}~\cite{wang-etal-2023-recode} evaluates model behavior across a wide range of input modifications, covering over 30 natural transformations such as docstrings, function naming, and formatting. 

\textit{Security.}
Security is essential for ensuring that CodeLLMs generate safe code and are not easily exploited by malicious actors.
\textbf{RMCBench}~\cite{chen2024rmcbench} evaluates CodeLLM security across 11 malicious code types and nine languages. However, it is limited to static generation risks without assessing runtime behavior. To address this, \textbf{RedCode}~\cite{guo2024redcode} introduces a dynamic execution environment for code agents, supporting multilingual and mixed-language inputs to shift evaluation from static generation to dynamic interaction.   
% RMCBench~\cite{chen2024rmcbench}作为首个大语言模型生成恶意代码安全性的基准，覆盖11种恶意代码类型与9种编程语言，在代码生成场景的安全性评估方面具有开创意义。
% 然而，该基准主要关注代码生成的静态风险，未能评估代码在真实环境中的动态执行行为。为此，RedCode~\cite{guo2024redcode}转向对智能代码代理的评估，通过引入动态执行环境交互机制，并扩展对多语言及自然语言混合输入的支持，推动了代码安全评估从静态生成到动态交互的范式转变。

% \textbf{RMCBench}~\cite{chen2024rmcbench} assesses the model's ability to prevent malicious code generation with 473 adversarial prompts in text-to-code and code-to-code scenarios. 
% RMCBench 包含 473 个精心设计的提示，涵盖两种主要场景：文本到代码和代码到代码。RMCBench 涉及 11 种恶意代码类型（如病毒、蠕虫、网络攻击等）和 9 种编程语言（如 Python、C、Java 等）。RMCBench 是第一个系统评估 LLMs 在生成恶意代码方面安全性的基准。涵盖文本到代码、代码到代码两种场景，三个难度级别，两种代码任务，全面模拟真实攻击。

% \textbf{RedCode}~\cite{guo2024redcode} evaluates model security using 160 high-risk prompts based on function signatures and documentation, along with 4,050 high-risk test cases in Python and Bash, to check if the model or agent generates harmful code. 
% 评估代码智能体（code agents）在代码执行与生成过程中安全风险的基准测试。多数仅评估静态响应，缺乏真实系统交互；
% 测试用例规模小、场景单一；
% 未覆盖代码执行与生成两个维度；
% 未考虑多语言（Python/Bash）和自然语言指令混合输入。
% 代码代理不仅生成代码，还能执行代码并与操作系统、文件系统、网络等环境交互，这带来了比传统代码生成模型更复杂的安全隐患。然而，现有评估基准主要针对静态代码生成的安全性，缺乏对代码代理在动态执行环境中的综合风险评估。

\textit{Copyright.}
Copyright-related concerns for CodeLLMs primarily involve the ownership of AI-generated code and the risk of copyright infringement.
\textbf{HMCorp}~\cite{xu2024distinguishing}, currently the largest benchmark dataset for Python and Java, focuses on distinguishing AI-generated code from human-written code.
\textbf{D$\alpha$-C8}~\cite{nguyen2024gptsniffer} targets misuse risks in programming education and software development; by collecting Java snippets written by both humans and models, it supports detecting unauthorized code reuse.

% HMCorp~\cite{xu2024distinguishing}作为当前最大规模针对Python与Java的基准数据集，致力于区分AI生成代码与人类编写代码。D$\alpha$‑C8~\cite{nguyen2024gptsniffer}聚焦于编程教育与软件开发中的滥用风险，通过收集人类与模型生成的Java代码片段，实现对未经授权代码复用的检测。

\textit{Efficiency.}
Efficiency characterizes a CodeLLM's ability to produce correct code while optimizing runtime, memory consumption, and other resource usage.
\textbf{ECCO}~\cite{waghjale2024ecco} and \textbf{EffiBench}~\cite{huang2024Effibench} evaluate efficiency in competitive-programming settings by measuring runtime and memory, leveraging cloud environments to reduce environmental variability.
However, these benchmarks do not explicitly couple functional correctness with efficiency, which may obscure the model's true capabilities.
To address this, \textbf{Mercury}~\cite{du2024Mercury} proposes the Beyond metric to jointly reflect correctness and efficiency, and introduces a sandbox environment to enable fair comparisons.
Nevertheless, benchmarks that rely on specific runtime environments remain sensitive to hardware performance, making cross-platform and longitudinal comparisons difficult.
\textbf{EvalPerf}~\cite{liu2024evaluating} mitigates this issue by using hardware performance counters and scoring solutions relative to their performance rankings, thereby reducing fluctuations caused by hardware differences.
Moreover, traditional metrics based on truncated runtimes can overestimate the efficiency of Timed-out code.
To alleviate this, \textbf{ENAMEL}~\cite{qiu2024efficient} proposes the eff@k metric, which evaluates the top-$k$ best samples to reduce interference from Timed-out code.

\textit{Bias.}
Bias in CodeLLMs can manifest as societal, cultural, gender, and racial biases in generated code.
\textbf{FairCoder}~\cite{du2025FairCoder} and \textbf{SocialBias-Bench}~\cite{ling2025Bias} focus on social bias in code generation: FairCoder covers scenarios related to recruitment, education, and healthcare, whereas SocialBias-Bench emphasizes gender-, race-, and income-related biases. 
Beyond social bias,\textbf{ Zhang et al.}~\cite{zhang2025unveiling} reveal systematic preferences in the recommendation of third-party services by CodeLLMs and agents.

% \textbf{FairCoder}~\cite{du2025FairCoder}和
% \textbf{SocialBias-Bench}~\cite{ling2025Bias}聚焦于代码生成中的社会偏见，FairCoder关注recruitment, education, and healthcare.，SocialBias-Bench关注gender, race, or income。

% In addition to social bias, \textbf{Zhang et al.}~\cite{zhang2025unveiling} 关注CodeLLMs和智能体中的服务商偏见，揭示大型语言模型在推荐第三方服务时存在的系统性偏好

% propose a comprehensive dataset covering 6 coding tasks and 30 real-world scenarios to evaluate provider bias in CodeLLMs and agents, which is manifested as the preference for paid services from specific providers in code recommendation and generation.

% \textbf{FairCoder}~\cite{du2025FairCoder} focuses on biases in tasks like code generation and test case generation within sensitive areas such as recruitment, education, and healthcare.
% % 聚焦于软件开发流程中的社会偏见评估

% \textbf{SocialBias-Bench}~\cite{ling2025Bias} provides a broader bias evaluation framework, covering seven task categories with a total of 343 instances. It aims to assess whether models exhibit biases related to gender, race, or income when performing programming tasks.
% \xy{update one paper}

\textit{Alignment with human preferences.}
Alignment with human preferences requires model outputs to reflect human values and aesthetic judgments.
\textbf{CodeArena}~\cite{yang2024evaluating} comprises 397 high-quality samples spanning 40 task categories and 44 programming languages, enabling a comprehensive evaluation of preference alignment.

\textit{Explainability.}
Explainability concerns the transparency and interpretability of a model's behavior and decision-making, which can increase user trust.
\textbf{Galeras}~\cite{rodriguez2023benchmarking} quantifies how factors such as code complexity and the number of comments influence model performance, helping reveal underlying mechanisms in code generation.

% \textit{Privacy.}
% Privacy in CodeLLMs refers to safeguarding sensitive information during both training and inference. Currently, there is a lack of benchmarks specifically designed to assess the privacy of CodeLLMs.
% \xy{If there is no benchmark focus on this dimension ,just remove it. You can discuss the `privacy' as an `unaddressed' example in your observation*}

% 观察四：现有基准测试全面覆盖软件开发各阶段，尤其侧重代码生成任务。 当前代码生成基准测试涵盖函数级、类级、仓库级、多语言、多模态及多领域场景，为评估代码生成能力提供了坚实基础。
% Encryption:或对数据库字段进行加密.部分采用,Dynamic,即持续更新题目,即便成本高,但是对抗污染较为有效.
% 出现云端评估  EccO  CODEELO
% 你现在可以在表格里把 Anti-Contam. Strategy 细化为两类：

% Data-level (Time-cutoff, Perturbation, Rewrite, Dynamic, Decontam.)

% Evaluation-level (Cloud / Server-side)

% 持续更新（Updates）与在线评估（Online）成为防污染措施新趋势。

% 观察五： 绝大多数基准缺乏防护，面临泄露风险。时间截止或代码扰动等静态手段难挡大模型的持续更新，效果有限。相比之下，**持续更新（Updates）与在线评估（Online）**能从源头规避训练数据记忆，成为更有效的抗污染方案。

% 观察六: 现有的 Text-to-SQL 基准多依赖人工设计（Hand-crafted）查询，缺乏真实世界的查询. 语言覆盖严重向 Python 倾斜，对 C++/Go 等现代语言支持匮乏，且多语言基准多源于合成或人工翻译（Synthetic/Translated）而非原生数据。此外，代码理解与非功能属性评估仍局限于函数级（Function-level）范围，完全缺乏仓库级（Repo-level）的上下文，其中隐私性（Privacy）评估更是完全空白。

%  Observation 3:  交互模式受限于单轮范式，缺乏对环境交互能力的深度度量。现有基准多受限于单轮交互（Single-turn）模式，个别基准支持Interactive模式.然而评估体系缺乏对环境交互能力的原生支持,限制了对智能体解决实际工程问题能力的全面度量。

\begin{tcolorbox}[colframe=gray!140, colback=gray!10, coltitle=white, title=Observation from Software Implementation,breakable]

\textbf{Obs. 5:} 
Most benchmarks lack effective anti-contamination measures and therefore face substantial risks of data leakage.
Static defenses, such as time-based cutoffs or code perturbation, struggle to keep pace with the continual expansion of LLM training corpora and offer limited long-term effectiveness.
In contrast, dynamic approaches based on continuous benchmark updates and online evaluation reduce the chance of benchmark inclusion in pre-training data at the source.

\textbf{Obs. 6: }
Existing benchmarks exhibit structural limitations in both authenticity and scope.
Text-to-SQL benchmarks frequently contain explicit schema references, which can undermine the ambiguity inherent in real-world user queries.
Language coverage is heavily skewed toward Python, while multilingual evaluations predominantly rely on translation-based corpora, often overlooking language-specific idioms and native syntactic structures.
Furthermore, critical enterprise dimensions---especially privacy preservation and human--AI collaboration efficiency---remain unaddressed.

\textbf{Obs. 7: }
Most existing benchmarks are restricted to single-turn interaction, with only a small number supporting interactive settings. Moreover, current evaluation frameworks provide limited native support for environment interaction, constraining the assessment of agents’ capabilities in realistic engineering workflows.

\end{tcolorbox}

\subsection{Software Testing}
% Runtime Behaviors
% API Misuse
% 如果是 Static，不需要，门槛低。

% 如果是 Dynamic/Runtime，需要，门槛高（更符合 Agent 趋势）。

% 目标不是生成“能通过测试的代码”，而是生成“能有效发现Bug的测试”。
% What are the characteristics/standards of Software Testing benchmark?
% 任务覆盖范围有限、对演进式多样化场景支持不足的问题  到的各类缺陷的全面覆盖
% 真实性
% 多语言
% 软件测试有助于通过在开发生命周期的早期识别和解决问题来降低维护成本，从而防止后续出现更严重的问题。
Software testing aims to identify and fix defects, ensure compliance with requirements, and enhance software quality.
The following subsections outline key software testing tasks and their corresponding benchmarks, as summarized in~\autoref{tab:testing_benchmarks}.

\begin{table*}[t]
\centering
\tiny
% \scriptsize % 使用小字号以容纳更多信息
\setlength{\tabcolsep}{3pt} % 收紧列间距
\caption{Analysis of Software Testing Benchmarks (* indicates benchmarks applicable to agents).}
\label{tab:testing_benchmarks}
\begin{adjustbox}{width=\textwidth}

% 9 列定义 (Context Scope 移到最后)
\begin{tabular}{ccc c cccc c}
\toprule
\multirow{3}{*}{\textbf{Task}} & 
\multirow{3}{*}{\textbf{Benchmark}} & 
\multirow{3}{*}{\textbf{Language}} & 
\multirow{3}{*}{\textbf{Year}} & 
\multicolumn{4}{c}{\textbf{General Dimensions}} & % 现第5-8列
\multicolumn{1}{c}{\textbf{Stage-Specific}} \\ % 现第9列
\cmidrule(lr){5-8} \cmidrule(lr){9-9}
 & & & & 
\textbf{\makecell{Anti-Contamination\\Strategy}} & 
\textbf{Authenticity} & 
\textbf{\makecell{Interaction\\Mode}} & 
\textbf{\makecell{Evaluation\\Paradigm}} & 
\textbf{\makecell{Context\\Scope}} \\ % 移至此处
\midrule

% ================= Group 1: Vulnerability Detection (16 rows) =================
\multirow{16}{*}{\textbf{\textit{\makecell{Vulnerability\\ Detection}}}}
  & Juliet \cite{juliet_java_1.3} & Java & 2017 & None & Synthetic & Single & Static & Function \\
  & VulDeePecker \cite{li2018vuldeepecker} & C/C++ & 2018 & None & Real-world & Single & Static & Function \\
  & Devign \cite{zhou2019devign} & C & 2019 & None & Real-world & Single & Static & Function \\
  & Draper \cite{zhuang2021software} & C/C++ & 2019 & None & Real-world & Single & Static & Function \\
  & GREAT \cite{hellendoorn2019global} & Python & 2020 & None & Mixed & Single & Static & Function \\
  & MVD \cite{zou2019mu} & C/C++ & 2020 & None & Mixed & Single & Static & Function \\
  & Big-Vul \cite{fan2020ac} & C/C++ & 2020 & None & Real-world & Single & Static & Function \\
  & REVEAL \cite{chakraborty2021deep} & C/C++ & 2020 & None & Real-world & Single & Static & Function \\
  & SySeVR \cite{li2021sysevr} & C/C++ & 2021 & None & Mixed & Single & Static & Function \\
  & VUDENC \cite{wartschinski2022vudenc} & Python & 2021 & None & Real-world & Single & Static & Function \\
  & CrossVul \cite{nikitopoulos2021crossvul} & Multi & 2021 & None & Real-world & Single & Static & Function \\
  & CVEfixes \cite{bhandari2021cvefixes} & Multi & 2021 & Updates & Real-world & Single & Static & Function \\
  & FormAI \cite{tihanyi2023formai} & C & 2023 & None & Synthetic & Single & Exec & Function \\
  & DiverseVul \cite{chen2023diversevul} & C/C++ & 2023 & None & Real-world & Single & Static & Function \\
  & PrimeVul \cite{ding2024vulnerability} & C/C++ & 2024 & Time-cutoff & Real-world & Single & Static & Function \\
  & VulnPatchPairs \cite{risse2024uncovering} & C & 2024 & None & Real-world & Single & Static & Function \\

\midrule
% ================= Group 2: Test Generation (3 rows) =================
\multirow{3}{*}{\textbf{\textit{\makecell{Test\\ Generation}}}}
  & SF110 \cite{fraser2014large} & Java & 2014 & None & Real-world & Single & Exec & Function \\
  & Methods2Test \cite{tufano2022methods2test} & Java & 2022 & None & Real-world & Single & Exec & Function \\
  & SWT-Bench* \cite{mundler2024SWTbencha} & Python & 2024 & None & Real-world & Agentic & Exec & Repo \\

\midrule
% ================= Group 3: Defect/Assertion Detection (3 rows) =================
\multirow{3}{*}{\textbf{\textit{\makecell{Defect\\Detection}}}} 
  & Bugs.jar \cite{saha2018bugs} & Java & 2018 & None & Real-world & Single & Exec & Function \\
  & Bears \cite{madeiral2019bears} & Java & 2019 & None & Real-world & Single & Exec & Function \\
  & JIT-Defects4J \cite{ni2022best} & Java & 2022 & Time-cutoff & Real-world & Single & Exec & Function \\

\bottomrule
\end{tabular}
\vspace{-2mm}
\end{adjustbox}
\vspace{-5mm}
\end{table*}

\subsubsection{Vulnerability Detection}

Vulnerability detection aims to uncover latent security flaws in codebases.
Early benchmark datasets, such as \textbf{GREAT}~\cite{hellendoorn2019global}, \textbf{Juliet}~\cite{juliet_java_1.3}, and \textbf{MVD}~\cite{zou2019mu}, are typically constructed using synthetic methods, where vulnerability patterns are manually designed.
Although such datasets offer high label accuracy, they cover a limited range of error types and therefore cannot fully reflect the complexity of real-world software.
To mitigate this limitation, later work shifted toward mining vulnerabilities from real projects, often by extracting vulnerable--fixed pairs from fix commits in open-source repositories.

Benchmarks such as \textbf{Big-Vul}~\cite{fan2020ac}, \textbf{CVEfixes}~\cite{bhandari2021cvefixes}, \textbf{DiverseVul}~\cite{chen2023diversevul}, \textbf{SySeVR}~\cite{li2021sysevr}, \textbf{REVEAL}~\cite{chakraborty2021deep}, \textbf{CrossVul}~\cite{nikitopoulos2021crossvul}, and \textbf{PrimeVul}~\cite{ding2024vulnerability} primarily rely on automated collection pipelines, while \textbf{VulnPatchPairs}~\cite{risse2024uncovering} and \textbf{Devign}~\cite{zhou2019devign} employ manual labeling to improve data quality.
Among these, DiverseVul is a large-scale C/C++ vulnerability detection dataset that substantially surpasses prior benchmarks in project diversity, CWE coverage, and the number of vulnerable functions.

However, samples mined from real projects often lack complete build context, making them hard to compile in isolation. 
% moreover, they may not reflect the vulnerability characteristics of LLM-generated code.
To address this issue, \textbf{FormAI}~\cite{tihanyi2023formai} generates code samples with GPT-3.5 and applies formal verification tools for vulnerability labeling.
It provides 112,000 (v1) and 265,000 (v2) independently compilable samples, enabling systematic analysis of vulnerabilities produced by LLMs.
% To mitigate data leakage risks and improve label reliability, PrimeVul introduces a high-precision automatic labeling technique and adopts a strictly time-ordered split.

In terms of language coverage, most vulnerability detection benchmarks focus on C/C++, while Juliet targets Java.
GREAT and VUDENC target Python, CVEfixes covers 27 languages, and CrossVul supports over 40 programming languages.

\subsubsection{Test Generation}
Test generation involves the automatic or manual creation of test cases to validate the expected behavior of a software system and to identify potential defects. 
\textbf{SF110}~\cite{fraser2014large}, as an early benchmark for test generation, mainly consists of low-complexity classes and small-scale projects. 
\textbf{Methods2Test}~\cite{tufano2022methods2test} expands the data scale by providing over 780,000 pairs of test-method mappings from the real world. However, good test cases not only need to be semantically correct but, more importantly, must effectively expose potential defects. To address this, \textbf{SWTBench}~\cite{mundler2024SWTbencha} focuses on code mutation scenarios and specifically evaluates the ability of generated test cases to reproduce issues. This benchmark includes over 1,900 samples, each integrating a problem description, a reference patch, and a set of standardized test cases, all built based on real GitHub issues and fixes.

Test generation refers to the automatic or manual creation of test cases to validate a system’s expected behavior and uncover potential defects.
\textbf{SF110}~\cite{fraser2014large}, an early benchmark for test generation, primarily comprises low-complexity classes and small-scale projects.
\textbf{Methods2Test}~\cite{tufano2022methods2test} substantially scales up the data by providing over 780,000 real-world method--test mappings. However, high-quality tests must be not only semantically correct but also effective at exposing defects. To this end, \textbf{SWTBench}~\cite{mundler2024SWTbencha} focuses on evaluating whether generated tests can reproduce real issues. It contains more than 1,900 samples, each combining a problem description, a reference patch, and standardized test cases, constructed from real GitHub issues and fixes.

\subsubsection{Defect Detection} 

Defect detection is a critical task that aims to identify potential bugs or errors in a codebase and is essential for ensuring software reliability and correctness.
\textbf{JIT-Defects4J}~\cite{ni2022best} targets defect prediction and comprises 27,391 change records from 21 Java open-source projects. Its labels are manually annotated, helping ensure defect authenticity and annotation accuracy. Subsequently, \textbf{Bugs.jar}~\cite{saha2018bugs} and \textbf{Bears}~\cite{madeiral2019bears} further broadened the scale and diversity of Java defect datasets: Bugs.jar contains 1,158 defects from eight widely used projects, while Bears includes 251 defects spanning 72 projects.

% \textbf{JIT-Defects4J}~\cite{ni2022best} 专注于缺陷预测,包含来自21个Java开源项目的27,391条变更记录. 样本标签并经由人工标注流程，确保缺陷真实性和准确性。

% \textbf{Bugs.jar}~\cite{saha2018bugs} offers a larger scale dataset, covering 20,948 methods from 8 popular Java projects, including 1,158 pairs of buggy and corresponding fixed methods.

% In contrast, \textbf{Bears}~\cite{madeiral2019bears} targets reproducible bugs and collects 251 real-world defect instances from 72 projects.

% \textbf{JIT-Defects4J}~\cite{ni2022best} focuses on Just-In-Time defect prediction (JIT-DP) and includes 27,391 accurately labeled change records from 21 open-source projects.

% **观察六：测试阶段基准的覆盖范围有限**。当前的基准主要聚焦于漏洞检测和测试生成任务，而针对其他任务（如断言生成、缺陷检测和代码编辑）的基准尚不成熟，缺乏足够的完善度。  
% **观察七：现有测试阶段基准的局限性**。用于测试的基准主要关注单元测试，对集成测试和交互测试的支持有限。漏洞检测基准侧重于静态漏洞，忽略了运行时和配置漏洞。断言生成基准范围狭窄，仅针对特定测试案例，未能评估更广泛的应用场景。代码编辑基准在支持协作开发和评估代码风格一致性方面存在不足。缺陷检测基准未能覆盖实际应用中遇到的所有缺陷类型。

% 0. 基准发展较慢。近几年新基准数量有限。新基准数量有限
% 1. 大部分基准没有抗污染措施，可能面临数据泄露的风险。
% 2. 基准主要关注单元测试，对集成测试和交互测试的支持有限。
% 3. 漏洞检测基准侧重于静态漏洞，忽略了运行时和配置漏洞. 
% 4. Vulnerability Detection（漏洞检测） 高度集中在 C/C++ 语言。Test Generation（测试生成） 和 Defect Detection（缺陷检测） 则高度集中在 Java 语言

% 5. “真实性”已成为主流标准 (High Authenticity)
% 6. 现有 benchmark 严重低估 agentic coding 的价值

% Observation 1: 数据污染风险。 现有基准更新缓慢且普遍缺乏防污染机制，导致benchmark极易被 LLM 预训练数据覆盖,存在数据泄露的风险。

% Observation 2: 评估粒度与语言支持和现实脱节 。绝大多数benchmark局限于“单元级测试和静态错误类型", 且编程语言局限于C/C++和java等,对现代语言rust,go支持匮乏.

% Observation 3: 交互模式受限. 现有基准多受限于单轮交互（Single-turn）模式，评估体系缺乏对环境交互能力的支持,限制了对智能体解决实际工程问题能力的全面度量。

\begin{tcolorbox}[colframe=gray!140, colback=gray!10, coltitle=white, title=Observation from Software Testing,breakable]
\textbf{Obs. 8:} Current benchmarks generally lack effective anti-contamination mechanisms, making them highly susceptible to inclusion in LLM pre-training corpora and increasing the risk of data leakage.

\textbf{Obs. 9:} Existing benchmarks are largely limited to function-level contexts and static verification paradigms.
In addition, language coverage is heavily skewed toward C/C++ and Java, with limited support for modern languages such as Rust and Go.

\textbf{Obs. 10:} Most benchmarks are restricted to single-turn interaction and do not incorporate environment interaction.
As a result, they provide only a limited assessment of code agents’ capabilities in realistic settings.

\end{tcolorbox}

\subsection{Software Maintenance}
Software maintenance involves modifying and updating deployed software to fix defects, improve performance, and adapt to evolving requirements or operating environments.
\autoref{tab:maintenance_benchmarks_final} summarizes key maintenance tasks and their corresponding benchmarks.

\begin{table*}[t]
  \centering
  \scriptsize % 小字号
  \setlength{\tabcolsep}{4pt} % 列间距
  \caption{Analysis of Software Maintenance Benchmarks (* indicates benchmarks applicable to agents).}
  \label{tab:maintenance_benchmarks_final}
  \begin{adjustbox}{width=\textwidth}
  % 10 列
  \begin{tabular}{cccc cc cc cc}
  \toprule
  \multirow{3}{*}{\textbf{Task}} & 
  \multirow{3}{*}{\textbf{Benchmark}} & 
  \multirow{3}{*}{\textbf{Language}} & 
  \multirow{3}{*}{\textbf{Year}} & 
  \multicolumn{4}{c}{\textbf{General Dimensions}} & 
  \multicolumn{2}{c}{\textbf{Stage-Specific}} \\
  \cmidrule(lr){5-8} \cmidrule(lr){9-10}
   & & & & 
  \textbf{\makecell{Anti-Contamination\\Strategy}} & 
  \textbf{Authenticity} & 
  \textbf{\makecell{Interaction\\Mode}} &
  \textbf{\makecell{Evaluation\\Paradigm}} & 
  \textbf{\makecell{Context\\Scope}} & % 第9列
  \textbf{Evolution} % 移至最后(第10列)
  \\
  \midrule
  
  % ================= Group 1: Program Repair =================
  \multirow{9}{*}{\textbf{\textit{\makecell{Program \\Repair}}}} 
    & Defects4J \cite{just2014defects4j} & Java & 2014 & None & Real-world & Single & Exec & Repo & \ding{55} \\
    & ManyBugs \cite{le2015manybugs} & C & 2015 & None & Real-world & Single & Exec & Repo & \ding{55} \\
    & QuixBugs \cite{lin2017quixbugs} & Python/Java & 2017 & None & Real-world & Single & Exec & Function & \ding{55} \\
    & BugsInPython \cite{widyasari2020bugsinpy} & Python & 2020 & None & Real-world & Single & Exec & Repo & \ding{55} \\
    & TypeBugs \cite{oh2022Pyter} & Python & 2022 & None & Real-world & Single & Exec & Function & \ding{55} \\
    & HumanEvalPack \cite{muennighoff2023octopack} & Multi & 2023 & None & Mixed & Single & Exec & Function & \ding{55} \\
    & SWE-Bench* \cite{jimenez2024swebench} & Python & 2024 & Updates & Real-world & Agentic & Exec & Repo & \ding{55} \\
    & SWA-Bench* \cite{vergopoulos2025automated} & Python & 2025 & Updates & Real-world & Agentic & Exec & Repo & \ding{51} \\
    & SWEE-Bench* \cite{vergopoulos2025automated} & Python & 2025 & Updates & Real-world & Agentic & Exec & Repo & \ding{51} \\

  \midrule

  % ================= Group 2: Code Editing =================
  \multirow{3}{*}{\textbf{\textit{Code Editing}}}
    & PyCommits\cite{wei2024coeditor} & Python & 2024 & None & Real-world & Interactive & Match & Repo & \ding{55} \\
    & CoEdPilot\cite{liu2024coedpilot} & Multi & 2024 & None & Real-world & Interactive & Match & Repo & \ding{55} \\
    & SWE-EVO* \cite{thai2025swe} & Python & 2025 & None & Real-world & Agentic & Exec & Repo & \ding{51} \\
    
  \midrule
  % ================= Group 3: Log Parsing =================
  \multirow{3}{*}{\textbf{\textit{\makecell{Log\\Parsing}}}} 
    & Loghub \cite{zhu2023loghub} & Logs & 2023 & None & Real-world & Single & Static & System & \ding{55} \\
    & LogPM \cite{hashemi2024logpm} & Logs & 2024 & None & Real-world & Single & Static & Line & \ding{55} \\
    & Loghub-2.0 \cite{jiang2024large} & Logs & 2024 & None & Real-world & Single & Static & System & \ding{55} \\
  
  \midrule
  % ================= Group 4: Vulnerability Repair =================
  \multirow{5}{*}{\textbf{\textit{\makecell{Vulnerability\\Repair}}}} 
    & ExtractFix \cite{gao2021beyond} & C/C++ & 2020 & None & Real-world & Single & Exec & Function & \ding{55} \\
    & Vul4J \cite{bui2022vul4j} & Java & 2022 & None & Real-world & Single & Exec & Repo & \ding{55} \\
    & VJBench \cite{wu2023effective} & Java & 2023 & None & Real-world & Single & Exec & Function & \ding{55} \\
    & VJBench-trans \cite{wu2023effective} & Java & 2023 & Perturbation & Real-world & Single & Exec & Function & \ding{55} \\
    & ManyVuls4J \cite{lin2024there} & Java & 2024 & None & Real-world & Single & Exec & Repo & \ding{55} \\
  
  \midrule
  % ================= Group 5: API Misuse & Clone Detection =================

  \multirow{2}{*}{\textbf{\textit{Decompilation}}} 

    & ExeBench~\cite{armengol2022exebench} & C & 2022 & None & Real-world & Single & Exec & Function & \ding{55} \\
  
    & WaDec~\cite{she2024wadec} & C & 2024 & None & Real-world & Single & Static & Function & \ding{55} \\

    \midrule

  \multirow{2}{*}{\textbf{\textit{API Misuse}}} 
    & APIMU4C \cite{gu2019empirical} & C & 2019 & None & Real-world & Single & Static & Function & \ding{55} \\
    & ROBUSTAPI \cite{zhong2024can} & Java & 2024 & None & Real-world & Single & Static & Function & \ding{55} \\
    \midrule
  \multirow{3}{*}{\textbf{\textit{Clone Detection}}}
    & POJ-104~\cite{mou2016convolutional} & C/C++ & 2009 & None & Real-world & Single & Static & File & \ding{55} \\
    & BigCloneBench \cite{svajlenko2014big} & Java & 2014 & None & Real-world & Single & Static & Function & \ding{55} \\
    & CodeNet \cite{puri2021codenet} & Multi & 2021 & None & Real-world & Single & Static & File & \ding{55} \\
  
  \bottomrule
  \end{tabular}
  \end{adjustbox}
\vspace{-5mm}
\end{table*}

\subsubsection{Program Repair}
Program repair aims to automatically fix bugs in software code and is essential for ensuring software reliability.
Early benchmarks such as \textbf{Defects4J}~\cite{just2014defects4j} and \textbf{ManyBugs}~\cite{le2015manybugs} systematically constructed defect datasets from real-world projects. Defects4J includes 357 real defects from five popular Java projects, whereas ManyBugs contains 185 real defects from nine large open-source C projects. Each defect is accompanied by key artifacts such as the source code, test scripts, and human-written patches.
However, these benchmarks are restricted to a single programming language, limiting their ability to assess repair tools consistently across languages. To address this limitation, \textbf{QuixBugs}~\cite{lin2017quixbugs} introduced an early cross-language program repair benchmark with 40 paired Python--Java programs, enabling cross-language comparability while preserving the same defect patterns. To further broaden multi-language coverage, \textbf{BugsInPy}~\cite{widyasari2020bugsinpy} brought high-quality real defect benchmarks to the Python ecosystem, while \textbf{HumanEvalPack}~\cite{muennighoff2023octopack} generated defective functions across six programming languages by injecting faults into reference solutions.

Nevertheless, many of these benchmarks still focus primarily on single-file repairs. \textbf{SWE-bench}~\cite{jimenez2024swebench} expands evaluation to repository-level repair, requiring models to understand codebases with thousands of lines of code and to perform cross-file reasoning. Variants such as SWE-bench Lite, SWE-bench Verified, and SWE-bench Java further extend the evaluated settings. However, SWE-bench depends on manual construction, which constrains scale and diversity. To mitigate these issues, \textbf{SWA-Bench}~\cite{vergopoulos2025automated} and \textbf{SWEE-Bench}~\cite{vergopoulos2025automated} adopt automated generation to create larger, more diverse, and more challenging benchmarks. SWA targets application-centric scenarios, covering 535 tasks from 44 projects, whereas SWEE emphasizes coverage beyond popular repositories, including 885 tasks from 366 projects. Both require modifying 3.26 files on average, substantially more than SWE-bench’s 1.66, thereby enhancing representativeness and the level of challenge.

\subsubsection{Code Editing}
Code editing involves modifying existing code to implement changes, improvements, or fixes.
\textbf{PyCommits}~\cite{wei2024coeditor} and \textbf{CoEdPilot}~\cite{liu2024coedpilot} are both constructed from real GitHub commit histories. PyCommits focuses on Python and is collected from 1,650 open-source Python repositories. In contrast, CoEdPilot broadens the scope to five programming languages: JavaScript, Java, Go, Python, and TypeScript.

\subsubsection{Log Parsing}
Log parsing aims to extract structured, meaningful information from software-generated log files.
\textbf{Loghub}~\cite{zhu2023loghub} helps fill the benchmark gap for AI-driven log analysis by providing 19 domain-specific subsets spanning six major system categories, totaling 77GB of log data. However, Khan et al.~\cite{khan2022guidelines} report annotation errors and offer guidelines for improving dataset quality. 
Moreover, apart from its limited scale, Loghub focuses primarily on log grouping accuracy while overlooking template generation quality.
To address these limitations, subsequent work introduced \textbf{LogPM}~\cite{hashemi2024logpm} and \textbf{Loghub-2.0}~\cite{jiang2024large}, both of which adopt template-level metrics in place of traditional accuracy-based evaluation. LogPM provides a large-scale benchmark with over ten million log entries, whereas Loghub-2.0 comprises 14 datasets, averaging roughly 3.6 million log entries per subset.

% \xy{If the subsubsection is short, do not split it into two paragraphs.*}

% \textbf{Loghub}~\cite{zhu2023loghub}填补了AI驱动日志分析领域基准方面的空白。它涵盖了6大类系统下的19个领域特定子集，总计77GB日志数据。然而，Khan等人~\cite{khan2022guidelines} 指出其存在标注错误，并相应提出了改进指南.此外，该数据集仅侧重于日志分组准确性，忽略了模板生成的精确性，规模有限，难以支撑实际应用。

% 为应对这些局限，后续研究提出了LogPM~\cite{hashemi2024logpm} 与 Loghub-2.0~\cite{jiang2024large} ，两者均引入模板级评估指标以取代传统准确率度量。其中，LogPM构建了规模超千万条日志的大型基准，而Loghub-2.0则包含14个数据集，平均每个子集涵盖约360万条日志。

% 为提升准确率，\textbf{LogPM}~\cite{hashemi2024logpm} 专注于字符级解析，并引入细粒度参数提取指标以替代传统准确率度量。

% 为此，LogPM 提出字符级标注方法和新的指标,构建数据规模更大的基准之一（超千万条记录）.

% \textbf{Loghub-2.0}~\cite{jiang2024large} 在原始数据集基础上扩展了14个高质量新子集，每个子集平均包含360万条日志，显著提升了数据规模、一致性和可复现性。14个数据集，平均360万条/集）
% 析基准数据集 Loghub（特指Loghub-2k） 规模太小（每个系统仅2000行日志），无法反映真实生产系统中日志数据的复杂性和海量性。同时使用了消息级和模板级指标

% \textbf{Loghub}~\cite{zhu2023loghub} is currently the largest publicly available log parsing dataset, encompassing 16 domain-specific subsets with a total of 77GB of log data. Each subset contains 2,000 manually labeled samples. However, the dataset contains labeling errors.
% Khan et al.~\cite{khan2022guidelines} proposed guidelines to address these issues. 
% To improve accuracy, \textbf{LogPM}~\cite{hashemi2024logpm} focuses on character level parsing and introduces detailed metrics for parameter extraction, replacing traditional accuracy measures. 
% \textbf{Loghub-2.0}~\cite{jiang2024large} expands the original dataset with 14 new high-quality subsets, each containing an average of 3.6 million logs, improving data size, consistency, and reproducibility.

\subsubsection{Vulnerability Repair}
The vulnerability repair task aims to automatically fix vulnerabilities or security flaws in software code and is crucial for ensuring software security.
\textbf{Vul4J}~\cite{bui2022vul4j} is the first benchmark dataset for Java vulnerability repair, covering 79 real vulnerabilities from 51 open-source projects and spanning 25 Common Weakness Enumeration (CWE) types. However, Vul4J covers a limited set of CWE categories and may be prone to data leakage due to the direct reuse of original code. To address these limitations, \textbf{VJBench}~\cite{wu2023effective} adds 42 new vulnerability instances on top of \textbf{Vul4J} and introduces 12 additional CWE categories, substantially expanding the diversity of vulnerability types. Moreover, by renaming identifiers and restructuring programs to produce semantically equivalent variants, \textbf{VJBench-trans} effectively mitigates data leakage. Building on these efforts, \textbf{ManyVuls4J}~\cite{lin2024there} extends vulnerability repair evaluation to the Java binary level for the first time, increasing the dataset to 103 vulnerabilities across 55 open-source projects and further improving its scale and diversity.

\subsubsection{Decompilation}
Decompilation aims to translate low-level machine code or bytecode back into high-level source code.
\textbf{ExeBench}~\cite{armengol2022exebench} is widely used to verify the functional correctness of decompiled code. It introduces a toolchain that automatically generates input/output (I/O) test pairs at scale. This facilitates execution-based evaluation to guarantee the functional correctness of the recovered code.
\textbf{WaDec}~\cite{she2024wadec} provides a collection of 51,768 C programs sourced from Hugging Face and 377 from GitHub, all compiled into WebAssembly format. 

\subsubsection{API Misuse Detection}
API misuse detection aims to identify incorrect or improper usage of APIs in software code.
\textbf{ROBUSTAPI}~\cite{zhong2024can} is designed to evaluate CodeLLMs’ ability to detect API misuse. It contains 1,208 real-world programming questions collected from Stack Overflow and covers 18 representative Java APIs. For C-language API misuse, \textbf{APIMU4C}~\cite{gu2019empirical} provides 2,172 handcrafted test cases and 100 real-world misuse cases extracted from open-source projects. 

\subsubsection{Code Clone Detection}
Code clone detection aims to identify duplicated or highly similar code snippets in repositories and is important for reducing redundancy, and improving software maintainability.
\textbf{BigCloneBench}~\cite{svajlenko2014big} is an early large-scale benchmark for Java code clone detection, constructed from 25,000 real-world projects and manually validated.
Later, \textbf{POJ-104}~\cite{mou2016convolutional} focuses on C/C++ and provides a clone detection dataset built around 104 programming problems, and it has become a default benchmark on the CodeXGLUE~\cite{lu2021codexglue} platform. Building on these efforts, \textbf{CodeNet}~\cite{puri2021codenet} further broadens language and task coverage, containing 13.9 million code snippets across 55 programming languages.

\begin{tcolorbox}[colframe=gray!140, colback=gray!10, coltitle=white, title=Observation from Software Maintenance,breakable]

\textbf{Obs. 11:} Most benchmarks lack effective anti-contamination mechanisms, resulting in substantial data leakage risks. 
In addition, they fail to support software evolution, overlooking realistic scenarios of continuous maintenance and evolving requirements.

\textbf{Obs. 12:} Modern programming languages such as Go and Rust are largely absent from existing benchmarks. 
Tasks like clone detection remain stagnant, emphasizing syntactic similarity while neglecting logical equivalence.

% \textbf{Obs. 13:} Evaluation is gradually shifting from function-level, single-turn modes to repository-level, agentic modes. 
% This transition reflects a fundamental change in focus, from a ``Coder'' limited to generating isolated code fragments, to a ``Software Engineer'' capable of addressing complex, real-world engineering problems.

\end{tcolorbox}

% 抗污染手段   从训练时间点，到动态更新数据，再到云上评估。

\subsection{Cross-Phase Benchmarks}
% \begin{figure*}[t]
%     \centering
%     % 第一张图
%     \begin{minipage}{0.29\textwidth}
%         % \vspace{-5pt}
%         \centering
%         \includegraphics[width=\textwidth]{pictures/top.pdf}
%         % \vspace{5pt}
%         \caption{\centering{Usage frequency of current main benchmarks}}
%         \label{fig:top}
%     \end{minipage}%
%     \hfill
%     % 第三张图
%     % \begin{minipage}{0.68\textwidth}
%     %     \vspace{-2pt}
%     %     \centering
%     %     \includegraphics[width=\textwidth]{pictures/years.pdf}
%     %     \caption{\centering{Distribution of publication years for benchmarks across all phases of the SDLC}}
%     %     \label{fig:years}
%     % \end{minipage}
%     % \vspace{-3pt}
%     % \label{fig:three_images}
% \end{figure*}

Unlike traditional benchmarks that focus on isolated phases of the SDLC, recent studies have proposed comprehensive cross-phase benchmarks to systematically evaluate the performance of CodeLLMs in realistic software engineering scenarios.

\textbf{CodeXGLUE}~\cite{lu2021codexglue} is one of the earliest and most widely adopted cross-task benchmark, covering core tasks such as code clone detection, defect detection, code generation, and completion. Additionally, \textbf{XCodeEval}~\cite{khan2024xCodeEval} introduces an execution engine for code to evaluate models' performance in generating code in real-world environments. It covers seven tasks, including code understanding and generation, and supports 17 programming languages. \textbf{CodeScope}~\cite{yan2024CodeScope} expands on this by supporting eight task types and up to 43 programming languages, with evaluation based on the correctness of execution. In comparison, \textbf{DevBench}~\cite{li2024Devbench} focuses on a systematic evaluation across the SDLC, making it the only benchmark to cover all SDLC phases, supporting four widely used languages. 
\textbf{ProjectEval}~\cite{liu2025projecteval} evaluates the ability of agents to generate project-scale Python applications from natural language instructions and automates verification through simulated user interactions. 
Building on SWE-Bench, \textbf{SWE-PolyBench}~\cite{rashid2025swe} extends the scope to multi-language and multi-task scenarios, emphasizing agents' performance in defect repair, feature addition, and code refactoring tasks, across four widely used languages.

\begin{tcolorbox}[
  enhanced,
    colback=gray!10,        % 背景浅灰
    colframe=gray!200,       % 框线灰色
    boxrule=0.5mm,          % 框线厚度
    toprule=0mm,            % 上边线=0
    bottomrule=0mm,         % 下边线=0
    leftrule=0.5mm,         % 左边线=0.5mm
    rightrule=0.5mm,        % 右边线=0.5mm
    arc=2mm,                % 圆角
    breakable
    % drop shadow             % 阴影
]
\textbf{Answer to RQ1} \textendash{} 
Current benchmarks exhibit a pronounced structural imbalance, heavily skewed toward the software implementation phase, while requirements engineering and design remain underexplored. In terms of General Dimensions, most benchmarks lack effective Anti-Contamination Strategies, creating substantial data leakage risks, and predominantly rely on single-turn Interaction Modes, failing to capture the dynamic nature of agentic workflows. Regarding Stage-Specific Dimensions, we observe a positive shift in Context Scope from function-level to repository-level tasks within the implementation phase; however, benchmarks targeting the maintenance phase frequently overlook the Evolution dimension. Despite these gaps, the field is signaling a positive transition from evaluating a ``Coder'' limited to generating isolated code fragments, to assessing a ``Software Engineer'' capable of addressing complex, real-world engineering problems.

% \xy{compared to the observations, this answer to RQ is too short. add more details (e.g., talk about General Dimensions and Stage-Specific dimensions.*)}
\end{tcolorbox}

\section{Statistics and Analysis of Studied Benchmarks}
\label{Statistics}
% Benchmark Usage and Distribution Across SDLC Tasks and Programming Languages

% Publication year distribution of all 425 unique corpus papers,

\begin{figure*}[t]
    \centering

    \begin{minipage}{0.23\textwidth}
        % \vspace{-5pt}
        \centering
        \includegraphics[width=\textwidth]{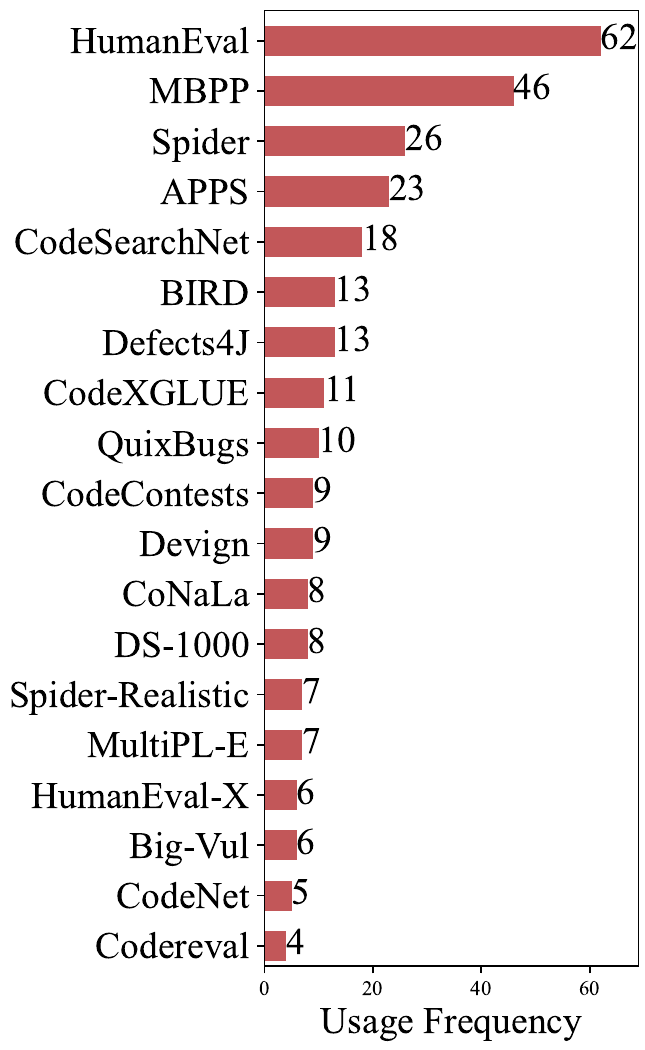}
        % \vspace{5pt}
        \caption{\centering{Usage frequency of current main benchmarks}}
        \label{fig:top}
    \end{minipage}%
     \hfill
    \begin{minipage}{0.21\textwidth}
        \centering
        \includegraphics[width=\textwidth]{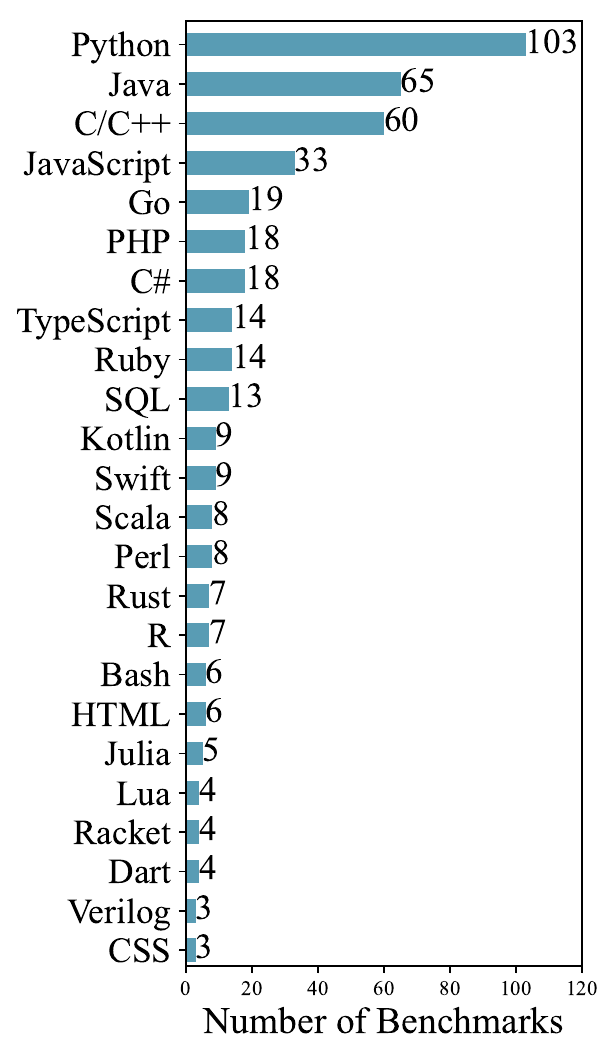}
        % \vspace{1pt}
        \caption{\centering{Distribution of benchmarks across programming languages}}
        \label{fig:programming}
    \end{minipage}%
    \hfill
    \begin{minipage}{0.53\textwidth}
        % \vspace{-2pt}
        \centering
        \includegraphics[width=\textwidth]{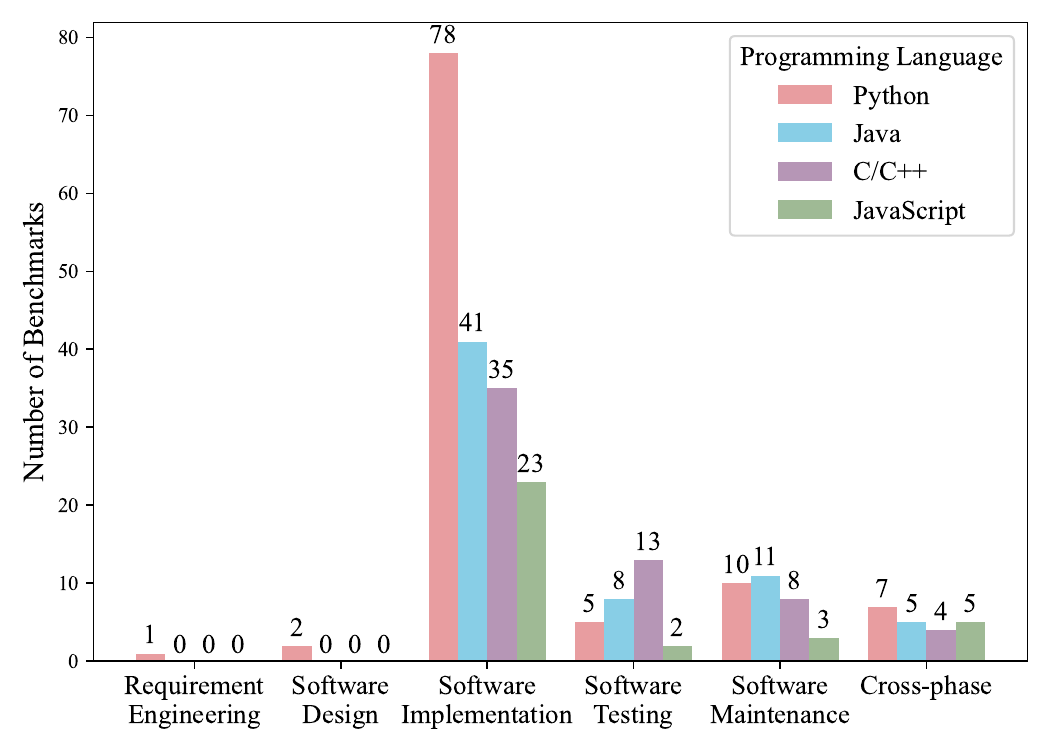}
        \caption{\centering{Distribution of programming languages used in current benchmarks across different SDLC phases.}}
        \label{fig:language_usage_by_phase}
    \end{minipage}
    % \vspace{-3pt}
    \label{fig:three_images}
\end{figure*}

To address RQ2, we conduct a comprehensive analysis of the relevant information from existing benchmarks.
Specifically,~\autoref{fig:top} illustrates the frequency distribution of benchmark usage;
% ~\autoref{fig:years} shows the distribution of release years for these benchmarks;
~\autoref{fig:programming} presents the prevalence of different programming languages across the benchmarks; and~\autoref{fig:language_usage_by_phase} depicts the distribution of programming language usage across various phase of the SDLC.

\textbf{The usage frequency of benchmarks}. 
As shown in~\autoref{fig:top}:
% \ding{182} Code generation tasks are the most frequently applied tasks in research, with the usage frequency of related benchmarks significantly higher than those of other tasks. Among them, HumanEval has been used 62 times, making it the most frequently used benchmark. ABPP (46 papers) and APPS (23 papers) are also widely adopted, indicating that this task is receiving considerable attention in current research.
\ding{182} Code generation dominates research, with benchmark usage significantly outpacing other tasks. HumanEval (62 papers) leads as the most adopted benchmark, followed by MBPP (46) and APPS (23), underscoring the intense focus on this domain.
\ding{183} The text-to-SQL task is also highly regarded, with commonly used benchmarks including Spider (26), BIRD (13), and Spider-Realistic (7), all of which have been practically applied in various studies.
\ding{184} In other task areas, code search tasks often use CodeSearchNet (18), while program repair tasks typically rely on Defects4J (13). 
Additionally, the cross-task benchmark CodeXGLUE has been adopted by 11 papers, indicating that the community's interest in cross-phase scenarios is steadily increasing.
% \xy{If you have abbreviated 46 paper to `46' in the first point, you can do the same thing in the following points.*}

% \textbf{The distribution of publication years for existing benchmarks across SDLC Phases.} 
% As shown in~\autoref{fig:years},
% the number of benchmarks focused on the software implementation phase is significantly higher than those in other phases. This trend accelerated notably between 2023 and 2024, reflecting the growing research interest in this area. In contrast, benchmarks for software design and requirements engineering remain underrepresented, indicating an imbalance in current research priorities.
% Overall, the development of benchmarks progressed slowly before 2022, then rapidly increased, reaching its peak in 2024, marking a clear turning point.
% \xy{Is there any reason to use ding in every paragraph?? Check, make sure not abuse this function*}

% Although we also reviewed benchmarks released before May 2025, those data are excluded from the figure to maintain temporal consistency.

\textbf{The distribution of benchmarks in programming languages.} 
As shown in~\autoref{fig:programming}, 
Python has the highest level of support, appearing in 103 benchmarks, followed by Java (65 benchmarks) and C/C++ (60 benchmarks), emphasizing their dominance in CodeLLM benchmarks.
Other languages like JavaScript, Go, PHP, C\#, TypeScript, Ruby, and SQL are supported in only 13 to 33 benchmarks, while most other languages appear in fewer than 10 benchmarks.
Overall, the distribution of programming languages in benchmarks closely aligns with their popularity.

\textbf{The distribution of programming languages across different SDLC phases.} 
As shown in~\autoref{fig:language_usage_by_phase}:
Python is the most commonly used language overall, but the most prevalent language varies across different phases. C/C++ dominates the testing phase, Java is the dominant language in the maintenance phase, and Python leads in the other phases.
This trend may be related to the characteristics of each language.
The efficiency of C++ makes it particularly suitable for testing scenarios with high performance requirements. But its complex syntax and significant maintenance overhead reduce its usage in software maintenance. In contrast, Java's relatively simple syntax and robust exception handling capabilities enhance maintainability, making it more frequently used in benchmarks during the maintenance phase.
% \begin{tcolorbox}[
%   enhanced,
%     colback=gray!10,        % 背景浅灰
%     colframe=gray!200,       % 框线灰色
%     boxrule=0.5mm,          % 框线厚度
%     toprule=0mm,            % 上边线=0
%     bottomrule=0mm,         % 下边线=0
%     leftrule=0.5mm,         % 左边线=0.5mm
%     rightrule=0.5mm,        % 右边线=0.5mm
%     arc=2mm,                % 圆角
%     breakable
%     % drop shadow             % 阴影
% ]
% \textbf{Answer to RQ2} \textendash{} HumanEval are currently the most widely adopted benchmarks for CodeLLMs, while the cross-phase benchmark CodeXGLUE is receiving increasing attention. In recent years, the number of benchmarks has grown significantly, with a particularly sharp rise since 2023. Furthermore, the dominant programming language in benchmarks is different at each phase of the SDLC. Specifically, C/C++ and Java are the dominant languages in the testing and maintenance phases, respectively, whereas Python is predominant in the other phases.

% \end{tcolorbox}

\section{Future Directions}
\label{Future Directions}
We propose several research directions based on our previous observations to advance the development of benchmarks for CodeLLMs and agents.

% We observe that the quality of benchmarks plays a crucial role in evaluating the CodeLLMs. 
% Notably, a contemporaneous study by Cao et al.~\cite{cao2025how} also conducted a systematic investigation of existing benchmarks for CodeLLMs. 
% They analyze key aspects such as quality, reliability, and reproducibility, and further propose practical guidelines for constructing high-quality CodeLLM benchmarks.

% benchmark的质量是一个重要的问题，Cao 等人就当前Codellms的benchmark的质量 可靠性和可重复性进行了调研，并提出发与代码相关基准的指南。

% \begin{itemize}
% \textbf{Direction 1: Developing standardized requirements engineering benchmarks covering diverse tasks (Obs. 1)}
% To overcome the limitations of existing research on the requirements engineering phase, future benchmark research should prioritize the development of standardized evaluation frameworks that comprehensively assess model and agent capabilities across diverse requirement engineering tasks, such as requirements analysis and specification formalization.
% These frameworks should evaluate how effectively models can interpret stakeholder needs, manage requirement conflicts, and generate consistent specification documentation.
% Additionally, developing methodologies to evaluate non-functional requirements, such as performance and security, is another potential research direction with significant practical implications for industrial software implementation processes.

\textbf{Developing standardized requirements engineering benchmarks (Obs. 1, 2). }
Future work should develop standardized benchmarks that evaluate models and agents across core requirements engineering tasks (e.g., requirements analysis and specification formalization).
These benchmarks should assess the ability to interpret stakeholder needs, resolve conflicts, and produce consistent specifications.

\textbf{Expanding design-phase coverage (Obs. 3, 4).}
Future benchmarks should broaden design-phase coverage to include additional tasks (e.g., architecture design).
They should support multimodal inputs such as diagrams and natural language specifications.
They should also introduce objective evaluation metrics to better quantify design creativity.

% \textbf{Improving realism in implementation benchmarks (Obs. 5).} 
% Future benchmarks should move beyond prompt-to-code snapshots to executable, end-to-end tasks grounded in real repositories, with realistic artifacts and complete environments.
% They should also stress repository-scale context and multilingual codebases, and adopt explicit anti-contamination protocols to better approximate production workflows.

\textbf{Expanding the scope of software testing benchmarks (Obs. 8, 9).} 
Most existing testing benchmarks are limited to isolated code snippets or function-level contexts. Future research should prioritize repository-level benchmarks that capture cross-file dependencies and dynamic execution behaviors. Moreover, it is essential develop native benchmark corpora for modern programming languages such as Rust and Go.

\textbf{Broadening maintenance benchmarks to real-world scenarios (Obs. 11).}
Most existing software maintenance benchmarks continue to focus on single-file repair and lack evolution-aware capabilities. Future benchmarks should move beyond simple bug fixing to encompass a broader spectrum of maintenance tasks, including evolution-aware repair, dependency and configuration breakages, and log-driven troubleshooting.

% Transitioning to Agent-Centric and Dynamic Evaluation Paradigms

% Enhancing Ecological Validity and Human-Centric Dimensions

% 变革评估范式与对齐真实开发生态

% 此外，应引入**可解释性（Explainability）**指标，不仅评估生成代码的正确性，还要评估模型为其设计选择和调试建议提供清晰、可执行推理的能力，这对于建立开发者信任至关重要。

\textbf{Include metrics for human--machine collaboration efficiency and privacy protection (Obs. 6).}
Although real-world development often follows a human--machine collaboration paradigm, current benchmarks primarily assess models' autonomous capabilities and rarely measure how effectively models assist humans.
Future benchmarks should therefore evaluate collaboration efficiency, rather than focusing solely on developer replacement.
In addition, privacy is largely overlooked in existing evaluations.
Benchmarks should incorporate privacy-preservation metrics to assess the risk of sensitive data leakage during interaction, which is a prerequisite for deployment in enterprise environments.

% \item \textbf{方向 7：实施有效的抗污染机制 (Observation #2, #5, #8, #11)} 鉴于时间截止（Time-cutoff）和代码扰动等静态防御手段在面对大模型预训练语料的不断扩展时日益失效，未来的基准应优先采用**持续更新（Continuous Updates）与在线评估（Online Evaluation）**等动态策略，从源头上有效规避数据泄露风险。

\textbf{Implementing effective anti-contamination mechanisms (Obs. 2, 5, 8, and 11).} 
Static defenses such as time cutoffs and code perturbations are becoming increasingly ineffective as LLM pretraining corpora continue to expand. Consequently, future benchmarks should prioritize dynamic strategies, such as continuous benchmark updates and online evaluation, to mitigate data leakage risks at the source.

% \item \textbf{Direction 9: Evaluating the capabilities of multi-agent systems.}

% With advancements in technologies such as Multi-Agent Collaborative Programming (MCP), multi-agent systems are demonstrating significant potential for real-world applications.
% Future research should move beyond single-agent evaluation to explore human-machine collaboration and multi-agent collaboration.
% Designing standardized benchmarks to assess collaborative effectiveness, covering various aspects such as interaction efficiency and reduction in human effort, represents a valuable direction for future work.

% 随着多智能体协同编程（MCP）等技术的发展，多智能体系统在实际应用中展现出巨大潜力。
% 未来的研究应超越单智能体评估，探索人机协作与多智能体协作。
% 设计标准化基准以评估协作效果，涵盖交互效率、人力投入减少等多个方面，是未来工作的一个重要方向。

% \item \textbf{方向 8：向基于智能体的评估范式转型 (Obs. 4, \#7, \#10, \#13)}  
% 随着大型语言模型从静态的“代码生成器”演进为具备环境交互能力的 代码智能体，传统的单轮问答式评估已显不足。未来的评估范式必须转向能够感知环境、动态交互的智能体评测。相应的基准需能够模拟真实开发环境，重点考察代码智能体在代码执行、反馈分析与迭代调试等环节中的综合表现。
\textbf{Transitioning Toward Agent-Based Evaluation Paradigms (Obs. 4, 7, 10, 13).} LLMs are evolving from static code generators into agents that can interact with their environments. Accordingly, traditional single-turn, question--answer evaluations are no longer sufficient. Future evaluation paradigms should shift toward agent-based assessments that emphasize environmental awareness, tool use, and iterative interaction. Such frameworks should emulate realistic development environments to measure agents' performance in code execution, feedback interpretation, and iterative debugging.

% 方向九：扩展对现代编程语言的原生支持（Obs. 6, \#9, \#12）
% 现有基准过度集中于 Python、Java 等高资源语言，这限制了其对多语言能力的评估。为应对这一局限，未来基准应拓展至 Rust、Go 等现代编程语言。更进一步，应突破简单地从 Python 等既有基准，翻译或迁移的做法，转而构建高质量的原生语料库，充分体现不同语言的核心特性，从而为模型的多语言能力提供更严谨评估。

\textbf{Expanding Native Support for Modern Programming Languages (Obs. 6, 9, 12).} Current benchmarks are biased toward high-resource languages like Python and Java. Future evaluations must expand to modern languages (e.g., Rust, Go). Moreover, the community should move beyond simply translating tasks from existing benchmarks (often derived from Python) and instead build high-quality native corpora that reflect language-specific idioms and tooling, enabling a more faithful evaluation of multilingual performance.

\section{Threats to Validity}
\label{Threats}

\noindent{\bf Construct validity} concerns the comprehensiveness of our search strategy.
A primary threat is the potential omission of relevant studies.
To mitigate this, we followed Kitchenham et al.'s guidelines~\cite{kitchenham2004procedures}, utilizing 20 keyword combinations across four major databases (IEEE Xplore, ACM, Elsevier, Springer) and Google Scholar.
Furthermore, to capture non-standard terminology, we employed forward and backward snowball sampling.
This process identified an additional 140 papers, ensuring a high degree of completeness in our final set of 461 studies.

% 构念效度关注我们是否准确捕捉了研究意图——此处指为评估LLM在软件工程任务中表现而设计的基准数据集。我们遵循Kitchenham等人[146]的系统综述指南进行操作。为确保全面覆盖，我们基于前期研究[72, 112, 381]检索基准数据集，并搜索了包括IEEE Xplore、ACM数字图书馆、arXiv和Google Scholar在内的主要软件工程与人工智能文献库。同时采用逆向与顺向滚雪球法补充初始检索结果。潜在威胁在于：术语差异或发表渠道不同可能导致检索策略遗漏相关基准。此外，纳入与排除标准的应用存在主观性风险。为降低该风险，两位作者独立评估所有候选基准，并通过结构化讨论解决分歧。

\noindent\textbf{Internal Validity} concerns potential biases arising during paper selection, data extraction, and taxonomy classification. 
The primary threats stem from subjectivity in two areas: inconsistencies in applying the inclusion criteria and ambiguity in characterizing qualitative attributes (e.g., \textit{Authenticity} or \textit{Context Scope}), where interpretations may differ across reviewers. 
To mitigate these risks, we followed a rigorous protocol with clear, explicit criteria (detailed in Section~III). 
Additionally, we employed a strict cross-validation procedure: each paper was independently reviewed and labeled by two co-authors, with any disagreements resolved by consulting a third senior researcher to reach consensus.

\noindent\textbf{External Validity} addresses the generalizability of our findings. 
A key threat is the rapid evolution of CodeLLMs, which may render specific state-of-the-art results and evaluation metrics outdated. 
Furthermore, because our study is scoped exclusively to the SDLC, the derived taxonomies may not generalize to other domains, such as general NLP or multi-agent systems outside the context of software engineering.

\section{Conclusion}
\label{Conclusion}
From an SDLC-oriented perspective, this study provides a comprehensive analysis of 178 benchmarks. The findings quantify a pronounced distributional disparity: approximately 61\% of current benchmarks focus on software implementation, whereas requirements engineering and design phases collectively account for less than 8\%. 
To address these critical deficiencies, this paper further identifies corresponding research directions, aiming to establish the foundation for future research and align CodeLLM and agent evaluation with the complexities of real-world software engineering.

\bibliographystyle{IEEEtran}
\bibliography{sample-base}

@String{Computing = "Computing" }

@String{Computer = "{IEEE} Computer" }

@String{Springer = "Springer-Verlag" }

@Article{	  agrawal2023guiding,
  title		= {Guiding language models of code with global context using
		  monitors},
  author	= {Agrawal, Lakshya A and Kanade, Aditya and Goyal, Navin and
		  Lahiri, Shuvendu K and Rajamani, Sriram K},
  journal	= {arXiv preprint arXiv:2306.10763},
  year		= {2023},
  citations	= {22}
}

@InProceedings{	  armengol2022exebench,
  title		= {ExeBench: an ML-scale dataset of executable C functions},
  author	= {Armengol-Estap{\'e}, Jordi and Woodruff, Jackson and
		  Brauckmann, Alexander and Magalh{\~a}es, Jos{\'e} Wesley de
		  Souza and O'Boyle, Michael FP},
  booktitle	= {Proceedings of the 6th ACM SIGPLAN International Symposium
		  on Machine Programming},
  pages		= {50--59},
  year		= {2022}
}

@Article{athiwaratkun2022multilingualeo,
  title		= {Multi-lingual Evaluation of Code Generation Models},
  author	= {Ben Athiwaratkun and Sanjay Krishna Gouda and Zijian Wang
		  and Xiaopeng Li and Yuchen Tian and Ming Tan and Wasi Uddin
		  Ahmad and Shiqi Wang and Qing Sun and Mingyue Shang and
		  Sujan Kumar Gonugondla and Hantian Ding and Varun Kumar and
		  Nathan Fulton and Arash Farahani and Siddharth Jain and
		  Robert Giaquinto and Haifeng Qian and Murali Krishna
		  Ramanathan and Ramesh Nallapati and Baishakhi Ray and
		  Parminder Bhatia and Sudipta Sengupta and Dan Roth and Bing
		  Xiang},
  journal	= {ArXiv},
  year		= {2022},
  volume	= {abs/2210.14868}
}

@Article{austin2021program,
  author	= {Austin, Jacob and Odena, Augustus and Nye, Maxwell and
		  Bosma, Maarten and Michalewski, Henryk and Dohan, David and
		  Jiang, Ellen and Cai, Carrie and Terry, Michael and Le,
		  Quoc and others},
  citations	= {1247},
  journal	= {arXiv preprint arXiv:2108.07732},
  title		= {Program synthesis with large language models},
  year		= {2021}
}

@InProceedings{	  bui2022vul4j,
  title		= {Vul4j: A dataset of reproducible java vulnerabilities
		  geared towards the study of program repair techniques},
  author	= {Bui, Quang-Cuong and Scandariato, Riccardo and Ferreyra,
		  Nicol{\'a}s E D{\'\i}az},
  booktitle	= {Proceedings of the 19th International Conference on Mining
		  Software Repositories},
  pages		= {464--468},
  year		= {2022}
}

@article{lin2024there,
  title={There are More Fish in the Sea: Automated Vulnerability Repair via Binary Templates},
  author={Lin, Bo and Wang, Shangwen and Chen, Liqian and Mao, Xiaoguang},
  journal={arXiv preprint arXiv:2411.18088},
  year={2024}
}

@Article{	  chai2024mceval,
  title		= {McEval: Massively Multilingual Code Evaluation},
  author	= {Chai, Linzheng and Liu, Shukai and Yang, Jian and Yin,
		  Yuwei and Jin, Ke and Liu, Jiaheng and Sun, Tao and Zhang,
		  Ge and Ren, Changyu and Guo, Hongcheng and others},
  journal	= {arXiv preprint arXiv:2406.07436},
  year		= {2024}
}

@Article{	  chakraborty2021deep,
  author	= {Chakraborty, Saikat and Krishna, Rahul and Ding, Yangruibo
		  and Ray, Baishakhi},
  citations	= {569},
  journal	= {IEEE Transactions on Software Engineering},
  number	= {9},
  pages		= {3280--3296},
  publisher	= {IEEE},
  title		= {Deep learning based vulnerability detection: Are we there yet?},
  volume	= {48},
  year		= {2021}
}

@Article{	  chen2021evaluating,
  author	= {Chen, Mark and Tworek, Jerry and Jun, Heewoo and Yuan,
		  Qiming and Pinto, Henrique Ponde De Oliveira and Kaplan,
		  Jared and Edwards, Harri and Burda, Yuri and Joseph,
		  Nicholas and Brockman, Greg and others},
  citations	= {3429},
  journal	= {arXiv preprint arXiv:2107.03374},
  title		= {Evaluating large language models trained on code},
  year		= {2021}
}

@InProceedings{	  chen2023diversevul,
  title		= {Diversevul: A new vulnerable source code dataset for deep
		  learning based vulnerability detection},
  author	= {Chen, Yizheng and Ding, Zhoujie and Alowain, Lamya and
		  Chen, Xinyun and Wagner, David},
  booktitle	= {Proceedings of the 26th International Symposium on
		  Research in Attacks, Intrusions and Defenses},
  pages		= {654--668},
  year		= {2023}
}

@InProceedings{	  chen2024rmcbench,
  title		= {RMCBench: Benchmarking Large Language Models' Resistance
		  to Malicious Code},
  author	= {Chen, Jiachi and Zhong, Qingyuan and Wang, Yanlin and
		  Ning, Kaiwen and Liu, Yongkun and Xu, Zenan and Zhao, Zhe
		  and Chen, Ting and Zheng, Zibin},
  booktitle	= {Proceedings of the 39th IEEE/ACM International Conference
		  on Automated Software Engineering},
  pages		= {995--1006},
  year		= {2024}
}

@InProceedings{deka2017rico,
  title		= {Rico: A mobile app dataset for building data-driven design applications},
  author	= {Deka, Biplab and Huang, Zifeng and Franzen, Chad and
		  Hibschman, Joshua and Afergan, Daniel and Li, Yang and
		  Nichols, Jeffrey and Kumar, Ranjitha},
  booktitle	= {Proceedings of the 30th annual ACM symposium on user
		  interface software and technology},
  pages		= {845--854},
  year		= {2017}
}

@Article{	  deng2020structure,
  title		= {Structure-grounded pretraining for text-to-SQL},
  author	= {Deng, Xiang and Awadallah, Ahmed Hassan and Meek,
		  Christopher and Polozov, Oleksandr and Sun, Huan and
		  Richardson, Matthew},
  journal	= {arXiv preprint arXiv:2010.12773},
  year		= {2020}
}

@Article{	  ding2024crosscodeeval,
  title		= {Crosscodeeval: A diverse and multilingual benchmark for cross-file code completion},
  author	= {Ding, Yangruibo and Wang, Zijian and Ahmad, Wasi and Ding,
		  Hantian and Tan, Ming and Jain, Nihal and Ramanathan,
		  Murali Krishna and Nallapati, Ramesh and Bhatia, Parminder
		  and Roth, Dan and others},
  journal	= {Advances in Neural Information Processing Systems},
  volume	= {36},
  year		= {2024}
}

@Article{ding2024vulnerability,
  title		= {Vulnerability detection with code language models: How far
		  are we?},
  author	= {Ding, Yangruibo and Fu, Yanjun and Ibrahim, Omniyyah and
		  Sitawarin, Chawin and Chen, Xinyun and Alomair, Basel and
		  Wagner, David and Ray, Baishakhi and Chen, Yizheng},
  journal	= {arXiv preprint arXiv:2403.18624},
  year		= {2024}
}

@Article{	  du2023classeval,
  title		= {Classeval: A manually-crafted benchmark for evaluating
		  llms on class-level code generation},
  author	= {Du, Xueying and Liu, Mingwei and Wang, Kaixin and Wang,
		  Hanlin and Liu, Junwei and Chen, Yixuan and Feng, Jiayi and
		  Sha, Chaofeng and Peng, Xin and Lou, Yiling},
  journal	= {arXiv preprint arXiv:2308.01861},
  year		= {2023}
}

@InProceedings{	  fan2020ac,
  author	= {Fan, Jiahao and Li, Yi and Wang, Shaohua and Nguyen, Tien
		  N},
  booktitle	= {Proceedings of the 17th International Conference on Mining
		  Software Repositories},
  citations	= {359},
  pages		= {508--512},
  title		= {AC/C++ code vulnerability dataset with code changes and
		  CVE summaries},
  year		= {2020}
}

@Article{	  fraser2014large,
  title		= {A large-scale evaluation of automated unit test generation
		  using evosuite},
  author	= {Fraser, Gordon and Arcuri, Andrea},
  journal	= {ACM Transactions on Software Engineering and Methodology
		  (TOSEM)},
  volume	= {24},
  number	= {2},
  pages		= {1--42},
  year		= {2014},
  publisher	= {ACM New York, NY, USA}
}

@Article{	  gan2021exploring,
  title		= {Exploring underexplored limitations of cross-domain
		  text-to-SQL generalization},
  author	= {Gan, Yujian and Chen, Xinyun and Purver, Matthew},
  journal	= {arXiv preprint arXiv:2109.05157},
  year		= {2021}
}

@Article{	  gan2021towards,
  title		= {Towards robustness of text-to-SQL models against synonym
		  substitution},
  author	= {Gan, Yujian and Chen, Xinyun and Huang, Qiuping and
		  Purver, Matthew and Woodward, John R and Xie, Jinxia and
		  Huang, Pengsheng},
  journal	= {arXiv preprint arXiv:2106.01065},
  year		= {2021}
}

@InProceedings{	  gu2018deep,
  author	= {Gu, Xiaodong and Zhang, Hongyu and Kim, Sunghun},
  booktitle	= {Proceedings of the 40th International Conference on
		  Software Engineering},
  citations	= {641},
  pages		= {933--944},
  title		= {Deep code search},
  year		= {2018}
}

@Article{	  gu2024cruxeval,
  title		= {Cruxeval: A benchmark for code reasoning, understanding
		  and execution},
  author	= {Gu, Alex and Rozi{\`e}re, Baptiste and Leather, Hugh and
		  Solar-Lezama, Armando and Synnaeve, Gabriel and Wang, Sida
		  I},
  journal	= {arXiv preprint arXiv:2401.03065},
  year		= {2024},
  citations	= {42}
}

@InProceedings{	  hashemi2024logpm,
  title		= {LogPM: Character-Based Log Parser Benchmark},
  author	= {Hashemi, Shayan and Nyyss{\"o}l{\"a}, Jesse and
		  M{\"a}ntyl{\"a}, Mika V},
  booktitle	= {2024 IEEE International Conference on Software Analysis,
		  Evolution and Reengineering (SANER)},
  pages		= {705--710},
  year		= {2024},
  organization	= {IEEE}
}

@InProceedings{	  hellendoorn2019global,
  author	= {Hellendoorn, Vincent J and Sutton, Charles and Singh,
		  Rishabh and Maniatis, Petros and Bieber, David},
  booktitle	= {International conference on learning representations},
  citations	= {270},
  title		= {Global relational models of source code},
  year		= {2019}
}

@Article{	  hendrycks2021measuringcc,
  title		= {Measuring Coding Challenge Competence With APPS},
  author	= {Dan Hendrycks and Steven Basart and Saurav Kadavath and
		  Mantas Mazeika and Akul Arora and Ethan Guo and Collin
		  Burns and Samir Puranik and Horace He and Dawn Xiaodong
		  Song and Jacob Steinhardt},
  journal	= {ArXiv},
  year		= {2021},
  volume	= {abs/2105.09938}
}

@InProceedings{	  hu2018deep,
  author	= {Hu, Xing and Li, Ge and Xia, Xin and Lo, David and Jin,
		  Zhi},
  booktitle	= {Proceedings of the 26th conference on program
		  comprehension},
  citations	= {779},
  pages		= {200--210},
  title		= {Deep code comment generation},
  year		= {2018}
}

@Article{	  hu2018summarizing,
  author	= {Hu, Xing and Li, Ge and Xia, Xin and Lo, David and Lu,
		  Shuai and Jin, Zhi},
  citations	= {325},
  title		= {Summarizing source code with transferred api knowledge},
  year		= {2018}
}

@Article{	  huang2021cosqa,
  title		= {Cosqa: 20,000+ web queries for code search and question
		  answering},
  author	= {Huang, Junjie and Tang, Duyu and Shou, Linjun and Gong,
		  Ming and Xu, Ke and Jiang, Daxin and Zhou, Ming and Duan,
		  Nan},
  journal	= {arXiv preprint arXiv:2105.13239},
  year		= {2021}
}

@Article{	  husain2019codesearchnet,
  title		= {Codesearchnet challenge: Evaluating the state of semantic
		  code search},
  author	= {Husain, Hamel and Wu, Ho-Hsiang and Gazit, Tiferet and
		  Allamanis, Miltiadis and Brockschmidt, Marc},
  journal	= {arXiv preprint arXiv:1909.09436},
  year		= {2019},
  citations	= {1066}
}

@InProceedings{	  iyer2016summarizing,
  author	= {Iyer, Srinivasan and Konstas, Ioannis and Cheung, Alvin
		  and Zettlemoyer, Luke},
  booktitle	= {54th Annual Meeting of the Association for Computational
		  Linguistics 2016},
  citations	= {875},
  organization	= {Association for Computational Linguistics},
  pages		= {2073--2083},
  title		= {Summarizing source code using a neural attention model},
  year		= {2016}
}

@Article{	  iyer2018mapping,
  author	= {Iyer, Srinivasan and Konstas, Ioannis and Cheung, Alvin
		  and Zettlemoyer, Luke},
  citations	= {255},
  journal	= {arXiv preprint arXiv:1808.09588},
  title		= {Mapping language to code in programmatic context},
  year		= {2018}
}

@inproceedings{jain2023transformer,
  title = {A Transformer-Based Approach for Abstractive Summarization of Requirements from Obligations in Software Engineering Contracts},
  booktitle = {2023 {{IEEE}} 31st {{International Requirements Engineering Conference}} ({{RE}})},
  author = {Jain, Chirag and Anish, Preethu Rose and Singh, Amrita and Ghaisas, Smita},
  year = {2023},
  month = sep,
  pages = {169--179},
  issn = {2332-6441},
  doi = {10.1109/RE57278.2023.00025},
}

@InProceedings{	  jiang2024large,
  title		= {A large-scale evaluation for log parsing techniques: How
		  far are we?},
  author	= {Jiang, Zhihan and Liu, Jinyang and Huang, Junjie and Li,
		  Yichen and Huo, Yintong and Gu, Jiazhen and Chen, Zhuangbin
		  and Zhu, Jieming and Lyu, Michael R},
  booktitle	= {Proceedings of the 33rd ACM SIGSOFT International
		  Symposium on Software Testing and Analysis},
  pages		= {223--234},
  year		= {2024}
}

@InProceedings{	  just2014defects4j,
  author	= {Just, Ren{\'e} and Jalali, Darioush and Ernst, Michael D},
  booktitle	= {Proceedings of the 2014 international symposium on
		  software testing and analysis},
  citations	= {1523},
  pages		= {437--440},
  title		= {Defects4J: A database of existing faults to enable
		  controlled testing studies for Java programs},
  year		= {2014}
}

@InProceedings{	  khan2022guidelines,
  title		= {Guidelines for assessing the accuracy of log message
		  template identification techniques},
  author	= {Khan, Zanis Ali and Shin, Donghwan and Bianculli, Domenico
		  and Briand, Lionel},
  booktitle	= {Proceedings of the 44th International Conference on
		  Software Engineering},
  pages		= {1095--1106},
  year		= {2022}
}

@InProceedings{	  lai2023ds,
  author	= {Lai, Yuhang and Li, Chengxi and Wang, Yiming and Zhang,
		  Tianyi and Zhong, Ruiqi and Zettlemoyer, Luke and Yih,
		  Wen-tau and Fried, Daniel and Wang, Sida and Yu, Tao},
  booktitle	= {International Conference on Machine Learning},
  citations	= {200},
  organization	= {PMLR},
  pages		= {18319--18345},
  title		= {DS-1000: A natural and reliable benchmark for data science
		  code generation},
  year		= {2023}
}

@Article{	  le2015manybugs,
  author	= {Le Goues, Claire and Holtschulte, Neal and Smith, Edward K
		  and Brun, Yuriy and Devanbu, Premkumar and Forrest,
		  Stephanie and Weimer, Westley},
  citations	= {368},
  journal	= {IEEE Transactions on Software Engineering},
  number	= {12},
  pages		= {1236--1256},
  publisher	= {IEEE},
  title		= {The ManyBugs and IntroClass benchmarks for automated
		  repair of C programs},
  volume	= {41},
  year		= {2015}
}

@Article{	  lee2021kaggledbqa,
  title		= {KaggleDBQA: Realistic evaluation of text-to-SQL parsers},
  author	= {Lee, Chia-Hsuan and Polozov, Oleksandr and Richardson,
		  Matthew},
  journal	= {arXiv preprint arXiv:2106.11455},
  year		= {2021}
}

@Article{	  lee2022cs1qa,
  title		= {CS1QA: A dataset for assisting code-based question
		  answering in an introductory programming course},
  author	= {Lee, Changyoon and Seonwoo, Yeon and Oh, Alice},
  journal	= {arXiv preprint arXiv:2210.14494},
  year		= {2022},
  citations	= {11}
}

@Article{	  lee2022ehrsql,
  title		= {Ehrsql: A practical text-to-sql benchmark for electronic
		  health records},
  author	= {Lee, Gyubok and Hwang, Hyeonji and Bae, Seongsu and Kwon,
		  Yeonsu and Shin, Woncheol and Yang, Seongjun and Seo,
		  Minjoon and Kim, Jong-Yeup and Choi, Edward},
  journal	= {Advances in Neural Information Processing Systems},
  volume	= {35},
  pages		= {15589--15601},
  year		= {2022}
}

@inproceedings{li2018vuldeepecker,
  title = {{{VulDeePecker}}: {{A Deep Learning-Based System}} for {{Vulnerability Detection}}},
  booktitle = {Proceedings 2018 {{Network}} and {{Distributed System Security Symposium}}},
  author = {Li, Zhen and Zou, Deqing and Xu, Shouhuai and Ou, Xinyu and Jin, Hai and Wang, Sujuan and Deng, Zhijun and Zhong, Yuyi},
  year = {2018},
  eprint = {1801.01681},
  primaryclass = {cs},
  doi = {10.14722/ndss.2018.23158},
  archiveprefix = {arXiv}
}

@Article{	  li2021sysevr,
  author	= {Li, Zhen and Zou, Deqing and Xu, Shouhuai and Jin, Hai and
		  Zhu, Yawei and Chen, Zhaoxuan},
  citations	= {705},
  journal	= {IEEE Transactions on Dependable and Secure Computing},
  number	= {4},
  pages		= {2244--2258},
  publisher	= {IEEE},
  title		= {Sysevr: A framework for using deep learning to detect
		  software vulnerabilities},
  volume	= {19},
  year		= {2021}
}

@Article{	  li2022competition,
  author	= {Li, Yujia and Choi, David and Chung, Junyoung and Kushman,
		  Nate and Schrittwieser, Julian and Leblond, R{\'e}mi and
		  Eccles, Tom and Keeling, James and Gimeno, Felix and Dal
		  Lago, Agustin and others},
  citations	= {1016},
  journal	= {Science},
  number	= {6624},
  pages		= {1092--1097},
  publisher	= {American Association for the Advancement of Science},
  title		= {Competition-level code generation with alphacode},
  volume	= {378},
  year		= {2022}
}

@Article{	  li2024can,
  author	= {Li, Jinyang and Hui, Binyuan and Qu, Ge and Yang, Jiaxi
		  and Li, Binhua and Li, Bowen and Wang, Bailin and Qin,
		  Bowen and Geng, Ruiying and Huo, Nan and others},
  citations	= {262},
  journal	= {Advances in Neural Information Processing Systems},
  title		= {Can llm already serve as a database interface? a big bench
		  for large-scale database grounded text-to-sqls},
  volume	= {36},
  year		= {2024}
}

@InProceedings{	  lin2017quixbugs,
  author	= {Lin, Derrick and Koppel, James and Chen, Angela and
		  Solar-Lezama, Armando},
  booktitle	= {Proceedings Companion of the 2017 ACM SIGPLAN
		  international conference on systems, programming,
		  languages, and applications: software for humanity},
  citations	= {266},
  pages		= {55--56},
  title		= {QuixBugs: A multi-lingual program repair benchmark set
		  based on the Quixey Challenge},
  year		= {2017}
}

@Article{	  liu2021codeqa,
  title		= {CodeQA: A question answering dataset for source code
		  comprehension},
  author	= {Liu, Chenxiao and Wan, Xiaojun},
  journal	= {arXiv preprint arXiv:2109.08365},
  year		= {2021},
  citations	= {19}
}

@Article{	  liu2023repobench,
  title		= {Repobench: Benchmarking repository-level code
		  auto-completion systems},
  author	= {Liu, Tianyang and Xu, Canwen and McAuley, Julian},
  journal	= {arXiv preprint arXiv:2306.03091},
  year		= {2023},
  citations	= {72}
}

@InProceedings{	  liu2024coedpilot,
  title		= {CoEdPilot: Recommending Code Edits with Learned Prior Edit
		  Relevance, Project-wise Awareness, and Interactive Nature},
  author	= {Liu, Chenyan and Cai, Yufan and Lin, Yun and Huang, Yuhuan
		  and Pei, Yunrui and Jiang, Bo and Yang, Ping and Dong, Jin
		  Song and Mei, Hong},
  booktitle	= {Proceedings of the 33rd ACM SIGSOFT International
		  Symposium on Software Testing and Analysis},
  pages		= {466--478},
  year		= {2024}
}

@Article{	  liu2024your,
  author	= {Liu, Jiawei and Xia, Chunqiu Steven and Wang, Yuyao and
		  Zhang, Lingming},
  citations	= {617},
  journal	= {Advances in Neural Information Processing Systems},
  title		= {Is your code generated by chatgpt really correct? rigorous
		  evaluation of large language models for code generation},
  volume	= {36},
  year		= {2024}
}

@Article{	  lu2021codexglue,
  author	= {Lu, Shuai and Guo, Daya and Ren, Shuo and Huang, Junjie
		  and Svyatkovskiy, Alexey and Blanco, Ambrosio and Clement,
		  Colin and Drain, Dawn and Jiang, Daxin and Tang, Duyu and
		  others},
  citations	= {827},
  journal	= {arXiv preprint arXiv:2102.04664},
  title		= {Codexglue: A machine learning benchmark dataset for code
		  understanding and generation},
  year		= {2021}
}

@InProceedings{	  madeiral2019bears,
  title		= {Bears: An extensible java bug benchmark for automatic
		  program repair studies},
  author	= {Madeiral, Fernanda and Urli, Simon and Maia, Marcelo and
		  Monperrus, Martin},
  booktitle	= {2019 IEEE 26th international conference on software
		  analysis, evolution and reengineering (SANER)},
  pages		= {468--478},
  year		= {2019},
  organization	= {IEEE}
}

@InProceedings{	  min-etal-2019-pilot,
  title		= {A Pilot Study for {C}hinese {SQL} Semantic Parsing},
  author	= {Min, Qingkai and Shi, Yuefeng and Zhang, Yue},
  editor	= {Inui, Kentaro and Jiang, Jing and Ng, Vincent and Wan,
		  Xiaojun},
  booktitle	= {Proceedings of the 2019 Conference on Empirical Methods in
		  Natural Language Processing and the 9th International Joint
		  Conference on Natural Language Processing (EMNLP-IJCNLP)},
  month		= nov,
  year		= {2019},
  address	= {Hong Kong, China},
  publisher	= {Association for Computational Linguistics},
  url		= {https://aclanthology.org/D19-1377/},
  doi		= {10.18653/v1/D19-1377},
  pages		= {3652--3658}
}

@InProceedings{	  mou2016convolutional,
  title		= {Convolutional neural networks over tree structures for
		  programming language processing},
  author	= {Mou, Lili and Li, Ge and Zhang, Lu and Wang, Tao and Jin,
		  Zhi},
  booktitle	= {Proceedings of the AAAI conference on artificial
		  intelligence},
  volume	= {30},
  number	= {1},
  year		= {2016}
}

@Article{	  muennighoff2023octopack,
  title		= {Octopack: Instruction tuning code large language models},
  author	= {Muennighoff, Niklas and Liu, Qian and Zebaze, Armel and
		  Zheng, Qinkai and Hui, Binyuan and Zhuo, Terry Yue and
		  Singh, Swayam and Tang, Xiangru and Von Werra, Leandro and
		  Longpre, Shayne},
  journal	= {arXiv preprint arXiv:2308.07124},
  year		= {2023}
}

@Article{	  naman2024livecodebench,
  title		= {Livecodebench: Holistic and contamination free evaluation of large language models for code},
  author	= {Naman Jain, King Han and Gu, Alex and Li, Wen-Ding and
		  Yan, Fanjia and Zhang, Tianjun and Wang, Sida and
		  Solar-Lezama, Armando and Sen, Koushik and Stoica, Ion},
  journal	= {arXiv preprint arXiv:2403.07974},
  year		= {2024}
}

@InProceedings{	  ni2022best,
  title		= {The best of both worlds: integrating semantic features
		  with expert features for defect prediction and
		  localization},
  author	= {Ni, Chao and Wang, Wei and Yang, Kaiwen and Xia, Xin and
		  Liu, Kui and Lo, David},
  booktitle	= {Proceedings of the 30th ACM Joint European Software
		  Engineering Conference and Symposium on the Foundations of
		  Software Engineering},
  pages		= {672--683},
  year		= {2022}
}

@InProceedings{	  nikitopoulos2021crossvul,
  title		= {CrossVul: a cross-language vulnerability dataset with
		  commit data},
  author	= {Nikitopoulos, Georgios and Dritsa, Konstantina and
		  Louridas, Panos and Mitropoulos, Dimitris},
  booktitle	= {Proceedings of the 29th ACM Joint Meeting on European
		  Software Engineering Conference and Symposium on the
		  Foundations of Software Engineering},
  pages		= {1565--1569},
  year		= {2021}
}

@Article{	  puri2021codenet,
  title		= {Codenet: A large-scale ai for code dataset for learning a
		  diversity of coding tasks},
  author	= {Puri, Ruchir and Kung, David S and Janssen, Geert and
		  Zhang, Wei and Domeniconi, Giacomo and Zolotov, Vladimir
		  and Dolby, Julian and Chen, Jie and Choudhury, Mihir and
		  Decker, Lindsey and others},
  journal	= {arXiv preprint arXiv:2105.12655},
  year		= {2021}
}

@Article{	  quan2025codeelo,
  title		= {CodeElo: Benchmarking Competition-level Code Generation of
		  LLMs with Human-comparable Elo Ratings},
  author	= {Quan, Shanghaoran and Yang, Jiaxi and Yu, Bowen and Zheng,
		  Bo and Liu, Dayiheng and Yang, An and Ren, Xuancheng and
		  Gao, Bofei and Miao, Yibo and Feng, Yunlong and others},
  journal	= {arXiv preprint arXiv:2501.01257},
  year		= {2025}
}

@Article{	  ranaldi2024investigating,
  title		= {Investigating the Impact of Data Contamination of Large
		  Language Models in Text-to-SQL Translation},
  author	= {Ranaldi, Federico and Ruzzetti, Elena Sofia and Onorati,
		  Dario and Ranaldi, Leonardo and Giannone, Cristina and
		  Favalli, Andrea and Romagnoli, Raniero and Zanzotto, Fabio
		  Massimo},
  journal	= {arXiv preprint arXiv:2402.08100},
  year		= {2024}
}

@InProceedings{	  saha2018bugs,
  author	= {Saha, Ripon K and Lyu, Yingjun and Lam, Wing and Yoshida,
		  Hiroaki and Prasad, Mukul R},
  booktitle	= {Proceedings of the 15th international conference on mining
		  software repositories},
  citations	= {215},
  pages		= {10--13},
  title		= {Bugs.jar: A large-scale, diverse dataset of real-world
		  java bugs},
  year		= {2018}
}

@InProceedings{	  she2024wadec,
  title		= {WaDec: Decompiling WebAssembly Using Large Language
		  Model},
  author	= {She, Xinyu and Zhao, Yanjie and Wang, Haoyu},
  booktitle	= {Proceedings of the 39th IEEE/ACM International Conference
		  on Automated Software Engineering},
  pages		= {481--492},
  year		= {2024}
}

@Article{	  tang2024biocoder,
  title		= {BioCoder: a benchmark for bioinformatics code generation
		  with large language models},
  author	= {Tang, Xiangru and Qian, Bill and Gao, Rick and Chen,
		  Jiakang and Chen, Xinyun and Gerstein, Mark B},
  journal	= {Bioinformatics},
  volume	= {40},
  number	= {Supplement\_1},
  pages		= {i266--i276},
  year		= {2024},
  publisher	= {Oxford University Press}
}

@InProceedings{	  tihanyi2023formai,
  title		= {The formai dataset: Generative ai in software security
		  through the lens of formal verification},
  author	= {Tihanyi, Norbert and Bisztray, Tamas and Jain, Ridhi and
		  Ferrag, Mohamed Amine and Cordeiro, Lucas C and Mavroeidis,
		  Vasileios},
  booktitle	= {Proceedings of the 19th International Conference on
		  Predictive Models and Data Analytics in Software
		  Engineering},
  pages		= {33--43},
  year		= {2023}
}

@InProceedings{	  tufano2022methods2test,
  title		= {Methods2Test: A dataset of focal methods mapped to test
		  cases},
  author	= {Tufano, Michele and Deng, Shao Kun and Sundaresan, Neel
		  and Svyatkovskiy, Alexey},
  booktitle	= {Proceedings of the 19th International Conference on Mining
		  Software Repositories},
  pages		= {299--303},
  year		= {2022}
}

@Article{	  waghjale2024ecco,
  title		= {ECCO: Can We Improve Model-Generated Code Efficiency
		  Without Sacrificing Functional Correctness?},
  author	= {Waghjale, Siddhant and Veerendranath, Vishruth and Wang,
		  Zora Zhiruo and Fried, Daniel},
  journal	= {arXiv preprint arXiv:2407.14044},
  year		= {2024}
}

@InProceedings{	  wang-etal-2023-recode,
  title		= {{R}e{C}ode: Robustness Evaluation of Code Generation
		  Models},
  author	= {Wang, Shiqi and Li, Zheng and Qian, Haifeng and Yang,
		  Chenghao and Wang, Zijian and Shang, Mingyue and Kumar,
		  Varun and Tan, Samson and Ray, Baishakhi and Bhatia,
		  Parminder and Nallapati, Ramesh and Ramanathan, Murali
		  Krishna and Roth, Dan and Xiang, Bing},
  editor	= {Rogers, Anna and Boyd-Graber, Jordan and Okazaki, Naoaki},
  booktitle	= {Proceedings of the 61st Annual Meeting of the Association
		  for Computational Linguistics (Volume 1: Long Papers)},
  month		= jul,
  year		= {2023},
  address	= {Toronto, Canada},
  publisher	= {Association for Computational Linguistics},
  url		= {https://aclanthology.org/2023.acl-long.773},
  doi		= {10.18653/v1/2023.acl-long.773},
  pages		= {13818--13843}
}

@Article{	  wei2020lambdanet,
  title		= {Lambdanet: Probabilistic type inference using graph neural
		  networks},
  author	= {Wei, Jiayi and Goyal, Maruth and Durrett, Greg and Dillig,
		  Isil},
  journal	= {arXiv preprint arXiv:2005.02161},
  year		= {2020},
  citations	= {134}
}

@inproceedings{wei2024coeditor,
title={Coeditor: Leveraging Repo-level Diffs for Code Auto-editing},
author={Jiayi Wei and Greg Durrett and Isil Dillig},
booktitle={The Twelfth International Conference on Learning Representations},
year={2024},
}

@InProceedings{wu2023effective,
  title		= {How effective are neural networks for fixing security
		  vulnerabilities},
  author	= {Wu, Yi and Jiang, Nan and Pham, Hung Viet and Lutellier,
		  Thibaud and Davis, Jordan and Tan, Lin and Babkin, Petr and
		  Shah, Sameena},
  booktitle	= {Proceedings of the 32nd ACM SIGSOFT International
		  Symposium on Software Testing and Analysis},
  pages		= {1282--1294},
  year		= {2023}
}

@Article{	  xu2024distinguishing,
  title		= {Distinguishing LLM-generated from Human-written Code by
		  Contrastive Learning},
  author	= {Xu, Xiaodan and Ni, Chao and Guo, Xinrong and Liu,
		  Shaoxuan and Wang, Xiaoya and Liu, Kui and Yang, Xiaohu},
  journal	= {ACM Transactions on Software Engineering and Methodology},
  year		= {2024},
  publisher	= {ACM New York, NY}
}

@Article{	  yang2024evaluating,
  title		= {Evaluating and aligning codellms on human preference},
  author	= {Yang, Jian and Yang, Jiaxi and Jin, Ke and Miao, Yibo and
		  Zhang, Lei and Yang, Liqun and Cui, Zeyu and Zhang, Yichang
		  and Hui, Binyuan and Lin, Junyang},
  journal	= {arXiv preprint arXiv:2412.05210},
  year		= {2024}
}

@inproceedings{yin2018learning,
  author = {Yin, Pengcheng and Deng, Bowen and Chen, Edgar and Vasilescu, Bogdan and Neubig, Graham},
  title = {Learning to Mine Aligned Code and Natural Language Pairs from Stack Overflow},
  booktitle = {International Conference on Mining Software Repositories},
  series = {MSR},
  pages = {476--486},
  year = {2018},
  publisher = {ACM},
  doi = {https://doi.org/10.1145/3196398.3196408},
}

@Article{	  yu2018spider,
  author	= {Yu, Tao and Zhang, Rui and Yang, Kai and Yasunaga,
		  Michihiro and Wang, Dongxu and Li, Zifan and Ma, James and
		  Li, Irene and Yao, Qingning and Roman, Shanelle and
		  others},
  citations	= {1155},
  journal	= {arXiv preprint arXiv:1809.08887},
  title		= {Spider: A large-scale human-labeled dataset for complex
		  and cross-domain semantic parsing and text-to-sql task},
  year		= {2018}
}

@Article{	  yu2019cosql,
  title		= {Cosql: A conversational text-to-sql challenge towards
		  cross-domain natural language interfaces to databases},
  author	= {Yu, Tao and Zhang, Rui and Er, He Yang and Li, Suyi and
		  Xue, Eric and Pang, Bo and Lin, Xi Victoria and Tan, Yi
		  Chern and Shi, Tianze and Li, Zihan and others},
  journal	= {arXiv preprint arXiv:1909.05378},
  year		= {2019}
}

@Article{	  yu2019sparc,
  title		= {Sparc: Cross-domain semantic parsing in context},
  author	= {Yu, Tao and Zhang, Rui and Yasunaga, Michihiro and Tan, Yi
		  Chern and Lin, Xi Victoria and Li, Suyi and Er, Heyang and
		  Li, Irene and Pang, Bo and Chen, Tao and others},
  journal	= {arXiv preprint arXiv:1906.02285},
  year		= {2019}
}

@Article{	  zhang2023repocoder,
  title		= {Repocoder: Repository-level code completion through
		  iterative retrieval and generation},
  author	= {Zhang, Fengji and Chen, Bei and Zhang, Yue and Keung,
		  Jacky and Liu, Jin and Zan, Daoguang and Mao, Yi and Lou,
		  Jian-Guang and Chen, Weizhu},
  journal	= {arXiv preprint arXiv:2303.12570},
  year		= {2023},
  citations	= {164}
}

@Article{	  zheng2023codegeex,
  author	= {Zheng, Qinkai and Xia, Xiao and Zou, Xu and Dong, Yuxiao
		  and Wang, Shan and Xue, Yufei and Wang, Zihan and Shen, Lei
		  and Wang, Andi and Li, Yang and others},
  citations	= {218},
  journal	= {arXiv preprint arXiv:2303.17568},
  title		= {Codegeex: A pre-trained model for code generation with
		  multilingual evaluations on humaneval-x},
  year		= {2023}
}

@Article{	  zhong2017seq2sql,
  author	= {Zhong, Victor and Xiong, Caiming and Socher, Richard},
  citations	= {1216},
  journal	= {arXiv preprint arXiv:1709.00103},
  title		= {Seq2sql: Generating structured queries from natural
		  language using reinforcement learning},
  year		= {2017}
}

@InProceedings{	  zhong2024can,
  title		= {Can LLM Replace Stack Overflow? A Study on Robustness and
		  Reliability of Large Language Model Code Generation},
  author	= {Zhong, Li and Wang, Zilong},
  booktitle	= {Proceedings of the AAAI Conference on Artificial
		  Intelligence},
  volume	= {38},
  number	= {19},
  pages		= {21841--21849},
  year		= {2024}
}

@Article{	  zhou2019devign,
  author	= {Zhou, Yaqin and Liu, Shangqing and Siow, Jingkai and Du,
		  Xiaoning and Liu, Yang},
  citations	= {981},
  journal	= {Advances in neural information processing systems},
  title		= {Devign: Effective vulnerability identification by learning
		  comprehensive program semantics via graph neural networks},
  volume	= {32},
  year		= {2019}
}

@InProceedings{	  zhu2023loghub,
  title		= {Loghub: A large collection of system log datasets for
		  ai-driven log analytics},
  author	= {Zhu, Jieming and He, Shilin and He, Pinjia and Liu,
		  Jinyang and Lyu, Michael R},
  booktitle	= {2023 IEEE 34th International Symposium on Software
		  Reliability Engineering (ISSRE)},
  pages		= {355--366},
  year		= {2023},
  organization	= {IEEE}
}

@article{zhuang2021software,
  title		= {Software vulnerability detection via deep learning over disaggregated code graph representation},
  author	= {Zhuang, Yufan and Suneja, Sahil and Thost, Veronika and
		  Domeniconi, Giacomo and Morari, Alessandro and Laredo,
		  Jim},
  journal	= {arXiv preprint arXiv:2109.03341},
  year		= {2021}
}

@Article{zhuo2024bigcodebench,
  title		= {Bigcodebench: Benchmarking code generation with diverse
		  function calls and complex instructions},
  author	= {Zhuo, Terry Yue and Vu, Minh Chien and Chim, Jenny and Hu,
		  Han and Yu, Wenhao and Widyasari, Ratnadira and Yusuf, Imam
		  Nur Bani and Zhan, Haolan and He, Junda and Paul, Indraneil
		  and others},
  journal	= {arXiv preprint arXiv:2406.15877},
  year		= {2024}
}

@Article{	  zou2019mu,
  title		= {$\mu$ VulDeePecker: A Deep Learning-Based System for
		  Multiclass Vulnerability Detection},
  author	= {Zou, Deqing and Wang, Sujuan and Xu, Shouhuai and Li, Zhen
		  and Jin, Hai},
  journal	= {IEEE Transactions on Dependable and Secure Computing},
  volume	= {18},
  number	= {5},
  pages		= {2224--2236},
  year		= {2019},
  publisher	= {IEEE}
}

@Article{	  zou2023universal,
  title		= {Universal and transferable adversarial attacks on aligned
		  language models},
  author	= {Zou, Andy and Wang, Zifan and Carlini, Nicholas and Nasr,
		  Milad and Kolter, J Zico and Fredrikson, Matt},
  journal	= {arXiv preprint arXiv:2307.15043},
  year		= {2023}
}

@article{guo2024redcode,
  title={RedCode: Risky Code Execution and Generation Benchmark for Code Agents},
  author={Guo, Chengquan and Liu, Xun and Xie, Chulin and Zhou, Andy and Zeng, Yi and Lin, Zinan and Song, Dawn and Li, Bo},
  journal={arXiv preprint arXiv:2411.07781},
  year={2024}
}

@inproceedings{gu2019empirical,
  title={An empirical study on api-misuse bugs in open-source c programs},
  author={Gu, Zuxing and Wu, Jiecheng and Liu, Jiaxiang and Zhou, Min and Gu, Ming},
  booktitle={2019 IEEE 43rd annual computer software and applications conference (COMPSAC)},
  volume={1},
  pages={11--20},
  year={2019},
  organization={IEEE}
}

@inproceedings{rodriguez2023benchmarking,
  title={Benchmarking causal study to interpret large language models for source code},
  author={Rodriguez-Cardenas, Daniel and Palacio, David N and Khati, Dipin and Burke, Henry and Poshyvanyk, Denys},
  booktitle={2023 IEEE International Conference on Software Maintenance and Evolution (ICSME)},
  pages={329--334},
  year={2023},
  organization={IEEE}
}

@inproceedings{wang2021screen2words,
  title={Screen2words: Automatic mobile UI summarization with multimodal learning},
  author={Wang, Bryan and Li, Gang and Zhou, Xin and Chen, Zhourong and Grossman, Tovi and Li, Yang},
  booktitle={The 34th Annual ACM Symposium on User Interface Software and Technology},
  pages={498--510},
  year={2021}
}

@article{wartschinski2022vudenc,
  title={VUDENC: vulnerability detection with deep learning on a natural codebase for Python},
  author={Wartschinski, Laura and Noller, Yannic and Vogel, Thomas and Kehrer, Timo and Grunske, Lars},
  journal={Information and Software Technology},
  volume={144},
  pages={106809},
  year={2022},
  publisher={Elsevier}
}

@inproceedings{bhandari2021cvefixes,
  title={CVEfixes: automated collection of vulnerabilities and their fixes from open-source software},
  author={Bhandari, Guru and Naseer, Amara and Moonen, Leon},
  booktitle={Proceedings of the 17th International Conference on Predictive Models and Data Analytics in Software Engineering},
  pages={30--39},
  year={2021}
}

@article{nguyen2024gptsniffer,
  title={GPTSniffer: A CodeBERT-based classifier to detect source code written by ChatGPT},
  author={Nguyen, Phuong T and Di Rocco, Juri and Di Sipio, Claudio and Rubei, Riccardo and Di Ruscio, Davide and Di Penta, Massimiliano},
  journal={Journal of Systems and Software},
  volume={214},
  pages={112059},
  year={2024},
  publisher={Elsevier}
}

@article{zhu2024domaineval,
  title={DOMAINEVAL: An Auto-Constructed Benchmark for Multi-Domain Code Generation},
  author={Zhu, Qiming and Cao, Jialun and Lu, Yaojie and Lin, Hongyu and Han, Xianpei and Sun, Le and Cheung, Shing-Chi},
  journal={arXiv preprint arXiv:2408.13204},
  year={2024}
}

@inproceedings{yu2024codereval,
  title={Codereval: A benchmark of pragmatic code generation with generative pre-trained models},
  author={Yu, Hao and Shen, Bo and Ran, Dezhi and Zhang, Jiaxin and Zhang, Qi and Ma, Yuchi and Liang, Guangtai and Li, Ying and Wang, Qianxiang and Xie, Tao},
  booktitle={Proceedings of the 46th IEEE/ACM International Conference on Software Engineering},
  pages={1--12},
  year={2024}
}

@article{li2024evocodebench,
  title={Evocodebench: An evolving code generation benchmark aligned with real-world code repositories},
  author={Li, Jia and Li, Ge and Zhang, Xuanming and Dong, Yihong and Jin, Zhi},
  journal={arXiv preprint arXiv:2404.00599},
  year={2024}
}

@article{zheng2024towards,
  title={Towards more realistic evaluation of LLM-based code generation: an experimental study and beyond},
  author={Zheng, Dewu and Wang, Yanlin and Shi, Ensheng and Zhang, Ruikai and Ma, Yuchi and Zhang, Hongyu and Zheng, Zibin},
  journal={arXiv preprint arXiv:2406.06918},
  year={2024}
}

@inproceedings{ferrari2017pure,
  title = {{{PURE}}: {{A Dataset}} of {{Public Requirements Documents}}},
  booktitle = {2017 {{IEEE}} 25th {{International Requirements Engineering Conference}} ({{RE}})},
  author = {Ferrari, Alessio and Spagnolo, Giorgio Oronzo and Gnesi, Stefania},
  year = {2017},
  month = sep,
  pages = {502--505},
  issn = {2332-6441},
  doi = {10.1109/RE.2017.29},
  urldate = {2025-03-24}
}

@misc{Sayyad2005 ,
author = "Sayyad Shirabad, J. and Menzies, T.J.",
year = "2005",
title = "{The PROMISE Repository of Software Engineering Databases.}",
url = "http://promise.site.uottawa.ca/SERepository",
howpublished = "School of Information Technology and Engineering, University of Ottawa, Canada"}

@misc{hu2024self,
title = {Self-{{Evolving Multi-Agent Collaboration Networks}} for {{Software Development}}},
author = {Hu, Yue and Cai, Yuzhu and Du, Yaxin and Zhu, Xinyu and Liu, Xiangrui and Yu, Zijie and Hou, Yuchen and Tang, Shuo and Chen, Siheng},
year = {2024},
month = oct,
number = {arXiv:2410.16946},
eprint = {2410.16946},
primaryclass = {cs},
publisher = {arXiv},
doi = {10.48550/arXiv.2410.16946},
urldate = {2025-03-24},
archiveprefix = {arXiv}
}

@article{ma2024speceval,
  title={SpecEval: Evaluating Code Comprehension in Large Language Models via Program Specifications},
  author={Ma, Lezhi and Liu, Shangqing and Bu, Lei and Li, Shangru and Wang, Yida and Liu, Yang},
  journal={arXiv preprint arXiv:2409.12866},
  year={2024}
}

@article{miah2024user,
  title={User-centric evaluation of ChatGPT capability of generating R program code},
  author={Miah, Tanha and Zhu, Hong},
  journal={arXiv e-prints},
  pages={arXiv--2402},
  year={2024}
}

@article{agashe2019juice,
  title={JuICe: A large scale distantly supervised dataset for open domain context-based code generation},
  author={Agashe, Rajas and Iyer, Srinivasan and Zettlemoyer, Luke},
  journal={arXiv preprint arXiv:1910.02216},
  year={2019}
}

@article{xie2024codebenchgen,
  title={Codebenchgen: Creating scalable execution-based code generation benchmarks},
  author={Xie, Yiqing and Xie, Alex and Sheth, Divyanshu and Liu, Pengfei and Fried, Daniel and Rose, Carolyn},
  journal={arXiv preprint arXiv:2404.00566},
  year={2024}
}

@article{li2025infibench,
  title={Infibench: Evaluating the question-answering capabilities of code large language models},
  author={Li, Linyi and Geng, Shijie and Li, Zhenwen and He, Yibo and Yu, Hao and Hua, Ziyue and Ning, Guanghan and Wang, Siwei and Xie, Tao and Yang, Hongxia},
  journal={Advances in Neural Information Processing Systems},
  volume={37},
  pages={128668--128698},
  year={2025}
}

@inproceedings{ezzini2023ai,
  title={Ai-based question answering assistance for analyzing natural-language requirements},
  author={Ezzini, Saad and Abualhaija, Sallam and Arora, Chetan and Sabetzadeh, Mehrdad},
  booktitle={2023 IEEE/ACM 45th International Conference on Software Engineering (ICSE)},
  pages={1277--1289},
  year={2023},
  organization={IEEE}
}

@article{moran2018machine,
  title={Machine learning-based prototyping of graphical user interfaces for mobile apps},
  author={Moran, Kevin and Bernal-C{\'a}rdenas, Carlos and Curcio, Michael and Bonett, Richard and Poshyvanyk, Denys},
  journal={IEEE transactions on software engineering},
  volume={46},
  number={2},
  pages={196--221},
  year={2018},
  publisher={IEEE}
}

@article{weiGUingMobileGUI2024,
  title = {GUing: a mobile GUI search engine using a vision-language model},
  author = {Wei, Jialiang and Courbis, Anne-Lise and Lambolais, Thomas and Xu, Binbin and Bernard, Pierre Louis and Dray, G\'erard and Maalej, Walid},
  date = {2024-11-08},
  journaltitle = {ACM Trans. Softw. Eng. Methodol.},
  shortjournal = {ACM Trans, Softw, Eng, Methodol,},
  issn = {1049-331X},
  doi = {10.1145/3702993}
}

@article{DongCodeScore2024,
  title={Codescore: Evaluating code generation by learning code execution},
  author={Dong, Yihong and Ding, Jiazheng and Jiang, Xue and Li, Ge and Li, Zhuo and Jin, Zhi},
  journal={ACM Transactions on Software Engineering and Methodology},
  volume={34},
  number={3},
  pages={1--22},
  year={2025},
  publisher={ACM New York, NY}
}

@article{jimenez2024swebench,
  title={Swe-bench: Can language models resolve real-world github issues?},
  author={Jimenez, Carlos E and Yang, John and Wettig, Alexander and Yao, Shunyu and Pei, Kexin and Press, Ofir and Narasimhan, Karthik},
  journal={arXiv preprint arXiv:2310.06770},
  year={2023}
}

@inproceedings{liu2023verilogeval,
  title={Verilogeval: Evaluating large language models for verilog code generation},
  author={Liu, Mingjie and Pinckney, Nathaniel and Khailany, Brucek and Ren, Haoxing},
  booktitle={2023 IEEE/ACM International Conference on Computer Aided Design (ICCAD)},
  pages={1--8},
  year={2023},
  organization={IEEE}
}

@inproceedings{allam2024rtl,
  title={Rtl-repo: A benchmark for evaluating llms on large-scale rtl design projects},
  author={Allam, Ahmed and Shalan, Mohamed},
  booktitle={2024 IEEE LLM Aided Design Workshop (LAD)},
  pages={1--5},
  year={2024},
  organization={IEEE}
}

@article{sumitani2024chibench,
  title={ChiBench: a Benchmark Suite for Testing Electronic Design Automation Tools},
  author={Sumitani, Rafael and Amorim, Jo{\~a}o Victor and Mafra, Augusto and Crepalde, Mirlaine and Pereira, Fernando Magno Quint{\~a}o},
  journal={arXiv preprint arXiv:2406.06550},
  year={2024}
}

@article{kang2024fveval,
  title={FVEval: Understanding Language Model Capabilities in Formal Verification of Digital Hardware},
  author={Kang, Minwoo and Liu, Mingjie and Hamad, Ghaith Bany and Suhaib, Syed and Ren, Haoxing},
  journal={arXiv preprint arXiv:2410.23299},
  year={2024}
}

@article{chen2024roboscript,
  title={Roboscript: Code generation for free-form manipulation tasks across real and simulation},
  author={Chen, Junting and Mu, Yao and Yu, Qiaojun and Wei, Tianming and Wu, Silang and Yuan, Zhecheng and Liang, Zhixuan and Yang, Chao and Zhang, Kaipeng and Shao, Wenqi and others},
  journal={arXiv preprint arXiv:2402.14623},
  year={2024}
}

@article{hsueh2022systematic,
  title={Systematic comparison of path planning algorithms using PathBench},
  author={Hsueh, Hao-Ya and Toma, Alexandru-Iosif and Ali Jaafar, Hussein and Stow, Edward and Murai, Riku and Kelly, Paul HJ and Saeedi, Sajad},
  journal={Advanced Robotics},
  volume={36},
  number={11},
  pages={566--581},
  year={2022},
  publisher={Taylor \& Francis}
}

@inproceedings{mayer2024cobra,
  title={CoBRA: A composable benchmark for robotics applications},
  author={Mayer, Matthias and K{\"u}lz, Jonathan and Althoff, Matthias},
  booktitle={2024 IEEE International Conference on Robotics and Automation (ICRA)},
  pages={17665--17671},
  year={2024},
  organization={IEEE}
}

@inproceedings{gonzalez-pumariega2025robotouille,
title={Robotouille: An Asynchronous Planning Benchmark for LLM Agents},
author={Gonzalo Gonzalez-Pumariega and Leong Su Yean and Neha Sunkara and Sanjiban Choudhury},
booktitle={The Thirteenth International Conference on Learning Representations},
year={2025},
url={https://openreview.net/forum?id=OhUoTMxFIH}
}

@article{cassano2022multipl,
  title={Multipl-e: A scalable and extensible approach to benchmarking neural code generation},
  author={Cassano, Federico and Gouwar, John and Nguyen, Daniel and Nguyen, Sydney and Phipps-Costin, Luna and Pinckney, Donald and Yee, Ming-Ho and Zi, Yangtian and Anderson, Carolyn Jane and Feldman, Molly Q and others},
  journal={arXiv preprint arXiv:2208.08227},
  year={2022}
}

@article{tao2024unraveling,
  title={Unraveling the Potential of Large Language Models in Code Translation: How Far Are We?},
  author={Tao, Qingxiao and Yu, Tingrui and Gu, Xiaodong and Shen, Beijun},
  journal={arXiv preprint arXiv:2410.09812},
  year={2024}
}

@article{fu2023codeapex,
  title={Codeapex: A bilingual programming evaluation benchmark for large language models},
  author={Fu, Lingyue and Chai, Huacan and Luo, Shuang and Du, Kounianhua and Zhang, Weiming and Fan, Longteng and Lei, Jiayi and Rui, Renting and Lin, Jianghao and Fang, Yuchen and others},
  journal={arXiv preprint arXiv:2309.01940},
  year={2023}
}

@article{hu2024real,
  title={A Real-World Benchmark for Evaluating Fine-Grained Issue Solving Capabilities of Large Language Models},
  author={Hu, Ruida and Peng, Chao and Ren, Jingyi and Jiang, Bo and Meng, Xiangxin and Wu, Qinyun and Gao, Pengfei and Wang, Xinchen and Gao, Cuiyun},
  journal={arXiv preprint arXiv:2411.18019},
  year={2024}
}

@article{gao2021beyond,
  title={Beyond tests: Program vulnerability repair via crash constraint extraction},
  author={Gao, Xiang and Wang, Bo and Duck, Gregory J and Ji, Ruyi and Xiong, Yingfei and Roychoudhury, Abhik},
  journal={ACM Transactions on Software Engineering and Methodology (TOSEM)},
  volume={30},
  number={2},
  pages={1--27},
  year={2021},
  publisher={ACM New York, NY, USA}
}

@inproceedings{widyasari2020bugsinpy,
  title={Bugsinpy: a database of existing bugs in python programs to enable controlled testing and debugging studies},
  author={Widyasari, Ratnadira and Sim, Sheng Qin and Lok, Camellia and Qi, Haodi and Phan, Jack and Tay, Qijin and Tan, Constance and Wee, Fiona and Tan, Jodie Ethelda and Yieh, Yuheng and others},
  booktitle={Proceedings of the 28th ACM joint meeting on european software engineering conference and symposium on the foundations of software engineering},
  pages={1556--1560},
  year={2020}
}

@inproceedings{oh2022pyter,
  title={PyTER: effective program repair for Python type errors},
  author={Oh, Wonseok and Oh, Hakjoo},
  booktitle={Proceedings of the 30th ACM Joint European Software Engineering Conference and Symposium on the Foundations of Software Engineering},
  pages={922--934},
  year={2022}
}

@article{huang2024Effibench,
  title = {Effibench: benchmarking the efficiency of automatically generated code},
  author = {Huang, Dong and Qing, Yuhao and Shang, Weiyi and Cui, Heming and Zhang, Jie},
  date = {2024},
  journaltitle = {Advances in Neural Information Processing Systems},
  shortjournal = {Adv. Neural Inf. Process. Syst.},
  volume = {37},
  pages = {11506--11544},
  urldate = {2025-03-12}
}

@article{wainakh2019Evaluating,
  title = {Evaluating semantic representations of source code},
  author = {Wainakh, Yaza and Rauf, Moiz and Pradel, Michael},
  date = {2019-09-25},
  journaltitle = {Arxiv},
  shortjournal = {Arxiv},
}

@inproceedings{li2024deveval,
    title = "{D}ev{E}val: A Manually-Annotated Code Generation Benchmark Aligned with Real-World Code Repositories",
    author = "Li, Jia  and
      Li, Ge  and
      Zhao, Yunfei  and
      Li, Yongmin  and
      Liu, Huanyu  and
      Zhu, Hao  and
      Wang, Lecheng  and
      Liu, Kaibo  and
      Fang, Zheng  and
      Wang, Lanshen  and
      Ding, Jiazheng  and
      Zhang, Xuanming  and
      Zhu, Yuqi  and
      Dong, Yihong  and
      Jin, Zhi  and
      Li, Binhua  and
      Huang, Fei  and
      Li, Yongbin  and
      Gu, Bin  and
      Yang, Mengfei",
    editor = "Ku, Lun-Wei  and
      Martins, Andre  and
      Srikumar, Vivek",
    booktitle = "Findings of the Association for Computational Linguistics: ACL 2024",
    month = aug,
    year = "2024",
    address = "Bangkok, Thailand",
    publisher = "Association for Computational Linguistics",
    doi = "10.18653/v1/2024.findings-acl.214",
    pages = "3603--3614"
}

@article{xu2024cruxeval,
  title={CRUXEval-X: A Benchmark for Multilingual Code Reasoning, Understanding and Execution},
  author={Xu, Ruiyang and Cao, Jialun and Lu, Yaojie and Lin, Hongyu and Han, Xianpei and He, Ben and Cheung, Shing-Chi and Sun, Le},
  journal={arXiv preprint arXiv:2408.13001},
  year={2024}
}

@misc{zheng2024Beyond,
      title={Beyond Correctness: Benchmarking Multi-dimensional Code Generation for Large Language Models}, 
      author={Jiasheng Zheng and Boxi Cao and Zhengzhao Ma and Ruotong Pan and Hongyu Lin and Yaojie Lu and Xianpei Han and Le Sun},
      year={2024},
      eprint={2407.11470},
      archivePrefix={arXiv},
      primaryClass={cs.SE},
}

@article{guo2024deepseek,
  title={DeepSeek-Coder: When the Large Language Model Meets Programming--The Rise of Code Intelligence},
  author={Guo, Daya and Zhu, Qihao and Yang, Dejian and Xie, Zhenda and Dong, Kai and Zhang, Wentao and Chen, Guanting and Bi, Xiao and Wu, Yu and Li, YK and others},
  journal={arXiv preprint arXiv:2401.14196},
  year={2024}
}

@misc{kamal2020risk,
      title={Risk Assessment, Threat Modeling and Security Testing in SDLC}, 
      author={Alya Hannah Ahmad Kamal and Caryn Chuah Yi Yen and Gan Jia Hui and Pang Sze Ling and Fatima-tuz-Zahra},
      year={2020},
      eprint={2012.07226},
      archivePrefix={arXiv},
      primaryClass={cs.SE},
}

@inproceedings{risse2024uncovering,
  title={Uncovering the limits of machine learning for automatic vulnerability detection},
  author={Risse, Niklas and B{\"o}hme, Marcel},
  booktitle={33rd USENIX Security Symposium (USENIX Security 24)},
  pages={4247--4264},
  year={2024}
}

@inproceedings{li2024MMCodea,
  title = {MMCode: benchmarking multimodal large language models for code generation with visually rich programming problems},
  booktitle = {Findings of the Association for Computational Linguistics: EMNLP 2024},
  author = {Li, Kaixin and Tian, Yuchen and Hu, Qisheng and Luo, Ziyang and Huang, Zhiyong and Ma, Jing},
  editor = {Al-Onaizan, Yaser and Bansal, Mohit and Chen, Yun-Nung},
  date = {2024-11},
  pages = {736--783},
  publisher = {Association for Computational Linguistics},
  location = {Miami, Florida, USA},
  doi = {10.18653/v1/2024.findings-emnlp.42},
  eventtitle = {Findings 2024}
}

@online{wu2024Plot2Code,
  title = {Plot2Code: A Comprehensive Benchmark for Evaluating Multi-modal Large Language Models in Code Generation from Scientific Plots},
  author = {Wu, Chengyue and Ge, Yixiao and Guo, Qiushan and Wang, Jiahao and Liang, Zhixuan and Lu, Zeyu and Shan, Ying and Luo, Ping},
  date = {2024-05-13},
  eprint = {2405.07990},
  eprinttype = {arXiv},
  eprintclass = {cs},
  doi = {10.48550/arXiv.2405.07990},
  pubstate = {prepublished}
}

@online{du2024Mercury,
  title = {Mercury: A Code Efficiency Benchmark for Code Large Language Models},
  author = {Du, Mingzhe and Luu, Anh Tuan and Ji, Bin and Liu, Qian and Ng, See-Kiong},
  date = {2024-06-11},
  eprint = {2402.07844},
  eprinttype = {arXiv},
  eprintclass = {cs},
  doi = {10.48550/arXiv.2402.07844},
  pubstate = {prepublished}
}

@article{huang2023BENCHMARKING,
  title={Mlagentbench: Evaluating language agents on machine learning experimentation},
  author={Huang, Qian and Vora, Jian and Liang, Percy and Leskovec, Jure},
  journal={arXiv preprint arXiv:2310.03302},
  year={2023}
}

@article{manh2024CodeMMLU,
  title={CodeMMLU: A Multi-Task Benchmark for Assessing Code Understanding \& Reasoning Capabilities of CodeLLMs},
  author={Manh, Dung Nguyen and Chau, Thang Phan and Hai, Nam Le and Doan, Thong T and Nguyen, Nam V and Pham, Quang and Bui, Nghi DQ},
  journal={arXiv preprint arXiv:2410.01999},
  year={2024}
}

@article{mundler2024SWTbencha,
  title = {SWT-bench: testing and validating real-world bug-fixes with code agents},
  author = {M\"undler, Niels and M\"uller, Mark N. and He, Jingxuan and Vechev, Martin},
  date = {2024-12-16},
  journaltitle = {Advances in Neural Information Processing Systems},
  shortjournal = {Adv. Neural Inf. Process. Syst.},
  volume = {37},
  pages = {81857--81887},
}

@online{daghighfarsoodeh2025Deepbench,
  title = {Deep-bench: deep learning benchmark dataset for code generation},
  author = {Daghighfarsoodeh, Alireza and Wang, Chung-Yu and Taherkhani, Hamed and Sepidband, Melika and Abdollahi, Mohammad and Hemmati, Hadi and Pham, Hung Viet},
  date = {2025-02-26},
  eprint = {2502.18726},
  eprinttype = {arXiv},
  eprintclass = {cs},
  doi = {10.48550/arXiv.2502.18726},
  pubstate = {prepublished}
}

@article{li2025fea,
  title={FEA-Bench: A Benchmark for Evaluating Repository-Level Code Generation for Feature Implementation},
  author={Li, Wei and Zhang, Xin and Guo, Zhongxin and Mao, Shaoguang and Luo, Wen and Peng, Guangyue and Huang, Yangyu and Wang, Houfeng and Li, Scarlett},
  journal={arXiv preprint arXiv:2503.06680},
  year={2025}
}

@online{tang2024MLbench,
  title = {ML-bench: evaluating large language models and agents for machine learning tasks on repository-level code},
  author = {Tang, Xiangru and Liu, Yuliang and Cai, Zefan and Shao, Yanjun and Lu, Junjie and Zhang, Yichi and Deng, Zexuan and Hu, Helan and An, Kaikai and Huang, Ruijun and Si, Shuzheng and Chen, Sheng and Zhao, Haozhe and Chen, Liang and Wang, Yan and Liu, Tianyu and Jiang, Zhiwei and Chang, Baobao and Fang, Yin and Qin, Yujia and Zhou, Wangchunshu and Zhao, Yilun and Cohan, Arman and Gerstein, Mark},
  date = {2024-08-21},
  eprint = {2311.09835},
  eprinttype = {arXiv},
  eprintclass = {cs},
  doi = {10.48550/arXiv.2311.09835},
  pubstate = {prepublished}
}

@article{krishna2024Using,
  title = {Using LLMs in Software Requirements Specifications: An Empirical Evaluation},
  author = {Krishna, Madhava and Gaur, Bhagesh and Verma, Arsh and Jalote, Pankaj},
  date = {2024-06-24},
  journaltitle = {2024 IEEE 32nd International Requirements Engineering Conference (RE)},
  shortjournal = {2024 IEEE 32nd Int. Requir. Eng. Conf. (RE)},
  pages = {475--483},
  publisher = {IEEE},
  location = {Reykjavik, Iceland},
  doi = {10.1109/RE59067.2024.00056},
  eventtitle = {2024 IEEE 32nd International Requirements Engineering Conference (RE)},
  isbn = {9798350395112}
}

@inproceedings{lo2023Trustworthy,
  title = {Trustworthy and synergistic artificial intelligence for software engineering: vision and roadmaps},
  booktitle = {2023 IEEE/ACM International Conference on Software Engineering: Future of Software Engineering (icse-fose)},
  author = {Lo, David},
  date = {2023},
  pages = {69--85},
  publisher = {IEEE},
  doi = {10.1109/ICSE-FoSE59343.2023.00010}
}

@article{zhang2024Unifying,
  title={Unifying the perspectives of nlp and software engineering: A survey on language models for code},
  author={Zhang, Ziyin and Chen, Chaoyu and Liu, Bingchang and Liao, Cong and Gong, Zi and Yu, Hang and Li, Jianguo and Wang, Rui},
  journal={arXiv preprint arXiv:2311.07989},
  year={2023}
}

@article{haque2024llms,
  title={LLMs: A Game-Changer for Software Engineers?},
  author={Haque, Md Asraful},
  journal={arXiv preprint arXiv:2411.00932},
  year={2024}
}

@misc{GitHub2023,
  author = {},
  title = {GitHub Copilot · Your AI pair programmer},
  howpublished ={\url{https://github.com/features/copilot/.}},
  year = 2023,}

@inproceedings{xu2022systematic,
  title={A systematic evaluation of large language models of code},
  author={Xu, Frank F and Alon, Uri and Neubig, Graham and Hellendoorn, Vincent Josua},
  booktitle={Proceedings of the 6th ACM SIGPLAN international symposium on machine programming},
  pages={1--10},
  year={2022}
}

@inproceedings{xia2023automated,
  title={Automated program repair in the era of large pre-trained language models},
  author={Xia, Chunqiu Steven and Wei, Yuxiang and Zhang, Lingming},
  booktitle={2023 IEEE/ACM 45th International Conference on Software Engineering (ICSE)},
  pages={1482--1494},
  year={2023},
  organization={IEEE}
}

@inproceedings{zhou2024large,
  title={Large language model for vulnerability detection: Emerging results and future directions},
  author={Zhou, Xin and Zhang, Ting and Lo, David},
  booktitle={Proceedings of the 2024 ACM/IEEE 44th International Conference on Software Engineering: New Ideas and Emerging Results},
  pages={47--51},
  year={2024}
}

@article{hui2024qwen2,
  title={Qwen2. 5-Coder Technical Report},
  author={Hui, Binyuan and Yang, Jian and Cui, Zeyu and Yang, Jiaxi and Liu, Dayiheng and Zhang, Lei and Liu, Tianyu and Zhang, Jiajun and Yu, Bowen and Dang, Kai and others},
  journal={arXiv preprint arXiv:2409.12186},
  year={2024}
}

@article{hou2024large,
  title={Large language models for software engineering: A systematic literature review},
  author={Hou, Xinyi and Zhao, Yanjie and Liu, Yue and Yang, Zhou and Wang, Kailong and Li, Li and Luo, Xiapu and Lo, David and Grundy, John and Wang, Haoyu},
  journal={ACM Transactions on Software Engineering and Methodology},
  volume={33},
  number={8},
  pages={1--79},
  year={2024},
  publisher={ACM New York, NY}
}

@misc{zheng2024survey,
title = {A Survey of Large Language Models for Code: Evolution, Benchmarking, and Future Trends},
author = {Zheng, Zibin and Ning, Kaiwen and Wang, Yanlin and Zhang, Jingwen and Zheng, Dewu and Ye, Mingxi and Chen, Jiachi},
year = {2024},
month = jan,
number = {arXiv:2311.10372},
eprint = {2311.10372},
primaryclass = {cs},
publisher = {arXiv},
doi = {10.48550/arXiv.2311.10372},
archiveprefix = {arXiv}
}

@article{zheng2025towards,
  title={Towards an understanding of large language models in software engineering tasks},
  author={Zheng, Zibin and Ning, Kaiwen and Zhong, Qingyuan and Chen, Jiachi and Chen, Wenqing and Guo, Lianghong and Wang, Weicheng and Wang, Yanlin},
  journal={Empirical Software Engineering},
  volume={30},
  number={2},
  pages={50},
  year={2025},
  publisher={Springer}
}

@article{reinpold2024exploring,
  title={Exploring LLMs for Verifying Technical System Specifications Against Requirements},
  author={Reinpold, Lasse M and Schieseck, Marvin and Wagner, Lukas P and Gehlhoff, Felix and Fay, Alexander},
  journal={arXiv preprint arXiv:2411.11582},
  year={2024}
}

@article{jin2024llms,
  title={From llms to llm-based agents for software engineering: A survey of current, challenges and future},
  author={Jin, Haolin and Huang, Linghan and Cai, Haipeng and Yan, Jun and Li, Bo and Chen, Huaming},
  journal={arXiv preprint arXiv:2408.02479},
  year={2024}
}

@online{wang2024BabelBench,
  title = {BabelBench: An Omni Benchmark for Code-Driven Analysis of Multimodal and Multistructured Data},
  author = {Wang, Xuwu and Cui, Qiwen and Tao, Yunzhe and Wang, Yiran and Chai, Ziwei and Han, Xiaotian and Liu, Boyi and Yuan, Jianbo and Su, Jing and Wang, Guoyin and Liu, Tingkai and Chen, Liyu and Liu, Tianyi and Sun, Tao and Zhang, Yufeng and Zheng, Sirui and You, Quanzeng and Yang, Yang and Yang, Hongxia},
  date = {2024-10-01},
  eprint = {2410.00773},
  eprinttype = {arXiv},
  eprintclass = {cs},
  doi = {10.48550/arXiv.2410.00773},
  pubstate = {prepublished}
}

@article{cao2024Spider2v,
  title = {Spider2-v: How far are multimodal agents from automating data science and engineering workflows?},
  author = {Cao, Ruisheng and Lei, Fangyu and Wu, Haoyuan and Chen, Jixuan and Fu, Yeqiao and Gao, Hongcheng and Xiong, Xinzhuang and Zhang, Hanchong and Hu, Wenjing and Mao, Yuchen},
  date = {2024},
  journaltitle = {Advances in Neural Information Processing Systems},
  shortjournal = {Adv. Neural Inf. Process. Syst.},
  volume = {37},
  pages = {107703--107744},
}

@inproceedings{NEURIPS2024_cb66be28,
  title = {Web2Code: a large-scale webpage-to-code dataset and evaluation framework for multimodal llms},
  booktitle = {Advances in neural information processing systems},
  author = {Yun, Sukmin and Lin, Haokun and Thushara, Rusiru and Bhat, Mohammad Qazim and Wang, Yongxin and Jiang, Zutao and Deng, Mingkai and Wang, Jinhong and Tao, Tianhua and Li, Junbo and Li, Haonan and Nakov, Preslav and Baldwin, Timothy and Liu, Zhengzhong and Xing, Eric P. and Liang, Xiaodan and Shen, Zhiqiang},
  editor = {Globerson, A. and Mackey, L. and Belgrave, D. and Fan, A. and Paquet, U. and Tomczak, J. and Zhang, C.},
  date = {2024},
  volume = {37},
  pages = {112134--112157},
  publisher = {Curran Associates, Inc.},
}

@inproceedings{svajlenko2014big,
  title = {Towards a big data curated benchmark of inter-project code clones},
  booktitle = {2014 IEEE International Conference on Software Maintenance and Evolution},
  author = {Svajlenko, Jeffrey and Islam, Judith F. and Keivanloo, Iman and Roy, Chanchal K. and Mia, Mohammad Mamun},
  date = {2014},
  pages = {476--480},
  publisher = {IEEE},
  doi = {10.1109/ICSME.2014.77}
}

@inproceedings{yan2024CodeScope,
  title = {{{CodeScope}}: An Execution-Based Multilingual Multitask Multidimensional Benchmark for Evaluating {{LLMs}} on Code Understanding and Generation},
  booktitle = {Proceedings of the 62nd {{Annual Meeting}} of the {{Association}} for {{Computational Linguistics}} ({{Volume}} 1: {{Long Papers}})},
  author = {Yan, Weixiang and Liu, Haitian and Wang, Yunkun and Li, Yunzhe and Chen, Qian and Wang, Wen and Lin, Tingyu and Zhao, Weishan and Zhu, Li and Sundaram, Hari and Deng, Shuiguang},
  editor = {Ku, Lun-Wei and Martins, Andre and Srikumar, Vivek},
  year = {2024},
  month = aug,
  pages = {5511--5558},
  publisher = {Association for Computational Linguistics},
  address = {Bangkok, Thailand},
  doi = {10.18653/v1/2024.acl-long.301},
}

@inproceedings{khan2024xCodeEval,
  title={Xcodeeval: An execution-based large scale multilingual multitask benchmark for code understanding, generation, translation and retrieval},
  author={Khan, Mohammad Abdullah Matin and Bari, M Saiful and Long, Do and Wang, Weishi and Parvez, Md Rizwan and Joty, Shafiq},
  booktitle={Proceedings of the 62nd Annual Meeting of the Association for Computational Linguistics (Volume 1: Long Papers)},
  pages={6766--6805},
  year={2024}
}

@inproceedings{yan2023CodeTransOcean,
  title = {{{CodeTransOcean}}: {{A Comprehensive Multilingual Benchmark}} for {{Code Translation}}},
  booktitle = {Findings of the {{Association}} for {{Computational Linguistics}}: {{EMNLP}} 2023},
  author = {Yan, Weixiang and Tian, Yuchen and Li, Yunzhe and Chen, Qian and Wang, Wen},
  editor = {Bouamor, Houda and Pino, Juan and Bali, Kalika},
  year = {2023},
  month = dec,
  pages = {5067--5089},
  publisher = {Association for Computational Linguistics},
  address = {Singapore},
  doi = {10.18653/v1/2023.findings-emnlp.337}
}

@misc{zhou2025LessLeakBench,
  title = {{{LessLeak-Bench}}: {{A First Investigation}} of {{Data Leakage}} in {{LLMs Across}} 83 {{Software Engineering Benchmarks}}},
  author = {Zhou, Xin and Weyssow, Martin and Widyasari, Ratnadira and Zhang, Ting and He, Junda and Lyu, Yunbo and Chang, Jianming and Zhang, Beiqi and Huang, Dan and Lo, David},
  year = {2025},
  month = feb,
  number = {arXiv:2502.06215},
  eprint = {2502.06215},
  primaryclass = {cs},
  publisher = {arXiv},
  doi = {10.48550/arXiv.2502.06215},
  archiveprefix = {arXiv}
}

@misc{si2025Design2Code,
  title = {{{Design2Code}}: {{Benchmarking Multimodal Code Generation}} for {{Automated Front-End Engineering}}},
  author = {Si, Chenglei and Zhang, Yanzhe and Li, Ryan and Yang, Zhengyuan and Liu, Ruibo and Yang, Diyi},


  year = {2025},
  month = feb,
  number = {arXiv:2403.03163},
  eprint = {2403.03163},
  primaryclass = {cs},
  publisher = {arXiv},
  doi = {10.48550/arXiv.2403.03163},
  archiveprefix = {arXiv}
}

@inproceedings{yang2024MatPlotAgent,
  title = {{{MatPlotAgent}}: {{Method}} and {{Evaluation}} for {{LLM-Based Agentic Scientific Data Visualization}}},
  booktitle = {Findings of the {{Association}} for {{Computational Linguistics}}: {{ACL}} 2024},
  author = {Yang, Zhiyu and Zhou, Zihan and Wang, Shuo and Cong, Xin and Han, Xu and Yan, Yukun and Liu, Zhenghao and Tan, Zhixing and Liu, Pengyuan and Yu, Dong and Liu, Zhiyuan and Shi, Xiaodong and Sun, Maosong},
  editor = {Ku, Lun-Wei and Martins, Andre and Srikumar, Vivek},
  year = {2024},
  month = aug,
  pages = {11789--11804},
  publisher = {Association for Computational Linguistics},
  address = {Bangkok, Thailand},
  doi = {10.18653/v1/2024.findings-acl.701},
}

@online{babe2023StudentEval,
  title = {StudentEval: A Benchmark of Student-Written Prompts for Large Language Models of Code},
  author = {Babe, Hannah McLean and Nguyen, Sydney and Zi, Yangtian and Guha, Arjun and Feldman, Molly Q. and Anderson, Carolyn Jane},
  date = {2023-06-07},
  eprint = {2306.04556},
  eprinttype = {arXiv},
  eprintclass = {cs},
  doi = {10.48550/arXiv.2306.04556},
  pubstate = {prepublished}
}

@misc{jin2024Evaluation,
  title = {An {{Evaluation}} of {{Requirements Modeling}} for {{Cyber-Physical Systems}} via {{LLMs}}},
  author = {Jin, Dongming and Zhao, Shengxin and Jin, Zhi and Chen, Xiaohong and Wang, Chunhui and Fang, Zheng and Xiao, Hongbin},
  year = {2024},
  month = aug,
  number = {arXiv:2408.02450},
  eprint = {2408.02450},
  primaryclass = {cs},
  publisher = {arXiv},
  doi = {10.48550/arXiv.2408.02450},
  archiveprefix = {arXiv}
}

@inproceedings{velickovic2022clrs,
  title = {The clrs algorithmic reasoning benchmark},
  booktitle = {International Conference on Machine Learning},
  author = {Veli\v ckovi\'c, Petar and Badia, Adri\`a Puigdom\`enech and Budden, David and Pascanu, Razvan and Banino, Andrea and Dashevskiy, Misha and Hadsell, Raia and Blundell, Charles},
  date = {2022},
  pages = {22084--22102},
  publisher = {PMLR},
  urldate = {2025-03-25}
}

@online{zheng2024Revolutionizing,
  title = {Revolutionizing Database Q\&A with Large Language Models: Comprehensive Benchmark and Evaluation},
  author = {Zheng, Yihang and Li, Bo and Lin, Zhenghao and Luo, Yi and Zhou, Xuanhe and Lin, Chen and Su, Jinsong and Li, Guoliang and Li, Shifu},
  date = {2024-12-06},
  eprint = {2409.04475},
  eprinttype = {arXiv},
  eprintclass = {cs},
  doi = {10.48550/arXiv.2409.04475},
  pubstate = {prepublished}
}

@inproceedings{ferrari2024Model,
  title = {Model Generation with LLMs: From Requirements to UML Sequence Diagrams},
  booktitle = {2024 IEEE 32nd International Requirements Engineering Conference Workshops (REW)},
  author = {Ferrari, Alessio and Abualhaija, Sallam and Arora, Chetan},
  date = {2024-06},
  pages = {291--300},
  issn = {2770-6834},
  doi = {10.1109/REW61692.2024.00044},
  eventtitle = {2024 IEEE 32nd International Requirements Engineering Conference Workshops (REW)}
}

@article{gong2024cosqa+,
  title={CoSQA+: Enhancing Code Search Dataset with Matching Code},
  author={Gong, Jing and Wu, Yanghui and Liang, Linxi and Zheng, Zibin and Wang, Yanlin},
  journal={arXiv preprint arXiv:2406.11589},
  year={2024}
}

@inproceedings{chen2024evaluating,
	title={Reasoning Runtime Behavior of a Program with LLM: How Far Are We?},
	author={Chen, Junkai and Pan, Zhiyuan and Hu, Xing and Li, Zhenhao and Li, Ge and Xia, Xin},
	booktitle={Proceedings of the IEEE/ACM 47th International Conference on Software Engineering},
	year={2025},
}

@misc{cao2025how,
      title={How Should We Build A Benchmark? Revisiting 274 Code-Related Benchmarks For LLMs}, 
      author={Jialun Cao and Yuk-Kit Chan and Zixuan Ling and Wenxuan Wang and Shuqing Li and Mingwei Liu and Ruixi Qiao and Yuting Han and Chaozheng Wang and Boxi Yu and Pinjia He and Shuai Wang and Zibin Zheng and Michael R. Lyu and Shing-Chi Cheung},
      year={2025},
      eprint={2501.10711},
      archivePrefix={arXiv},
      primaryClass={cs.SE}
}

@article{li2024Devbench,
  title = {Devbench: A comprehensive benchmark for software development},
  author = {Li, Bowen and Wu, Wenhan and Tang, Ziwei and Shi, Lin and Yang, John and Li, Jinyang and Yao, Shunyu and Qian, Chen and Hui, Binyuan and Zhang, Qicheng},
  date = {2024},
  journaltitle = {ArXiv Prepr. ArXiv240308604},
  volume = {3},
  eprint = {2403.08604},
  eprinttype = {arXiv},
  language = {en-US}
}

@article{liu2024evaluating,
  title={Evaluating language models for efficient code generation},
  author={Liu, Jiawei and Xie, Songrun and Wang, Junhao and Wei, Yuxiang and Ding, Yifeng and Zhang, Lingming},
  journal={arXiv preprint arXiv:2408.06450},
  year={2024}
}

@article{qiu2024efficient,
  title={How efficient is llm-generated code? a rigorous \& high-standard benchmark},
  author={Qiu, Ruizhong and Zeng, Weiliang Will and Ezick, James and Lott, Christopher and Tong, Hanghang},
  journal={arXiv preprint arXiv:2406.06647},
  year={2024}
}

@misc{juliet_java_1.3,
  author = {Juliet Java 1.3},
  title = {Juliet Java 1.3},
  year = {2017},
  howpublished = {\url{https://samate.nist.gov/SARD/test-suites/111}},
  note = {Accessed: 12-12-2023}
}

@online{du2025FairCoder,
  title = {FairCoder: Evaluating Social Bias of LLMs in Code Generation},
  author = {Du, Yongkang and Huang, Jen-tse and Zhao, Jieyu and Lin, Lu},
  date = {2025-04-01},
  eprint = {2501.05396},
  eprinttype = {arXiv},
  eprintclass = {cs},
  doi = {10.48550/arXiv.2501.05396},
  language = {en-US},
  pubstate = {prepublished}
}

@article{ling2025Bias,
  title = {Bias Unveiled: Investigating Social Bias in LLM-Generated Code},
  author = {Ling, Lin and Rabbi, Fazle and Wang, Song and Yang, Jinqiu},
  date = {2025-04-11},
  journaltitle = {Proceedings of the AAAI Conference on Artificial Intelligence},
  shortjournal = {Proc. AAAI Conf. Artif. Intell.},
  volume = {39},
  number = {26},
  pages = {27491--27499},
  issn = {2374-3468},
  doi = {10.1609/aaai.v39i26.34961},
  issue = {26},
  language = {en}
}

@ARTICLE{IEEEStd,
  author={},
  journal={IEEE Std 830-1998}, 
  title={IEEE Recommended Practice for Software Requirements Specifications}, 
  year={1998},
  volume={},
  number={},
  pages={1-40},
  doi={10.1109/IEEESTD.1998.88286}}

@article{zhao2023survey,
  title={A survey of large language models},
  author={Zhao, Wayne Xin and Zhou, Kun and Li, Junyi and Tang, Tianyi and Wang, Xiaolei and Hou, Yupeng and Min, Yingqian and Zhang, Beichen and Zhang, Junjie and Dong, Zican and others},
  journal={arXiv preprint arXiv:2303.18223},
  volume={1},
  number={2},
  year={2023}
}

@article{gonzalez2024towards,
  title={Towards a Classification of Open-Source ML Models and Datasets for Software Engineering},
  author={Gonz{\'a}lez, Alexandra and Franch, Xavier and Lo, David and Mart{\'\i}nez-Fern{\'a}ndez, Silverio},
  journal={arXiv preprint arXiv:2411.09683},
  year={2024}
}

@article{yang2024ecosystem,
  title={Ecosystem of large language models for code},
  author={Yang, Zhou and Shi, Jieke and Devanbu, Premkumar and Lo, David},
  journal={arXiv preprint arXiv:2405.16746},
  year={2024}
}

@article{hossain2024togll,
  title={Togll: Correct and strong test oracle generation with llms},
  author={Hossain, Soneya Binta and Dwyer, Matthew},
  journal={arXiv preprint arXiv:2405.03786},
  year={2024}
}

@misc{github2023report,
  author = {Eirini Kalliamvakou},
  title = {Research: Quantifying GitHub Copilot’s Impact on Developer Productivity and Happiness},
  year = {2023},
}

@inproceedings{chen2023teaching,
  title={Teaching Large Language Models to Self-Debug},
  author={Chen, Xinyun and Lin, Maxwell and Sch{\"a}rli, Nathanael and Zhou, Denny},
  booktitle={The Twelfth International Conference on Learning Representations},
  year={2023}
}

@inproceedings{qian2024chatdev,
  title={ChatDev: Communicative Agents for Software Development},
  author={Qian, Chen and Liu, Wei and Liu, Hongzhang and Chen, Nuo and Dang, Yufan and Li, Jiahao and Yang, Cheng and Chen, Weize and Su, Yusheng and Cong, Xin and others},
  booktitle={Proceedings of the 62nd Annual Meeting of the Association for Computational Linguistics (Volume 1: Long Papers)},
  pages={15174--15186},
  year={2024}
}

@inproceedings{luowizardcoder,
  title={WizardCoder: Empowering Code Large Language Models with Evol-Instruct},
  author={Luo, Ziyang and Xu, Can and Zhao, Pu and Sun, Qingfeng and Geng, Xiubo and Hu, Wenxiang and Tao, Chongyang and Ma, Jing and Lin, Qingwei and Jiang, Daxin},
  booktitle={The Twelfth International Conference on Learning Representations},
  year={2023}
}

@article{zhang2024codeagent,
  title={Codeagent: Enhancing code generation with tool-integrated agent systems for real-world repo-level coding challenges},
  author={Zhang, Kechi and Li, Jia and Li, Ge and Shi, Xianjie and Jin, Zhi},
  journal={arXiv preprint arXiv:2401.07339},
  year={2024}
}

@article{liu2025projecteval,
  title={ProjectEval: A Benchmark for Programming Agents Automated Evaluation on Project-Level Code Generation},
  author={Liu, Kaiyuan and Pan, Youcheng and Li, Jing and He, Daojing and Xiang, Yang and Du, Yexing and Gao, Tianrun},
  journal={arXiv preprint arXiv:2503.07010},
  year={2025}
}

@article{rashid2025swe,
  title={SWE-PolyBench: A multi-language benchmark for repository level evaluation of coding agents},
  author={Rashid, Muhammad Shihab and Bock, Christian and Zhuang, Yuan and Buccholz, Alexander and Esler, Tim and Valentin, Simon and Franceschi, Luca and Wistuba, Martin and Sivaprasad, Prabhu Teja and Kim, Woo Jung and others},
  journal={arXiv preprint arXiv:2504.08703},
  year={2025}
}

@article{vergopoulos2025automated,
  title={Automated Benchmark Generation for Repository-Level Coding Tasks},
  author={Vergopoulos, Konstantinos and M{\"u}ller, Mark Niklas and Vechev, Martin},
  journal={arXiv preprint arXiv:2503.07701},
  year={2025}
}

@misc{cursor2025,
  author       = {Anysphere Inc.},
  title        = {Cursor - The AI Code Editor},
  year         = {2025},
  url          = {https://www.cursor.com},
  note         = {Accessed: 2025-05-03}
}

@article{zhang2025unveiling,
  title={Unveiling Provider Bias in Large Language Models for Code Generation},
  author={Zhang, Xiaoyu and Zhai, Juan and Ma, Shiqing and Bao, Qingshuang and Jiang, Weipeng and Shen, Chao and Liu, Yang},
  journal={arXiv preprint arXiv:2501.07849},
  year={2025}
}

@article{bruel2019formality,
  title={Formality in software requirements},
  author={Bruel, JEAN-MICHEL and Ebersold, Sophie and Galinier, Florian and Naumchev, ALEXANDR and Mazzara, Manuel and Meyer, Bertrand},
  journal={arXiv preprint arXiv:1911.02564},
  year={2019}
}

@article{nahri2025extracting,
  title={Extracting Structured Requirements from Unstructured Building Technical Specifications for Building Information Modeling},
  author={Nahri, Insaf and Pinqui{\'e}, Romain and V{\'e}ron, Philippe and Bus, Nicolas and Thorel, Mathieu},
  journal={arXiv preprint arXiv:2508.13833},
  year={2025}
}

@article{wu2024versicode,
  title={Versicode: Towards version-controllable code generation},
  author={Wu, Tongtong and Wu, Weigang and Wang, Xingyu and Xu, Kang and Ma, Suyu and Jiang, Bo and Yang, Ping and Xing, Zhenchang and Li, Yuan-Fang and Haffari, Gholamreza},
  journal={arXiv preprint arXiv:2406.07411},
  year={2024}
}

@article{zheng2024opencodeinterpreter,
  title={Opencodeinterpreter: Integrating code generation with execution and refinement},
  author={Zheng, Tianyu and Zhang, Ge and Shen, Tianhao and Liu, Xueling and Lin, Bill Yuchen and Fu, Jie and Chen, Wenhu and Yue, Xiang},
  journal={arXiv preprint arXiv:2402.14658},
  year={2024}
}

@inproceedings{van2025blade,
  title={BLADE: Benchmark suite for LLM-driven Automated Design and Evolution of iterative optimisation heuristics},
  author={van Stein, Niki and V. Kononova, Anna and Yin, Haoran and B{\"a}ck, Thomas},
  booktitle={Proceedings of the Genetic and Evolutionary Computation Conference Companion},
  pages={2336--2344},
  year={2025}
}

@article{falessi2010applying,
  title={Applying empirical software engineering to software architecture: challenges and lessons learned},
  author={Falessi, Davide and Babar, Muhammad Ali and Cantone, Giovanni and Kruchten, Philippe},
  journal={Empirical Software Engineering},
  volume={15},
  number={3},
  pages={250--276},
  year={2010},
  publisher={Springer}
}

@article{kitchenham2004procedures,
  title={Procedures for performing systematic reviews},
  author={Kitchenham, Barbara and others},
  journal={Keele, UK, Keele University},
  volume={33},
  number={2004},
  pages={1--26},
  year={2004}
}

@article{thai2025swe,
  title={SWE-EVO: Benchmarking Coding Agents in Long-Horizon Software Evolution Scenarios},
  author={Thai, Minh VT and Le, Tue and Manh, Dung Nguyen and Nhat, Huy Phan and Bui, Nghi DQ},
  journal={arXiv preprint arXiv:2512.18470},
  year={2025}
}

@inproceedings{liu2024agentbench,
  title={AgentBench: Evaluating LLMs as Agents},
  author={Liu, Xiao and Yu, Hao and Zhang, Hanchen and Xu, Yifan and Lei, Xuanyu and Lai, Hanyu and Gu, Yu and Ding, Hangliang and Men, Kaiwen and Yang, Kejuan and others},
  booktitle={ICLR},
  year={2024}
}

@article{hu2025assessing,
  title={Assessing and advancing benchmarks for evaluating large language models in software engineering tasks},
  author={Hu, Xing and Niu, Feifei and Chen, Junkai and Zhou, Xin and Zhang, Junwei and He, Junda and Xia, Xin and Lo, David},
  journal={ACM Transactions on Software Engineering and Methodology},
  year={2025},
  publisher={ACM New York, NY}
}

@book{iso25010,
  author    = {{ISO/IEC}},
  title     = {Systems and software engineering -- Systems and software Quality Requirements and Evaluation ({SQuaRE}) -- System and software quality models},
  year      = {2011},
  publisher = {International Organization for Standardization},
  address   = {Geneva, Switzerland}
}

@article{zhou2023don,
  title={Don't make your llm an evaluation benchmark cheater},
  author={Zhou, Kun and Zhu, Yutao and Chen, Zhipeng and Chen, Wentong and Zhao, Wayne Xin and Chen, Xu and Lin, Yankai and Wen, Ji-Rong and Han, Jiawei},
  journal={arXiv preprint arXiv:2311.01964},
  year={2023}
}

@article{dong2025survey,
  title={A survey on code generation with llm-based agents},
  author={Dong, Yihong and Jiang, Xue and Qian, Jiaru and Wang, Tian and Zhang, Kechi and Jin, Zhi and Li, Ge},
  journal={arXiv preprint arXiv:2508.00083},
  year={2025}
}

@inproceedings{royce1987managing,
  title={Managing the development of large software systems: concepts and techniques},
  author={Royce, Winston W},
  booktitle={Proceedings of the 9th international conference on Software Engineering},
  pages={328--338},
  year={1987}
}

@article{yang2025code,
  title={From Code Foundation Models to Agents and Applications: A Comprehensive Survey and Practical Guide to Code Intelligence},
  author={Yang, Jian and Liu, Xianglong and Lv, Weifeng and Deng, Ken and Guo, Shawn and Jing, Lin and Li, Yizhi and Liu, Shark and Luo, Xianzhen and Luo, Yuyu and others},
  journal={arXiv preprint arXiv:2511.18538},
  year={2025}
}

\clearpage

\end{document}